\documentclass{aa}  

\usepackage{graphicx}
\usepackage{txfonts}
\usepackage[dvipsnames]{xcolor}
\usepackage[colorlinks=true,breaklinks,citecolor=blue]{hyperref}
\newcommand\mycom[2]{\genfrac{}{}{0pt}{}{#1}{#2}}

\usepackage{upgreek}

\newcommand{\be}{\begin{equation}}
\newcommand{\ee}{\end{equation}}
\newcommand{\bea}{\begin{eqnarray}}
\newcommand{\eea}{\end{eqnarray}}
\newcommand{\nn}{\nonumber}

\begin{document}

   \title{
   Simulating realistic self-interacting dark matter models including small and large-angle scattering
   }
   \titlerunning{Simulating realistic SIDM models}

   \author{Cenanda Arido\inst{\ref{inst:tum}}
          \and
          Moritz S. Fischer\inst{\ref{inst:usm},\ref{inst:origins}}
          \and 
          Mathias Garny\inst{\ref{inst:tum}}
          }
    \authorrunning{C.\ Arido, M.\ S.\ Fischer, M.\ Garny}

   \institute{Physik Department T31, Technische Universit\"at M\"unchen
              James-Franck-Stra\ss{}e 1, D-85748 Garching, Germany\label{inst:tum}\\
              \email{cenanda.arido@tum.de}\\
              \email{mathias.garny@tum.de}
         \and
             Fakult\"at für Physik, Universit\"ats-Sternwarte, Ludwig-Maximilians-Universit\"at M\"unchen, 
             Scheinerstr. 1, D-81679 München, Germany\label{inst:usm}\\
             \email{mfischer@usm.lmu.de}
         \and
            Excellence Cluster ORIGINS, Boltzmannstrasse 2, D-85748 Garching, Germany\label{inst:origins}
             }

   \date{Received  9 October, 2024  / Accepted 8 January, 2025}
 
  \abstract 
   {
   Dark matter (DM) self-interactions alter matter distribution on galactic scales and alleviate tensions with observations. A feature of the self-interaction cross section is its angular dependence, which influences offsets between galaxies and DM halos in merging galaxy clusters. While algorithms for modelling mostly forward-dominated or mostly large-angle scatterings exist, incorporating realistic angular dependencies within $N$-body simulations remains challenging.  
   }
   {To efficiently simulate models with a realistic angle dependence, such as light mediator models, we developed, validated, and applied a novel method.
   }
   {We combined existing approaches to describe small- and large-angle scattering regimes within a hybrid scheme. Below a critical angle, the scheme uses the effective description of small-angle scattering via a drag force combined with transverse momentum diffusion, while above the angle, it samples the dependence explicitly.
   }
   {
   We first verified the scheme using a test set-up with known analytical solutions, and we checked that our results are insensitive to the choice of the critical angle within an expected range. Next, we demonstrated that our scheme speeds up the computations by multiple orders of magnitude for realistic light mediator models. Finally, we applied the method to galaxy cluster mergers. We discuss the sensitivity of the offset between galaxies and DM to the angle dependence of the cross section. Our scheme ensures accurate offsets for mediator mass $m_\phi$ and DM mass $m_\chi$ within the range $0.1v/c\lesssim m_\phi/m_\chi\lesssim v/c$, while for larger (smaller) mass ratios, the offsets obtained for isotropic (forward-dominated) self-scattering are approached. Here, $v$ is the typical velocity scale. Equivalently, the upper condition can be expressed as $1.1\lesssim \sigma_{\rm tot}/\sigma_{\mathrm{\widetilde{T}}}\lesssim 10$ for the ratio of the total and momentum transfer cross sections, with the ratio being $1$ ($\infty$) in the isotropic (forward-dominated) limits. 
   }
   {}


   \maketitle
%
\section{Introduction}

While astrophysical observations ranging from galactic to cosmological scales support the existence of dark matter (DM) via its gravitational impact, it is plausible that DM also features non-gravitational interactions as any other measured particle species does. Extensive searches for non-gravitational interactions between DM and visible matter have not found an unambiguous signal so far, and such work has pushed the limits even beyond the scale of weak interactions in many cases. In contrast, interactions of DM with itself with a sizeable strength, comparable to that known from the strong interaction, remain a viable possibility that is testable by searching for its imprints on the dynamics of galaxies and galaxy clusters. 

Elastic $2\to 2$ DM self-interactions have been proposed by~\cite{Spergel_1999} in the context of the core-cusp problem, with early $N$-body simulations by~\cite{Burkert_2000} showing core formation followed by gravothermal collapse. Self-interacting DM (SIDM) has also been proposed for addressing other small-scale puzzles, such as the abundance and properties of satellite galaxy populations or the variability observed in galaxy rotation curves (see~\cite{Bullock_2017,Tulin_2018, Adhikari_2022} for reviews).

While SIDM on galactic scales typically requires a cross section (per DM mass) of the order of $\sigma/m\gtrsim 1\text{cm}^2\text{g}^{-1}$ to have a sizeable impact, galaxy cluster observations have been argued to require upper limits reaching down to $0.1\text{cm}^2\text{g}^{-1}$~\citep[see e.g.][]{Elbert_2016, Sagunski_2020, Andrade_2021, Despali_2022, Eckert_2022, Zhang_2024}. Given the very different typical velocities on cluster and galactic scales, this suggests a velocity-dependent cross section \citep[see for example][]{Kaplinghat_2015} such as occurs, for example, if the interaction is mediated by an exchange particle with mass $m_\phi$ that is light compared to the DM mass $m_\chi$~\citep{Feng_2009, Buckley_2010, Loeb_2011, Tulin_2013}.
Within this class of models, the scattering cross section also features a pronounced angular dependence, being strongly forward-dominated in the limit $m_\phi\to 0$, analogous to Rutherford scattering.

For isolated quasi-stationary halos, the impact of angular dependence on the density profile has been argued to be approximately captured by mapping simulations assuming isotropic scattering onto models with anisotropic differential cross sections by matching a suitable angle-averaged effective cross section, namely the viscosity cross section \citep[e.g.][]{Colquhoun_2020, Yang_2022D, Sabarish_2024}. 
However, for systems that are not quasi-stationary, such as during the evolution of satellite populations \citep{Kahlhoefer_2015, Fischer_2022, Fischer_2024a, Ragagnin_2024} or when approaching gravothermal collapse, and especially for strongly non-stationary processes, such as galaxy cluster mergers \citep[e.g.][]{Kahlhoefer_2014, Robertson_2017b, Fischer_2021a, Fischer_2021b, Fischer_2023, Sabarish_2024}, it is a priori less clear how they are affected by (strongly) anisotropic scattering. 

These reasons motivated us to develop efficient numerical algorithms for incorporating differential cross sections featuring a pronounced angular dependence, such as for light mediator models, in order to assess the accuracy of (and possibly refine) commonly employed approximate `mapping' prescriptions for a given observable and potentially identify new features that are specific to a given angular dependence. In this work, we develop such an algorithm, validate it, and apply it to study galaxy cluster mergers with angle-dependent cross sections that are characteristic for light mediator models.

Our work is based on a combination of two existing approaches, including one that is suitable for relatively rare individual self-scattering events, which feature large deflection angles \citep[dubbed rare SIDM or `rSIDM'; see Sect.~\ref{sec: Theoretical basis} and e.g.][for details]{Koda_2011, Vogelsberger_2012, Rocha_2013, Fry_2015, Robertson_2017b, Banerjee_2020}. In this first approach, SIDM can be modelled by collisions of $N$-body particles, with scattering angles sampled from the differential cross section. While being the most straightforward and direct, this method has the drawback of becoming prohibitively inefficient numerically for models in which the scattering rate is large but the momentum transfer in each individual scattering event is small.

The second approach is tailored precisely to efficiently describe such very frequent and strongly forward-dominated scatterings (dubbed frequent SIDM or `fSIDM'). The fSIDM approach formally corresponds to the limit in which the momentum transfer cross section is kept constant but the typical scattering angle approaches zero. Such frequent interactions are well known in the context of Coulomb interactions of energetic particles traversing a medium and can effectively be captured by a drag force complemented with transverse momentum diffusion. A drag force description has been derived by \cite{Kahlhoefer_2014} and applied to merging galaxy clusters. Building on that work, an algorithm for the fSIDM limit has been proposed in the context of $N$-body simulations of SIDM by \cite{Fischer_2021a} (respecting energy and momentum conservation). Their implementation has been developed further by \cite{Fischer_2021b, Fischer_2022, Fischer_2024a} and used by \cite{Fischer_2023, Fischer_2024b, Fischer_2024c, Sabarish_2024, Ragagnin_2024} to study the SIDM halo evolution, galaxy cluster merger dynamics, the evolution of cluster satellites, and dynamical friction as compared to the usually considered opposite limit of isotropic rSIDM. However, as mentioned above, realistic models feature both small- and large-angle scattering with a roughly comparable impact over a wide range of parameter space, and thus neither the fSIDM nor rSIDM scheme can be considered to describe viable particle models in general. 

Therefore, in this work we develop a new hybrid method that we dub the `hSIDM' scheme, and it is designed to be able to accurately describe (in principle) arbitrary differential cross sections. In particular, the hSIDM method can be used to efficiently simulate models featuring a strongly forward-dominated enhancement at low deflections angles as well as a given contribution with large-angle scattering. The hSIDM method is therefore particularly suited to study light mediator models, but it can easily be extended to other scenarios that have been proposed in the literature \citep[see e.g.][for an overview]{Tulin_2018}. The hSIDM algorithm takes advantage of the numerically much more efficient fSIDM approach for small scattering angles below a certain critical angle $\theta_{\mathrm{c}}$ and is complemented by the more conventional rSIDM treatment of large-angle scatterings.

After having developed the hSIDM scheme, we applied it to merging galaxy clusters. Systems such as the `Bullet Cluster' \citep[e.g.][]{Springel_2007, Mastropietro_2008, Lage_2014} or `El Gordo' \citep[e.g.][]{Donnert_2014, Molnar_2015, Zhang_2015, Kim_2021, Asencio_2020, Asencio_2023} have been studied and used to constrain DM self-interactions \citep[e.g.][]{Valdarnini_2024}. Besides other phenomena, such as the oscillations of the brightest cluster galaxy (BCG) at the late stages of the merger \citep[e.g.][]{Harvey_2019, Cross_2024}, offsets between the distribution of the DM and the cluster galaxies mainly shortly after the first pericentre passage have received a lot of attention \citep[e.g.][]{Harvey_2015, Robertson_2017a, Sirks_2024}. Theoretical studies to model the offsets that arise from the self-interactions effectively decelerating the two DM halos when they pass through each other while the galaxies are only indirectly affected via gravity have been conducted to gain insights into their evolution and allow for constraints from observed systems to be inferred. Offsets between the galaxy and DM distributions are measured employing strong gravitational lensing to infer the total mass distribution. These measurements are challenging, and previous claims of large offsets have been questioned \citep[e.g.][]{Bradac_2008, Dawson_2012, Dawson_2013, Jee_2014, Jee_2015, Harvey_2017, Peel_2017, Taylor_2017, Wittman_2018, Wittman_2023}. Until today, observed systems show no clear deviation from collisionless DM, and mainly upper bounds on the self-interaction cross section from DM-galaxy offsets exist.

This work is structured as follows. In Sect.~\ref{sec: Theoretical basis}, we first introduce the various exemplary differential cross sections considered in this work and review the rSIDM and fSIDM schemes for describing angle-dependent self-scattering. We introduce the new hSIDM scheme in Sect.~\ref{sec:hSIDM} and furthermore discuss an analytically solvable test set-up that we use for validation later on. The numerical implementation is discussed in Sect.~\ref{sec: implementation}, and we present our validation checks in Sect.~\ref{sec: Deflection test}. The hSIDM scheme is used to simulate galaxy cluster mergers for SIDM with angle dependence as predicted by light mediator models in Sect.~\ref{sec: Merger simulation and analysis}. Our main results for DM-galaxy offsets and their model-dependence is discussed in Sect.~\ref{sec:offsets}. Finally, we summarise and conclude in Sect.~\ref{sec: Conclusions}. Additional
information can be found in the Appendices.

\section{Review of angle dependence in self-interacting dark matter}
\label{sec: Theoretical basis}

In this section, we first briefly review the angle dependence of typical differential cross sections relevant for light mediator models of SIDM as well as various definitions of angle-averaged cross sections considered in this context. We then review the existing rSIDM and fSIDM approaches for describing rare large-angle and frequent small-angle scatterings, respectively, on which the hybrid scheme introduced in this work is based.

\subsection{Differential scattering cross sections}
\label{subsec: Scattering differential cross sections}

While the algorithm introduced in this work is capable of describing SIDM with an arbitrary differential cross section, we start by briefly reviewing some typical cases, that we  also use below for illustration.

\begin{figure}
    \centering
    \includegraphics[width=\columnwidth]{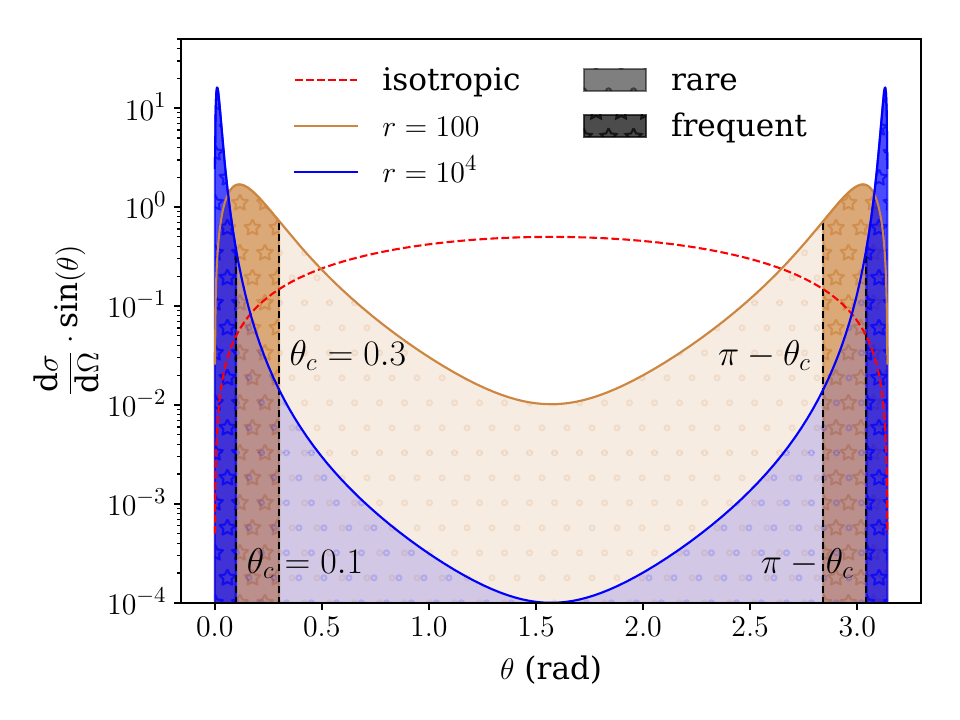}
    \caption{Illustration of the differential cross section for self-scattering of identical particles via a light mediator (referred to as M{\o}ller scattering in this work) for two values of the anisotropy parameter $r$ defined in Eq.~\eqref{eq:rdef} (solid lines).
    Noting the logarithmic $y$-axis, the strong enhancement for $\theta\to0$ and $\theta\to\uppi$ can be seen, and it becomes more pronounced the larger the $r$ (i.e.\ the lighter the mediator mass). We note that the isotropic case (red dashed) is not flat due to the scaling of the cross section with $\sin(\theta)$ (such that the lines represent the scattering rate within a given range $(\theta,\theta+\mathrm{d}\theta)$ of the azimuthal angle, for any value of the polar angle).
    We also illustrate the division into small and large-angle scattering by a critical angle $\theta_{\mathrm{c}}$, for two representative values, exemplifying the hybrid approach studied in this work.}
    \label{fig: Moeller differential cross section}
\end{figure}

Light mediator models are characterised by scattering via a Yukawa interaction, for which DM particles of mass $m_\chi$ scatter via the exchange of a mediator with mass $m_\phi$. One can further discriminate the cases in which the two incoming DM particles are identical (we refer to this case as `M{\o}ller scattering' in analogy to $e^-e^-$ scattering, noting however that the mediator particle is massive here), or distinguishable (being e.g.\ a particle and an antiparticle, referred to as `Rutherford scattering', but again representing the case of a massive mediator in this work). In the non-relativistic limit and in Born approximation, M{\o}ller scattering receives contributions from $t$- and $u$-channel diagrams, yielding
\begin{equation}
\label{eq: Moeller differential cross section}
    \begin{split}
    \left(\frac{\mathrm{d}\sigma}{\mathrm{d}\Omega}\right)_{\mathrm{M}} = \frac{1}{2} \frac{\sigma_{0}}{4 \uppi} &\left[ \frac{1}{\left(1+ r \sin^2\left( \frac{\theta}{2} \right)\right)^2} + \frac{1}{\left(1+ r \cos^2\left( \frac{\theta}{2} \right)\right)^2} \right. \\
    & \left. - \frac{1}{\left(1+ r \sin^2\left( \frac{\theta}{2} \right)\right)\left(1+ r \cos^2\left( \frac{\theta}{2} \right)\right)} \right],\\
\end{split}
\end{equation}
with the factor of one-half accounting for the double counting of the identical particles, while only the $t$-channel process contributes to Rutherford scattering, giving
\begin{equation}
\label{eq: Rutherford differential cross section}
\begin{split}
    \left(\frac{\mathrm{d}\sigma}{\mathrm{d}\Omega}\right)_{\mathrm{R}} = \frac{\sigma_{0}}{4 \uppi} & \frac{1}{\left(1+ r \sin^2\left( \frac{\theta}{2} \right)\right)^2}.\\
\end{split}
\end{equation}
Following the notations used in~\citet{Girmohanta_2022, Robertson_2017b}, we introduced a parameter $\sigma_0$ that is related to the Yukawa coupling strength $\alpha_\chi$ via $\sigma_0=\alpha_\chi^2m_\chi^2/m_\phi^4$. More importantly, the angle dependence is related to the parameter
\be\label{eq:rdef}
  r \equiv \left(\frac{\beta_{\mathrm{rel}} m_{\chi}}{m_{\phi}}\right)^2\,,
\ee
where $\beta_{\mathrm{rel}} = v_{\mathrm{rel}}/c$ is the relative velocity of the incoming particles. The value of $r$ characterises the dependence of the cross section on the deflection angle $\theta$, with the isotropic limit being approached for $r\ll 1$, while scattering becomes strongly forward-dominated (Coulomb-like) for large values of $r$. For illustration, we show the angle dependence of the M{\o}ller cross section in Fig.~\ref{fig: Moeller differential cross section} for $r=100$ and $10^4$, respectively, as well as an isotropic cross section for comparison. We emphasise the pronounced peaks towards small scattering angles (note the logarithmic $y$-axis). Since
the M{\o}ller cross section is symmetric under $\theta\to\uppi-\theta$, it is also enhanced for $\theta\to\uppi$, due to the fact that both scattering particles are indistinguishable.

Additionally, for validation purposes in Sect.~\ref{subsec: fixed-angle scattering implementation} we also consider a somewhat artificial case for which all scatterings proceed with a single, fixed deflection angle $\theta_{0}$, being formally given by a differential cross section
\begin{equation}
\label{eq: Delta differential cross section}
    \left(\frac{\mathrm{d}\sigma}{\mathrm{d}\Omega}\right)_{\mathrm{D}} = \frac{\sigma_{0}}{2 \uppi} \delta(\cos(\theta) - \cos(\theta_{0}))\,, 
\end{equation}
where $\delta(x)$ stands for the Dirac distribution.

\subsection{Angle-averaged cross sections}
\label{subsec: Angle dependent scattering-cross section}

In order to estimate the impact of DM self-scattering on various observables of interest, it is useful to define integrated, angle-averaged cross sections with specific weighting factors. These may also be used to map results obtained for a given angle dependence (e.g.\ isotropic) to other models, if the effect of interest is mainly sensitive to a particular averaged cross section. Here we briefly review the angle-averages that are often considered in the context of SIDM. We use them below to (i) characterise the relative strength of small- and large-angle scatterings in a given model,  and (ii) investigate in how far our results taking the precise angle dependence into account can indeed be mapped among models featuring distinct differential cross sections, but agreeing in the value of a certain angle-averaged cross section.

The cross section averaged over the complete range of deflection angles can in general be written as
\begin{equation}\label{eq:sigma_def}
    \sigma_{X} \equiv c_X \int \mathrm{d}\Omega  \,g_{X}(\theta) \,\frac{\mathrm{d}\sigma}{\mathrm{d}\Omega}
    = 2\uppi c_X \int_0^\uppi \mathrm{d}\theta\, g_{X}(\theta) \,\frac{\mathrm{d}\sigma}{\mathrm{d}\Omega}\,\sin(\theta)\,, 
\end{equation}
where we assumed independence of the polar angle in the second expression (being true in all cases we consider).
Here $g_X(\theta)$ is a weighting function that differs for each type of angle-average labelled by the index $X$, and $c_X$ is a normalisation constant chosen such that $c_X\int \mathrm{d}\Omega \,g_X(\theta)=4\uppi$. This ensures that all angle-averages agree with each other for an isotropic cross section.\footnote{We note that this normalisation condition differs from the definitions used in other works. For example our definition of the momentum transfer cross section differs by a factor of two from the one used by \cite{Fischer_2021a, Fischer_2021b, Fischer_2022, Fischer_2023}. However, our normalisation is for example in line with the work by \cite{Yang_2022D}.}

We consider the following cases:
\begin{enumerate}
    \item \textbf{Total cross section}: Determines the total scattering rate.
    \be
      g_{\rm tot}(\theta)\equiv 1,\quad c_{\rm tot}=1\,.
    \ee
    \item \textbf{Transfer cross section}: Weighting with momentum transfer for scattering of distinguishable particles.
    \be
      g_{\mathrm{T}}(\theta)\equiv 1-\cos(\theta),\quad c_{\mathrm{T}}=1\,.
    \ee
    \item \textbf{Modified transfer cross section}: Weighting with momentum transfer for scattering of identical particles.
    \be
      g_{\mathrm{\widetilde{T}}}(\theta)\equiv 1-|\cos(\theta)|,\quad c_{\mathrm{\widetilde{T}}}=2\,.
    \ee
    \item \textbf{Viscosity cross section}: Weighting with energy transfer (for both distinguishable and identical particles).
    \be
      g_{\mathrm{V}}(\theta)\equiv 1-\cos^2(\theta),\quad c_{\mathrm{V}}=\frac32\,.
    \ee
\end{enumerate}

The viscosity cross sections has been argued to be a good proxy for the impact of self-scattering on the distribution of DM in isolated halos, including core formation and collapse \citep[e.g.][]{Colquhoun_2020, Yang_2022D, Sabarish_2024}. The question whether an average cross section is a good description in mergers has also been investigated \citep[e.g.][]{Kahlhoefer_2014, Robertson_2017b, Fischer_2021a,Fischer_2021b}. Below we  scrutinise these findings for the evolution of DM and galaxy positions as well as their offset in cluster mergers, by comparing to a treatment capturing the exact angle dependence. The total cross section is mainly relevant for estimating the numerical effort (within the rSIDM scheme, see below) since it characterises the total scattering rate, irrespective of whether a significant amount of momentum is transferred between the scattering partners.

For reference, we give the angle-averages for the two extreme cases of {\it isotropic} scattering and {\it strongly forward-dominated} scattering, respectively,
\bea
  \sigma_{\rm tot} = \sigma_{\mathrm{T}} = \sigma_{\mathrm{\widetilde{T}}} = \sigma_{\mathrm{V}}\,,&& ({\rm isotropic})\nn\\
  \sigma_{\rm tot} \gg  \sigma_{\mathrm{T}} = \frac13\sigma_{\mathrm{V}}\,, && ({\rm forward-dom., \ distinguishable})\nn\\
  \sigma_{\rm tot} \gg  \sigma_{\mathrm{\widetilde{T}}} = \frac23\sigma_{\mathrm{V}}\,, && ({\rm forward-dom., \ identical})
\eea
where the last two lines refer to the cases of either distinguishable or identical particles, with `forward-dominated' meaning a strong enhancement for $\theta\to 0$ as well as $\theta\to\uppi$ in the latter case.

\subsection{Existing approaches for angle-dependent self-interacting dark matter}
\label{subsec: Existing approaches for angular regimes}

Here we briefly review the physical basis of the approaches that have been used to incorporate angle dependence into SIDM simulations. Technical details on the numerical implementation are provided in Sect.~\ref{sec: implementation}.

As mentioned above, we consider two existing approaches, both of which are combined in this work in order to be able to describe models with arbitrary differential cross sections. However, when used by themselves, each of the two methods is limited to a particular class of models.

The first method treats angle dependence explicitly for each scattering event, by sampling the deflection angle from a random distribution that follows from the differential cross section \citep[see e.g.][]{Robertson_2017b, Banerjee_2020}. This strategy is in practice numerically feasible for models for which an ${\cal O}(1)$ amount of energy and momentum is transferred in each scattering, thus being limited to large-angle scatterings, see~\cite{Robertson_2017b}. For these types of models the total and (modified) transfer cross sections are of comparable size, $\sigma_{\rm tot}/\sigma_{\mathrm{\widetilde{T}}} \sim {\cal O}(1)$ , which implies that even relatively `rare' scattering events can have a sizeable impact. We thus refer to this case as {\bf rare self-interacting dark matter (rSIDM)}.

The second method employs an effective treatment that is applicable if DM scatters very frequently, but the momentum transfer in each scattering is small, dubbed {\bf frequent self-interacting dark matter (fSIDM)} following~\cite{Fischer_2021a}. By itself, this method becomes exact for models approaching the formal limit $\sigma_{\rm tot}/\sigma_{\mathrm{\widetilde{T}}} \to\infty$, that is, very strongly forward-dominated scattering only. 

In this work we combine both methods, using them within the range of angles for which they work best. Before discussing this hybrid approach, we briefly review the physical basis of the fSIDM scheme \citep[see][for more details]{Kahlhoefer_2014, Kummer_2018}.
In analogy to the seminal description of Brownian motion
by~\cite{Einstein_1905} and \cite{Smoluchowski_1906}, the collective effect of frequent small-angle scattering can be described by a drag force that acts along the direction of motion (more precisely the relative velocity vector) as well as a stochastic force leading to diffusive motion. The drag force can be related to the (modified) transfer cross section, $F_{\rm drag}\propto \sigma_{\mathrm{\widetilde{T}}}$ (see Sect.~\ref{sec: implementation} for details), while the stochastic force can be included as a kick in a random direction transverse to the relative velocity vector, with magnitude consistent with momentum and energy conservation.

This set-up has been implemented by \cite{Fischer_2021a} and verified with multiple test problems including one for which the solution is known from~\cite{Moliere_1948}, being essentially a Gaussian spreading of an initially collimated beam traversing a medium of target particles.
In this work we generalise this test problem to models with arbitrary differential cross section in order to validate the hybrid scheme based on combining the rSIDM and fSIDM methods, which we turn to next.

\section{Hybrid scheme for angle dependence}
\label{sec:hSIDM}

In this section, we first present an algorithm that allows us to simulate DM self-scattering with arbitrary differential cross sections efficiently, dubbed {\bf hybrid self-interacting dark matter (hSIDM)}. Next, we review a test set-up featuring an exact analytical solution for any given differential cross section and then use it to assess under which conditions and for which choices of various technical parameters the hSIDM approach is expected to be valid.

\subsection{Hybrid approach to angle-dependent self-interacting dark matter}
\label{subsec: Hybrid self-interaction}

Here we introduce a new hybrid approach for taking the dependence of the DM self-interaction cross section on the deflection angle into account, that we refer to as hSIDM as stated above. The central idea is to combine the methods used previously to describe models with either only large (rSIDM) or only forward-dominated (fSIDM) scatterings, allowing for both regimes to be described efficiently within a single set-up. Since typical (e.g.\ light mediator) models can feature a very strong enhancement of the total scattering rate at small angles, (see e.g.\ Fig.~\ref{fig: Moeller differential cross section}) but also predict non-negligible large-angle scattering, it is necessary to capture both contributions in order to obtain accurate predictions for a given set of model parameters.

The central idea of the hybrid scheme is to introduce a critical angle, called $\theta_{\mathrm{c}}$, and use the effective drag force and transverse momentum diffusion (i.e.\ the fSIDM approach) for all scatterings with $\theta\leq\theta_{\mathrm{c}}$ or $\theta\geq \uppi-\theta_{\mathrm{c}}$. For angles within the interval $\theta_{\mathrm{c}}<\theta<\uppi-\theta_{\mathrm{c}}$, the rSIDM approach is used instead, i.e.\ a random sampling of deflection angles from the differential cross section. Thus, we write
\begin{equation}
\begin{split}
    \left(\frac{\mathrm{d}\sigma}{\mathrm{d}\Omega}\right)_{\mathrm{hSIDM}} = &\left(\frac{\mathrm{d}\sigma}{\mathrm{d}\Omega}\right)_{\mathrm{fSIDM}} \left[ \Theta(\theta_{\mathrm{c}} - \theta)+ \Theta(\theta - (\uppi - \theta_{\mathrm{c}})) \right] \\
    & + \left(\frac{\mathrm{d}\sigma}{\mathrm{d}\Omega}\right)_{\mathrm{rSIDM}} 
    \left[ 1-\Theta(\theta_{\mathrm{c}} - \theta)- \Theta(\theta - (\uppi - \theta_{\mathrm{c}})) \right],
\end{split}
\end{equation}
where $\Theta(\theta)$ is the Heaviside function. The splitting is also illustrated by the dark and light shaded regions in Fig.~\ref{fig: Moeller differential cross section} for two typical values $\theta_{\mathrm{c}}=0.1$ and $0.3$. In the following we refer to the range $\theta_{\mathrm{c}}<\theta<\uppi-\theta_{\mathrm{c}}$ as `large-angle' and to the ranges $\theta\leq\theta_{\mathrm{c}}$ or $\theta\geq \uppi-\theta_{\mathrm{c}}$ as `small-angle' regimes, noting that deflection angles close to $\uppi$ are also considered as being `small' for identical particles.\footnote{For the case of distinguishable particles, as e.g.\ for Rutherford scattering, a simpler splitting according to $\theta\leq\theta_{\mathrm{c}}$ (fSIDM) and $\theta>\theta_{\mathrm{c}}$ (rSIDM) is used. In the following we focus on the more relevant case of identical particles for brevity. The appropriate replacements in the angle-intervals are understood to be applied implicitly when considering Rutherford scattering (for certain test set-ups discussed below), while we assume identical particles in all other cases.}

Since the critical angle $\theta_{\mathrm{c}}$ is a purely technical parameter, all physical quantities should be independent of its choice, and we check this explicitly for all our results below. Theoretically, we expect this is the case provided $\theta_{\mathrm{c}}$ is chosen small enough such that the small-angle effective description underlying the fSIDM approach can be applied. As we see later on, the appropriate size of $\theta_{\mathrm{c}}$ depends on characteristics of the set-up, such as the characteristic dynamical time  and the considered time interval, in addition to the angle dependence of the cross section itself. We quantify the validity range for the choice of $\theta_{\mathrm{c}}$ within the hSIDM scheme in Sect.~\ref{subsec: Theory Deflection problem}, employing a test set-up that can be described analytically. In addition, we point to Sect.~\ref{subsec: Rutherford and Moeller scattering implementation} for a discussion of requirements constraining the choice of $\theta_{\mathrm{c}}$ within typical simulation set-ups, and to Sect.~\ref{subsec: hSIDM implementation performance} for a demonstration of the performance gain of the hybrid scheme compared to a direct implementation of angular dependence.

\begin{figure}
    \centering
    \includegraphics[width=\columnwidth]{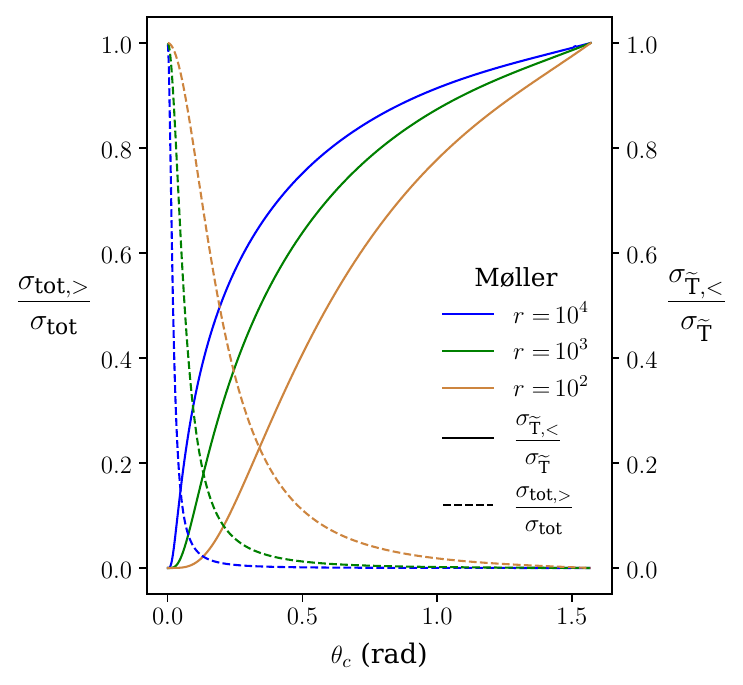}
    \caption{Ratio of the contribution to the {\it total} cross section from the {\it large}-angle regime (dashed lines) and of the modified {\it transfer} cross section from the {\it small}-angle regime versus the critical angle $\theta_{\mathrm{c}}$ used to separate small- and large-angle regions within the hSIDM scheme for the `M{\o}ller' cross section Eq.~\eqref{eq: Moeller differential cross section} and $r=10^2,10^3,10^4$, respectively. Solid lines characterise the expected physical relevance of small-angle scattering (in relative units). For moderate values of $\theta_{\mathrm{c}}\lesssim 0.5$, dashed lines conservatively estimate the expected ratio of the numerical effort of hSIDM compared to a naive implementation of angle-dependent scattering.}
    \label{fig: contribution of total and transfer cross section}
\end{figure}

The main reason for considering the hSIDM scheme is that the effective small-angle treatment is numerically much more efficient to capture the forward-dominated regime. We provide an explicit comparison of the speed-up when using hSIDM (as compared to a direct sampling of scattering angles over the entire range) in Sect.~\ref{subsec: hSIDM implementation performance} below. However, it is also possible to estimate the expected speed-up. For that purpose, it is convenient to consider the contributions to the various angle-averaged cross sections from Sect.~\ref{subsec: Angle dependent scattering-cross section} within the `small' and `large'-angle regimes depending on the choice of $\theta_{\mathrm{c}}$. We thus generalise Eq.~\eqref{eq:sigma_def} and define
\bea
 \label{eq: definitions of sigma<,>}
     \sigma_{X,<}(\theta_{\mathrm{c}}) &\equiv&  c_X \int\displaylimits_{ \theta \in I_{<} } \mathrm{d}\Omega \,g_{X}(\theta) \,\frac{\mathrm{d}\sigma}{\mathrm{d}\Omega}\,,
     \quad I_{<} = [0,\theta_{\mathrm{c}}] \cup [\uppi -\theta_{\mathrm{c}}, \uppi] \,, \nn\\
     \sigma_{X, >}(\theta_{\mathrm{c}}) &\equiv& c_X \int\displaylimits_{ \theta \in I_{>} } \mathrm{d}\Omega \,g_{X}(\theta) \,\frac{\mathrm{d}\sigma}{\mathrm{d}\Omega}\,,
     \quad I_{>} = (\theta_{\mathrm{c}}, \uppi - \theta_{\mathrm{c}})\,,
\eea
allowing us to decompose the total and (modified) transfer cross sections into the contributions treated with the fSIDM (<) and rSIDM (>) method within hSIDM, respectively,
\bea
  \sigma_{\rm tot} &=&  \sigma_{{\rm tot},<}(\theta_{\mathrm{c}}) + \sigma_{{\rm tot},>}(\theta_{\mathrm{c}})\,,\nn\\
  \sigma_{\mathrm{\widetilde{T}}} &=&  \sigma_{\mathrm{\widetilde T},<}(\theta_{\mathrm{c}}) + \sigma_{\mathrm{\widetilde T},>}(\theta_{\mathrm{c}})\,.
\eea
The drag force used for the small-angle regime is related to the (modified) transfer cross section.
Within hSIDM, only scatterings with $\theta\in I_<$ are treated in this way, and thus (see Sect.~\ref{sec: implementation} for details)
\be
  F_{\rm drag}(\theta_{\mathrm{c}}) \propto \sigma_{\mathrm{\widetilde T},<}(\theta_{\mathrm{c}}) \qquad ({\rm small-angle \ regime})\,.
\ee
The numerical costs depend on the size of the numerical time step. Within the fSIDM scheme, the time step is proportional to the drag force, $F_{\rm drag}$, and hence also proportional to the (modified) transfer cross section, $\sigma_{\mathrm{\widetilde T}}$. In contrast, the rSIDM scheme, for which individual collisions are modelled numerically as scattering of $N$-body particles (see Sect.~\ref{sec: implementation}), has a time step proportional to the total scattering rate, $R$, which is proportional to the {\it total} cross section of large-angle scattering, i.e.\ within the hSIDM scheme it is
\be
  R(\theta_{\mathrm{c}}) \propto \sigma_{{\rm tot},>}(\theta_{\mathrm{c}}) \qquad ({\rm large-angle \ regime})\,.
\ee
We note that for light mediator models (i.e.\ `M{\o}ller scattering'), the strong enhancement of the forward-scattering rate implies that for typical values of $\theta_{\mathrm{c}}\sim 0.1-0.3$ one has
$\sigma_{{\rm tot},>}(\theta_{\mathrm{c}}) \ll  \sigma_{\rm tot}$. The expected speed-up, $S$, within hSIDM as compared to a naive direct treatment of the angle dependence can thus be estimated as
\be\label{eq:effgain}
 S \propto \frac{\sigma_{\rm tot}}{\sigma_{{\rm tot},>}(\theta_{\mathrm{c}})}\,.
\ee
The dependence of this ratio on $\theta_{\mathrm{c}}$ for the M{\o}ller scattering cross section Eq.~\eqref{eq: Moeller differential cross section} is shown in Fig.~\ref{fig: contribution of total and transfer cross section} (dashed lines), for various values of the anisotropy parameter $r$ from Eq.~\eqref{eq:rdef}. Even for moderate choices $\theta_{\mathrm{c}}=0.1$ -- $0.3$, one has $\sigma_{{\rm tot},>}(\theta_{\mathrm{c}})/\sigma_{\rm tot}\ll 1$, indicating a large expected speed-up within hSIDM.  In practice, we find that the speed-up is even slightly larger than this estimate (see Sect.~\ref{subsec: hSIDM implementation performance}).

On the other hand, the {\it physical} relevance of small- versus large-angle scatterings can be estimated from comparing the (modified) {\it transfer} cross sections $\sigma_{\mathrm{\widetilde T},<}(\theta_{\mathrm{c}})$ and $\sigma_{\mathrm{\widetilde T},>}(\theta_{\mathrm{c}})$. The fraction $\sigma_{\mathrm{\widetilde T},<}(\theta_{\mathrm{c}})/\sigma_{\mathrm{\widetilde{T}}}$ related to the small-angle regime is shown by the solid lines in 
Fig.~\ref{fig: contribution of total and transfer cross section}. Thus, for typical values of $\theta_{\mathrm{c}}$ both small- and large-angle scatterings are expected to have a roughly comparable {\it physical} impact.

In summary, the main advantage of the hSIDM regime is that it allows us to treat very frequent small-angle scatterings much more efficiently as compared to a naive sampling of deflection angles, while in addition explicitly including the angle dependence of the physically roughly equally important contribution from large-angle scattering.
With the hSIDM scheme simulation of fairly anisotropic cross sections with significant contributions from both small and large angles that have not been feasible before, become possible thanks to the efficient treatment of small-angle scattering.

\subsection{Applicability of effective description for small-angle scattering: deflection test set-up} \label{subsec: Theory Deflection problem}

In order to gain insight on the expected range of validity of the hSIDM scheme, and especially on the choice of the critical angle $\theta_{\mathrm{c}}$, we investigate the applicability of the effective treatment of small-angle scattering. For that purpose, it is instructive to consider a simple test set-up that admits an analytical treatment. The purpose of considering this set-up is $(i)$ to derive analytical conditions on the angle dependence of the cross section, as well as the typical local dark matter density and velocity and dynamical timescale for which the effective treatment of small-angle scattering is valid (see Sect.~\ref{sec:fSIDMvalidity}), $(ii)$ translate these into validity conditions for the admissible range of values of $\theta_{\mathrm{c}}$ within the hSIDM scheme (see Sect.~\ref{sec:hSIDMvalidity}), $(iii)$ demonstrate the resulting speed-up of the numerical algorithm based on hSIDM, and $(iv)$ provide a stringent validation test for our numerical implementation (see Sect.~\ref{sec: Deflection test} below). Moreover, the test set-up can also be seen as a proxy how momentum and energy is transferred in a collision of two ``patches'' of the dark matter population within the two halos that collide in the initial stages of a galaxy cluster merger, provided the relative velocity is large compared to the velocity dispersion with each of the patches. 

\subsubsection{Deflection test set-up}

Following~\citet{Bethe_1953}, we consider a classic deflection problem of a beam of test particles traversing a homogeneous medium of target particles with number density $n$. All beam particles move initially in the same direction (say along the $z$-axis) with a given velocity $v$. After some time $t$, scatterings of the beam on target particles lead to a `broadening' of the beam. This can be described by a distribution function $f(t,\theta)$, such that $f(t,\theta)\cdot\sin(\theta)\,\mathrm{d}\theta$ measures the fraction of beam particles with azimuthal angle $\theta$ to the $z$-axis within the interval $(\theta,\theta+\mathrm{d}\theta)$ at time $t$ (assuming rotational symmetry around the $z$-axis).
The initial condition at $t=0$ is thus 
\be
  f(t=0,\theta)=\delta(\cos(\theta)-1)\,.
\ee
We study this deflection set-up in two variants, (i) in the limit for which no recoil is transmitted to the target particles (i.e.\ formally infinitely heavy target particles), and (ii) when including recoil and assuming equal masses for beam and target particles. We note that in this validation set-up scatterings occur only between beam and target particles, and that no gravitational force is included.

In the remainder of this subsection, we focus on case (i), for which the magnitude of the velocity is unchanged and only the direction of the test particles changes in each scattering.\footnote{This implies that the particles are distinguishable, as we assume for the rest of this subsection.} In this case, a full analytical solution for the Boltzmann equation determining the distribution function is known, going back to~\cite{Goudsmit_1940a, Goudsmit_1940b}. Following the more explicit derivation by~\citet{Bethe_1953}, one finds
\begin{equation}
\label{eq: Goudsmit-Saunderson}
\begin{split}
    f(t,\theta) = &\sum_{\ell=0}^{\infty} P_\ell\left(\cos(\theta)\right) \left(\ell+\frac{1}{2}\right)\\ 
    &\exp\left\{ -nvt \int \mathrm{d}\Omega' \frac{\mathrm{d}\sigma}{\mathrm{d}\Omega'} \left[ 1- P_\ell\left(\cos(\theta')\right)\right]  \right\}  \,,
\end{split}
\end{equation}
where $P_\ell(x)$ are Legendre polynomials, and $\mathrm{d}\sigma \rm /d\Omega$ is the differential scattering cross section for beam-target particle scattering. We note that the {\it total} number of scatterings each beam particle experiences on average between $t=0$ and time $t$  is given by the total opacity
\be
  \tau_{\rm tot}(t)\equiv nvt\sigma_{\rm tot}=nvt\int \mathrm{d}\Omega\frac{\mathrm{d}\sigma}{\mathrm{d}\Omega}\,,
\ee
and increases linearly with time $t$. 

Let us first discuss some general features. Using the properties $\int_{0}^\uppi \mathrm{d}\theta\,\sin(\theta) P_0(\cos(\theta))=2$ and
$\int_{0}^\uppi \mathrm{d}\theta\,\sin(\theta) P_\ell(\cos(\theta))=0$ for $\ell>0$ one sees that 
\be
  \int_{0}^\uppi \mathrm{d}\theta\,\sin(\theta)f(t,\theta)=1\,,
\ee
for all times $t$, i.e.\ the distribution is properly normalised at all times. Furthermore, the relation 
\be
  \sum_{\ell}P_\ell(x)(\ell+1/2)=\delta(x-1)\,, 
\ee
with $x \equiv \cos(\theta)$, ensures that the correct initial condition is reached for $t\to 0$. Additionally, since $P_\ell(\cos(\theta))\leq 1$ (and strictly less than unity for $\theta>0$ and $\ell>0$), all summands in Eq.~\eqref{eq: Goudsmit-Saunderson} are exponentially damped for $t\to\infty$ except for $\ell=0$, implying 
\be
  f(t\to\infty,\theta)\to 1/2\,,
\ee
approaches an isotropic distribution at late times, as expected after a large number of multiple scatterings.

The distribution thus evolves from a peaked to a flat shape, and the differential cross section determines how the intermediate evolution occurs. For example, for an {\it isotropic cross section}
one has
\bea
  f(t,\theta)\big|_{\rm iso} &=& \frac12+\sum_{\ell>0}P_\ell(\cos(\theta))\left(\ell+\frac12\right)e^{-\tau_{\rm tot}(t)}\nn\\
  &=& \frac12 (1-e^{-\tau_{\rm tot}(t)})+\delta\left(\cos(\theta)-1\right)e^{-\tau_{\rm tot}(t)}\,.
\eea
Thus for isotropic scattering the distribution is given by a superposition of a flat and a peaked component at any time, with time-dependent relative weighting factors $1-e^{-\tau_{\rm tot}}$ and $e^{-\tau_{\rm tot}}$, respectively. We note that the exponential damping factor in Eq.~\eqref{eq: Goudsmit-Saunderson} is identical for all $\ell>0$ in the isotropic case.

This changes completely when the differential cross section is strongly forward-dominated. In this case the Legendre polynomial inside the integral over $\mathrm{d}\sigma/\mathrm{d}\Omega'$ in Eq.~\eqref{eq: Goudsmit-Saunderson} can be approximated for small deflection angle as 
\be\label{eq:forwardapprox}
  1-P_\ell(x)=\frac12\ell(\ell+1)(1-x)+{\cal O}(1-x)^2\,,
\ee
which yields
\be\label{eq:deflectionforward}
  f(t,\theta)\bigg|_{\mycom{\rm forward}{\rm dominated}} = \sum_{\ell\geq 0}P_\ell(\cos(\theta))\left(\ell+\frac12\right)\exp\left[-\frac12\ell(\ell+1)\tau_{\mathrm{T}}(t)\right]\,,
\ee
where higher $\ell$ become more and more suppressed. The relevant quantity determining the evolution is the {\it transfer} cross section $\sigma_{\mathrm{T}}$, see Sect.~\ref{subsec: Angle dependent scattering-cross section}, that we use to define a {\it transfer} opacity via
\be\label{eq:tauT}
  \tau_{\mathrm{T}}(t)\equiv nvt\sigma_{\mathrm{T}}=nvt\int \mathrm{d}\Omega\,\left(1-\cos(\theta)\right)\,\frac{\mathrm{d}\sigma}{\mathrm{d}\Omega}\,.
\ee
This already indicates that for purely forward-dominated scatterings, the {\it transfer} and not the  total cross section is the relevant quantity for the dynamical evolution \citep[see][]{Kahlhoefer_2014}.\footnote{We note that it is the transfer and not the modified transfer cross section that appears here, since we presently consider a test set-up with distinguishable particles (beam versus target).}

The approximation given by Eq.~\eqref{eq:forwardapprox} is the essential assumption underlying the effective treatment of small-angle scattering entering the fSIDM approach. This can be seen more directly when assuming in addition that $\tau_{\mathrm{T}}$ is small enough such that sufficiently many $\ell$ contribute in the summation in Eq.~\eqref{eq:deflectionforward}. In this case the summation over $\ell$ can be replaced by a continuous integration using the Euler formula \citep[see e.g.][]{Bethe_1953}, and Eq.~\eqref{eq:deflectionforward} yields the Moli\`ere approximation \citep[compare with][]{Moliere_1948} 
\begin{equation}
\label{eq: Moliere}
    f(t,\theta) \approx \frac{1}{\tau_{\mathrm{T}}(t)} \exp\left[- \frac{\theta^2}{2\tau_{\mathrm{T}}(t)}\right]\,.
\end{equation}
The effect of multiple small-angle scattering thus yields a Gaussian broadening of the beam, as long as $\tau_{\mathrm{T}}(t)\lesssim {\cal O}(1)$. This is in direct correspondence to the effective fSIDM treatment of forward-dominated scatterings in terms of a transverse momentum diffusion process, which also leads to a Gaussian broadening when applied to the deflection set-up \citep[with identical width, see][]{Fischer_2021a}.\footnote{We note that the fSIDM approach is  more general than the Moli\`ere approximation, as the latter requires Eq.~\eqref{eq:forwardapprox} as well as $\tau_{\mathrm{T}}\lesssim {\cal O}(1)$, while for the former the validity of  Eq.~\eqref{eq:forwardapprox} is sufficient. In practice this means that for late times, for which the Gaussian width becomes comparable to $\uppi$, the Moli\`ere approximation breaks down, since the quantisation of angular momentum starts to play a role once the distribution function is populated over the entire finite allowed interval of angles between $0$ and $\uppi$. On the other hand, the fSIDM approach correctly yields an isotropic distribution at late times, as does Eq.~\eqref{eq:forwardapprox}.}

\subsubsection{Validity of the fSIDM approach}\label{sec:fSIDMvalidity}

Importantly, the analytically solvable deflection set-up allowed us to derive a general criterion for the validity of the effective forward-dominated approximation. In this subsection, we start by discussing the validity in the case of allowing for the entire range of possible scattering angles, i.e.\ the fSIDM approximation. Subsequently, we generalise the discussion to the hSIDM case, for which only angles below a critical angle are included in the effective small-angle approximation.

To obtain a quantitative estimate of the corrections to the forward-dominated approximation, we keep the next term in the Taylor expansion of Eq.~\eqref{eq:forwardapprox}, of order $(1-x)^2$ (where $x\equiv\cos(\theta)$), given by
\bea\label{eq:forwardapproxnlo}
   1-P_\ell(x) &=& \frac{\ell(\ell+1)}{2}(1-x)-\frac{(\ell-1)\ell(\ell+1)(\ell+2)}{16}(1-x)^2\nn\\
   && {} +{\cal O}(1-x)^3\,.
\eea
The fSIDM approach is valid if the contribution of the $(1-x)^2$ term to the distribution function $f(t,\theta)$ (entering via the exponential in Eq.~\eqref{eq: Goudsmit-Saunderson}) is small. This suggests to consider a `transfer-squared' cross section and a corresponding opacity given by
\be\label{eq:Tsq}
  \tau_{\mathrm{T}^2}(t)\equiv nvt\sigma_{\mathrm{T}^2},\qquad
  \sigma_{\mathrm{T}^2} \equiv \int \mathrm{d}\Omega\,\left(1-\cos(\theta)\right)^2\,\frac{\mathrm{d}\sigma}{\mathrm{d}\Omega}\,.
\ee
Including the correction term, the exponential factor in the forward-dominated approximation of Eq.~\eqref{eq:deflectionforward} becomes 
\be\label{eq:deflectionforwardnlo}
  \exp\left[-\frac12\ell(\ell+1)\tau_{\mathrm{T}}(t)+\frac{(\ell-1)\ell(\ell+1)(\ell+2)}{16}\tau_{\mathrm{T}^2}(t)(t)+\dots\right]\,,
\ee
where the ellipsis stand for even higher-order terms.
The effective small-angle approach is thus valid if the second summand in the exponential is small compared to the first one, i.e.\ $\tau_{\mathrm{T}^2}\ll 8\tau_{\mathrm{T}}/\ell^2$, for all values of $\ell$ that yield a sizeable contribution in the sum over $\ell$ in Eq.~\eqref{eq:deflectionforward}. Since the exponential factor becomes strongly suppressed when $\ell(\ell+1)\tau_{\mathrm{T}}/2\gg 1$, the relevant  range is up to $\ell^2\sim 2/\tau_{\mathrm{T}}$. This yields the validity condition
\be
  \tau_{\mathrm{T}^2}(t)\ll 4(\tau_{\mathrm{T}}(t))^2\,.
\ee
In practice, this means the small-angle treatment becomes valid only after a finite time $t_{\rm min}$, for which sufficient scatterings have taken place, given by
\be\label{eq:tmin}
  t \gg t_{\rm min}=\frac{1}{\rho v \frac{\sigma_{\mathrm{T}}}{m_\chi}}\frac{\sigma_{\mathrm{T}^2}}{4\sigma_{\mathrm{T}}} = t_{\rm dyn} \frac{\sigma_{\mathrm{T}^2}}{4\sigma_{\mathrm{T}}}\,,
\ee
where we expressed the number density in terms of the energy density using $\rho=m_\chi n$.
For the effective small-angle approach to be useful, $t_{\rm min}$ should be much smaller than the timescale on which the distribution reaches its late-time isotropic limit, which is of order of the typical dynamical evolution timescale $t_{\rm dyn}\equiv 1/(\rho v \frac{\sigma_{\mathrm{T}}}{m_\chi})$, where $v$ is the beam velocity. Thus, we arrive at the condition for the effective small-angle treatment to be applicable and useful, given by
\be
  \sigma_{\mathrm{T}^2} \ll \sigma_{\mathrm{T}}\,.
\ee
We note that this is the condition that is relevant when using the effective fSIDM approach for {\it all} possible deflection angles, being in practice limited to models that {\it exclusively} feature forward-dominated scattering. For realistic models, this is typically not the case. As an  example,  M{\o}ller scattering with $r=1,10,100,10^3,10^4$ yields $\sigma_{\mathrm{T}^2}/\sigma_{\mathrm{T}} = 0.32, 0.23, 0.14, 8.5\times10^{-2}, 5.9\times10^{-2}$ respectively. Analogously Rutherford scattering yields $1.2,0.82,0.51,0.33,0.24$. This motivates the use of the hybrid approach instead. We discuss its validity next.

\subsubsection{Validity of the hSIDM approach}\label{sec:hSIDMvalidity}

Let us now discuss how the validity of the effective treatment of forward-dominated scattering is changed within the hSIDM scheme, i.e.\ when only including deflection angles below a certain critical angle $\theta_{\mathrm{c}}$ (and above $\uppi-\theta_{\mathrm{c}}$ for indistinguishable particles) in the small-angle approximation. We start the discussion for the test set-up with distinguishable particles and no recoil (i.e.\ infinitely heavy target particles), for which the analytical solution of the deflection set-up has been given in Eq.~\eqref{eq: Goudsmit-Saunderson} above, and then generalise to the case of identical particles and with recoil (i.e.\ identical masses for beam and target particles). 

To obtain the hSIDM validity conditions, we replace the cross section integrated over all angles by the cross section integrated only over the small-angle regime, see Eq.~\eqref{eq: definitions of sigma<,>}, in the validity conditions obtained above. Thus, hSIDM validity requires
\be
  \tau_{\mathrm{T}^2,<}(t,\theta_{\mathrm{c}})\ll 4(\tau_{\mathrm{T},<}(t,\theta_{\mathrm{c}}))^2\,,
\ee
where $\tau_{X,<}(t,\theta_{\mathrm{c}})\equiv nvt\sigma_{X,<}(\theta_{\mathrm{c}})$ for $X=\mathrm{T}$ (transfer cross section) and $X=T^2$ (transfer-squared cross section, defined analogously to Eq.~\eqref{eq:Tsq} but integrated only over the angle-interval $I_<$). As above, this implies a minimal timescale after which the hSIDM approach is valid, given by
\be\label{eq:tminhSIDM}
  t \gg t_{\rm min}(\theta_{\mathrm{c}})=\frac{1}{\rho v \frac{\sigma_{\mathrm{T},<}}{m_\chi}}\frac{\sigma_{\mathrm{T}^2,<}}{4\sigma_{\mathrm{T},<}} = t_{\rm dyn} \frac{\sigma_{\mathrm{T}^2,<}}{4\sigma_{\mathrm{T},<}}\,.
\ee
Thus the hSIDM approach is valid and advantageous if
\be\label{eq:hSIDMvalidity}
  \sigma_{\mathrm{T}^2,<}(\theta_{\mathrm{c}}) \ll \sigma_{\mathrm{T},<}(\theta_{\mathrm{c}})\,.
\ee
The conditions from above can be easily generalised to the case of identical particles with equal mass by using the appropriate definition of the cross sections $\sigma_{X,<}(\theta_{\mathrm{c}})$ for that case, (a) integrated over the interval corresponding $I_{<} = [0,\theta_{\mathrm{c}}] \cup [\uppi -\theta_{\mathrm{c}}, \uppi]$ instead of $I_{<} = [0,\theta_{\mathrm{c}}]$, and (b) replacing the transfer ($X=\rm T$) by the modified transfer ($X=\rm {\widetilde T}$) cross section, see Sect.~\ref{subsec: Angle dependent scattering-cross section}, and analogously for $X=\mathrm{T}^2$.

As an illustrative example, for M{\o}ller (Rutherford) scattering with $r=10^3$, one obtains  $\sigma_{\mathrm{T}^2,<}(\theta_{\mathrm{c}})/\sigma_{\mathrm{T},<}(\theta_{\mathrm{c}})=1.3\times10^{-3} \,(2.6\times10^{-3})$ for $\theta_{\mathrm{c}}=0.1$. We provide a detailed comparison of hSIDM simulations to a full, explicit treatment of the angle dependence for various models in Sect.~\ref{sec: Deflection test}, where we also cross-check the theoretically expected validity conditions derived here.

We expect that the condition on the cross section Eq.~\eqref{eq:hSIDMvalidity} is valid also beyond the deflection set-up. We note that it is independent of the details of the test problem and only involves the angle dependence of the cross section itself, and thus Eq.~\eqref{eq:hSIDMvalidity} depends only on the properties of the particle physics model underlying the angle dependence of the self-scattering cross section. Moreover, as discussed above, the deflection test can be viewed as a proxy for the dynamics of two colliding patches of dark matter in a galaxy cluster merger, such that the main physical mechanism of self-interactions on the dark matter distribution is broadly similar. This supports the applicability of Eq.~\eqref{eq:hSIDMvalidity} in cluster merger simulation set-ups. We also expect that Eq.~\eqref{eq:tminhSIDM} can be generalised beyond the test problem when replacing $t_\text{dyn}$ on the right-hand side by the appropriate dynamical timescale on which the system of interest evolves.

\section{Implementation} \label{sec: implementation}

We implement the hSIDM scheme for efficiently describing DM self-interactions with arbitrary differential cross section in \textsc{OpenGadget3} \citep[see][and the references therein]{Groth_2023}, an upgraded version of the $N$-body code \textsc{gadget-2} \citep{Springel_2005}, following existing implementations for either purely forward-dominated (fSIDM) or large-angle scattering (rSIDM), respectively.
In Sect.~\ref{subsec: Existing Implementations} we briefly review existing implementations, and then introduce the numerical algorithm underlying this work in Sect.~\ref{subsec: Implementing hSIDM}. We discuss criteria for choosing the associated numerical time steps in Sect.~\ref{subsec: time step}.

\subsection{Existing implementations} \label{subsec: Existing Implementations}

In this section, we describe the numerical algorithm for the two existing methods reviewed in Sect.~\ref{subsec: Existing approaches for angular regimes}. We start with rSIDM for large-angle scattering and then turn to the purely forward-dominated fSIDM case.

\subsubsection{Algorithm for rare large-angle scattering: rSIDM} \label{subsubsec: Algorithm for rare large-angle scattering (rSIDM)}

$N$-body simulations of DM self-interactions via large-angle scatterings, dubbed rSIDM, are well-established. Most studies assume the simplest case of isotropic scattering, but also an explicit sampling of deflection angles from a given distribution has been considered (see e.g.~\citep{Robertson_2017a, Banerjee_2020}). Microscopic two-body interactions of DM particles are modelled by  collisions between numerical $N$-body particles within rSIDM. Various approaches along these lines exist, differing in the computation of scattering probabilities \citep[e.g.][]{Koda_2011, Vogelsberger_2012, Rocha_2013, Robertson_2017a}. Starting point is the interaction between two phase-space patches, represented by two numerical particles assumed to have equal mass. To simulate isotropic rSIDM, \cite{Fischer_2021a} utilised the  method  presented by \cite{Rocha_2013}. The kernel function, $W$, characterises the density distribution of DM for a numerical particle in configuration space. Specifically, the scattering probability of a numerical particle $i$ to scatter with a particle $j$ having the mass $m_j$ can be expressed as
\begin{equation}
\label{eq: Scattering probability in existing rSIDM method}
    P_i = \frac{\sigma_{\mathrm{tot}}}{m_{\chi}} m_j |\Delta \vec{v}_{ij}| \Delta t \Lambda_{ij},
\end{equation}
where $\Delta \vec{v}_{ij}$ is the relative velocity of the two numerical particles, $\Delta t$ denotes the time step, and $\Lambda_{ij} = \int \mathrm{d}V W(|\vec{x}-\vec{x}_i|,h_i) \, W(|\vec{x}-\vec{x}_j|,h_j) $ gives the kernel overlap, with $h$ being the kernel size. Following \cite{Fischer_2021a}, we employ a spline kernel \citep{Monaghan_1985} and choose $h$ adaptively such that it includes the $N_{\rm ngb}$ next neighbours. Based on the probability $P_i$, we determine whether two numerical particles scatter during the time step. If a randomly selected number $x$ from the interval $[0,1]$ satisfies $x \leq P_i$, the particles scatter. Within the rSIDM scheme, the scattering angle is chosen randomly according to the differential cross section. For that purpose, it is convenient to consider the cumulative probability \citep[as e.g.\ by][]{Robertson_2017a}
\begin{equation}
\label{eq: cumulative distribution function rSIDM}
    P(\theta) = \int_{0}^{\theta} \mathrm{d}\theta' \frac{2\uppi \sin(\theta')}{\sigma_{\mathrm{tot}}} \frac{\mathrm{d}\sigma}{\mathrm{d}\Omega'} \in [0,1]\quad {\rm for}\  \theta\in[0,\uppi]\,,
\end{equation} 
which gives the probability of scattering by an angle less than $\theta$. The deflection angle can then be determined by drawing another random number $y\in[0,1]$ with uniform distribution, and solving $P(\theta)=y$. We note that the deflection angle refers to the centre of mass frame, such that the computation of the vectorial velocities after the collision involves (i) a transformation of the initial velocity vectors into the centre of mass frame, (ii) application of the azimuthal deflection angle $\theta$ along with a random, uniform polar angle $\phi\in[0,2\uppi]$, and (iii) an inverse transformation of the modified velocity vectors back to the original `laboratory' frame. For the case of isotropic scattering, the velocity directions after the scattering can simply be chosen randomly in the centre of mass frame. Details on how the vectorial velocities are calculated in practice, are given in Appendix~\ref{appendix: Vector calculation for anisotropic scattering}.

\subsubsection{Algorithm for frequent small-angle scattering: fSIDM}\label{subsubsec: Algorithm for frequent small-angle scattering (fSIDM)}

The effective description of frequent, purely forward-dominated scattering has been implemented by~\cite{Fischer_2021a}. The self-interaction is described by a drag force and a diffusive random `kick'. As opposed to rSIDM, {\it all} pairs of numerical particles that are close enough to each other are affected by the interaction within a given time step. Here `close enough' means in practice that the kernel overlap is positive (i.e.~$\Lambda_{ij}>0$).
The individual `interaction' of two numerical particles $i$ and $j$ within fSIDM is then composed of two steps. Firstly, as already discussed in Sect.~\ref{subsec: Existing approaches for angular regimes}, a drag force proportional to the (modified) transfer cross section is applied, with magnitude given by~\citep[see Eq.~9 by][]{Fischer_2021a}\footnote{We note that the different normalisation of the cross section we use in this work, results in an additional factor of $1/2$ in Eq.~\eqref{eq:drag} compared to \cite{Fischer_2021a}.}
\begin{equation}\label{eq:drag}
    F_{\mathrm{drag}} = \frac{1}{4} |\Delta \vec{v}_{ij}|^2 \frac{\sigma_{\mathrm{\widetilde{T}}}}{m_{\chi}} m_i m_j \Lambda_{ij}.
\end{equation}
The drag force acts along the direction of the relative velocity vector $\Delta \vec{v}_{ij}$, and decelerates the numerical particles, reducing the velocity of particle $i$ by an amount $\Delta \vec{v}_{\rm drag}=(F_{\mathrm{drag}} / m_i) \Delta t$.
The second step involves a random `kick' in the direction transverse to the relative velocity vector, in a random direction within the transverse plane. This step corresponds to the diffusive part of the effective description of small-angle scattering. The magnitude of the transverse velocity kick is determined such that total energy is conserved, in other words, it makes up for the energy lost due to the drag force, given by
\begin{equation}\label{eq:dE}
    \frac{2 \Delta E}{m} =  |\Delta \vec{v}_{\mathrm{drag}}| \left( |\Delta \vec{v}_{ij}| - |\Delta \vec{v}_{\mathrm{drag}}|\right)\,.
\end{equation}
This two-step algorithm ensures that energy as well as three-momentum is explicitly conserved. We note that the deceleration due to the drag force and the transverse momentum kick are computed in the centre of mass frame, to determined the updated velocity vectors in the `laboratory' frame. 

\subsection{Implementation of the hybrid approach: hSIDM} \label{subsec: Implementing hSIDM}

In this work we implement the hybrid approach for DM self-interaction as described by an arbitrary differential cross section. For that purpose, we combine the approaches describe above, using the drag force and momentum diffusion method (fSIDM) for scattering with `small angle', and the explicit sampling technique (rSIDM) for `large-angle' scatterings. Both are separated by a critical angle $\theta_{\mathrm{c}}$. 

In each time step, for every pair with a positive kernel overlap $\Lambda_{ij}$, we  simulate the small-angle scattering, before we model large scattering angles. Moreover, the pairwise computation of the particle interactions is done in a consecutive manner to conserve energy explicitly. This means that a numerical particle cannot interact with multiple particles at the same time. Only after small- and large-angle scattering have been computed for the particles of a given pair, do further pairwise interactions with other particles follow. When the scatterings for all relevant pairs of numerical particles have been computed, the time step is complete and the next one can follow.

To model the small-angle scatters, we apply the drag force and transverse momentum kick to all relevant pairs of numerical particles. Compared to purely forward-dominated scattering (i.e.\ fSIDM), the cross section entering the drag force is replaced in Eq.~\eqref{eq:drag} as $\sigma_{\mathrm{\widetilde{T}}} \mapsto \sigma_{\mathrm{\widetilde T},<}(\theta_{\mathrm{c}})$, see Eq.~\eqref{eq: definitions of sigma<,>}. This takes into account the range of scattering angles treated with the effective `small-angle' approach. The analytical expression of the modified transfer cross section for the hSIDM scheme can be found in Appendix~\ref{appx: Implementation of frequent small-angle scattering for hSIDM}.

Next, we draw uniform random numbers $x,y\in[0,1]$ for each pair of numerical particles with $\Lambda_{ij}>0$ to determine whether they undergo `large-angle' scattering (if $x\leq P_i$) and, if yes, with which value of the scattering angle ($P(\theta)=y$). Compared to the pure rSIDM implementation, we replace
$\sigma_{\rm tot}\mapsto \sigma_{\rm tot,>}(\theta_{\mathrm{c}})$ in Eq.~\eqref{eq: Scattering probability in existing rSIDM method} and compute the cumulative probability as\footnote{This expression applies to scattering of identical particles. For distinguishable particles, we use the modified definition of $\sigma_{\mathrm{tot},>}$ compared to Eq.~\eqref{eq: definitions of sigma<,>} involving integration over $\theta\in[\theta_{\mathrm{c}},\uppi]$ such that $P(\theta)\in[0,1]$ for $\theta\in[\theta_{\mathrm{c}},\uppi]$ in that case.}
\begin{equation}
\label{eq: cumulative distribution function hSIDM}
    P(\theta) = \int_{\theta_{\mathrm{c}}}^{\theta} \mathrm{d}\theta' \frac{2\uppi \sin(\theta')}{\sigma_{\mathrm{tot},>}} \frac{\mathrm{d}\sigma}{\mathrm{d}\Omega'} \in [0,1]\quad {\rm for}\  \theta\in[\theta_{\mathrm{c}},\uppi-\theta_{\mathrm{c}}]\,.
\end{equation}
As before, the update of particle velocity vectors in the `laboratory frame'  is computed analogously as described below Eq.~\eqref{eq: cumulative distribution function rSIDM}. In Appendix~\ref{appx: Implementing large-angle anisotropic scattering in rSIDM scheme} we provide the analytical expression of the cumulative probability for models considered in this work.

We note that within hSIDM both the drag force $F_{\rm drag}=F_{\rm drag}(\theta_{\mathrm{c}})$ describing small-angle scattering as well as the probability $P_i=P_i(\theta_{\mathrm{c}})$ and cumulative distribution $P(\theta)=P(\theta;\theta_{\mathrm{c}})$ for large-angle scattering depend on the critical angle $\theta_{\mathrm{c}}$ that is used as a technical parameter to split between small- and large-angle regimes. We check below that all our results are independent of $\theta_{\mathrm{c}}$ within an expected range of validity. Finally, we stress that the potentially large (and numerically challenging) cross section $\sigma_{\mathrm{tot},<}(\theta_{\mathrm{c}})$ does {\it not} enter in hSIDM, but only the much smaller $\sigma_{\mathrm{\widetilde T},<}(\theta_{\mathrm{c}})$
as well as $\sigma_{\mathrm{tot},>}(\theta_{\mathrm{c}})$ and the differential angle-distribution $\mathrm{d}\sigma/\mathrm{d}\Omega$ for $\theta_{\mathrm{c}}\leq \theta\leq\uppi-\theta_{\mathrm{c}}$.

\subsection{Choice of the time step} \label{subsec: time step}

Compared to conventional $N$-body simulations taking only gravity into account, the maximal possible size of a single time step should be further constrained for SIDM to ensure accurate results, see for example~\cite{Vogelsberger_2012} and \cite{Fischer_2021b, Fischer_2024a}.
Next, we explain how we choose the time step.

When simulating hSIDM and gravity, the upper limit on the time step related to particle $i$ is determined as 
\be
  \Delta t_i={\rm min}(\Delta t_{{\rm grav.},i},\Delta t_{{\rm large-angle},i},\Delta t_{{\rm small-angle},i})\,,
\ee
where $\Delta t_{{\rm grav.},i}$ is set by the gravitational acceleration following the criterion given by \citet[see Eq.~34]{Springel_2005}.
The second one ($\Delta t_{{\rm large-angle},i}$) arises from the requirement that the scattering probability for large-angle scattering within a single time step needs to be sufficiently smaller than unity, $P_i\ll 1$,\footnote{This is the case even though our implementation can correctly describe multiple scatterings of a numerical particle per time step. In Appendix~\ref{appendix: Multiple scattering and time step}, we demonstrate and explain the required limitation of the scattering probability.} leading to
\begin{equation}
\label{eq: time steps in rSIDM}
    \Delta t_{{\rm large-angle}, i} \approx \frac{\kappa}{\max_j\left(|\Delta \vec{v}_{ij}|\right)} \frac{1}{m_j \,\Lambda_{ii}}\left(\frac{\sigma_{\mathrm{tot},>}(\theta_{\mathrm{c}})}{m_\chi}\right)^{-1}\,,
\end{equation}
where $\kappa$ is a dimensionless accuracy parameter (we use $\kappa= 0.01$), $m_j$ is the numerical particle mass, and $\Lambda_{ii}$ quantifies the maximum kernel overlap, determined from $\Lambda_{ij}$ with $j=i$.
The third criterion ($\Delta t_{{\rm large-angle},i}$) arises from the effective treatment of frequent small-angle scattering. The description via a drag force and momentum diffusion requires that the typical distribution of deflection angles generated after a single time step is still sufficiently narrow. In practice, this means that the contribution to the (modified) `transfer opacity' (compare to Eq.~\eqref{eq:tauT}) from scatterings between $t$ and $t+\Delta t$, given by $\Delta\tau_{\mathrm{\widetilde T},<}(\theta_{\mathrm{c}})=\rho v \Delta t \sigma_{\mathrm{\widetilde T},<}(\theta_{\mathrm{c}})/m_\chi$ for a particle moving though a medium with mass density $\rho$ and relative velocity $v$, needs to be sufficiently small, $\Delta\tau_{\mathrm{\widetilde T},<}(\theta_{\mathrm{c}})\ll {\cal O}(1)$. The origin of this criterion can be illustrated  within the deflection test set-up discussed above, by considering the Moli\`ere approximation from Eq.~\eqref{eq: Moliere}, which corresponds to the result obtained within the effective small-angle approach. Here the transfer opacity determines the amount of Gaussian broadening. Hence, within hSIDM the condition $\Delta\tau_{\mathrm{\widetilde T},<}(\theta_{\mathrm{c}})\ll {\cal O}(1)$ ensures that the contribution to the Gaussian width of the angular distribution from small-angle scatterings, and within a single time step, is sufficiently small (compared to the full range of the azimuthal angle). This leads to
\begin{equation}
\label{eq: time steps in fSIDM}
    \Delta t_{{\rm small-angle}, i} \approx \frac{\kappa}{\max_j\left(|\Delta \mathbf{v}_{ij}|\right)} \frac{1}{m_j \,\Lambda_{ii}}\left(\frac{\sigma_{\mathrm{\widetilde T},<}(\theta_{\mathrm{c}})}{m_\chi}\right)^{-1}\,,
\end{equation}
where $\kappa$ is again an accuracy parameter (we use $\kappa= 0.1$). This condition is analogous to the one considered in~\citet{Fischer_2024a} in the context of purely forward-dominated scattering (i.e.\ fSIDM, for which $\sigma_{\mathrm{\widetilde T},<}(\theta_{\mathrm{c}})$ is replaced by $\sigma_{\mathrm{\widetilde{T}}}$), see in particular Sect.~2.3 and Appendix~B therein.

As already emphasised, the virtue of the hSIDM scheme is that for models with an enhancement of the differential cross section in the forward direction, the effective treatment of such scatterings is much more efficient as compared to the case when applying the sampling method (rSIDM approach) to the entire range of deflection angles. Indeed, within pure rSIDM, the time step in Eq.~\eqref{eq: time steps in rSIDM} would contain $\sigma_{\rm tot}$, instead of $\sigma_{\mathrm{tot},>}(\theta_{\mathrm{c}})$ as for hSIDM. Due to $\sigma_{\rm tot}\gg \sigma_{\mathrm{tot},>}(\theta_{\mathrm{c}})$, this would lead to the requirement of excessively small time steps in order to be able to treat also small-angle scatterings as individual collisions of numerical particles within a given time step. Within hSIDM, only large-angle scatterings are treated in this way, while small-angle scatterings are treated separately by the effective drag force method. This relaxes the time step requirement by one to two orders of magnitude for typical light mediator models.

\section{Validation based on deflection test set-up}
\label{sec: Deflection test}

In this section we validate the hSIDM implementation described in the previous section, and demonstrate the numerical gain in efficiency of this scheme as compared to a naive sampling of deflection angles. 

\subsection{Simulation set-up and basic validation} \label{subsec: Simulation setup deflection test}

We consider the deflection test set-up discussed in Sect.~\ref{subsec: Theory Deflection problem}, consisting of a beam of test particles moving through a target. So we compare our numerical results to the analytical solution Eq.~\eqref{eq: Goudsmit-Saunderson} for the time evolution of the distribution function $f(t,\theta)$ of azimuthal angles of beam particles relative to the initial beam axis.

We set up an initial configuration with $8\,000$ test particles, initially moving with velocity $v = 2\,\mathrm{km\,s}^{-1}$ in a common beam axis direction, through a medium described by $92\,000$ target particles, with total mass $1\times10^{10}\,\mathrm{M}_{\sun}$ in a cube with side length of $14\,\mathrm{kpc}$, corresponding to a background density $\rho = 3.353\times10^6\,\mathrm{M}_{\sun}\,\mathrm{kpc}^{-3}$. Here we use $N_{\mathrm{ngb}}=64$ neighbouring particles. Beam particles can scatter off target particles with a given differential cross section $\mathrm{d}\sigma/\mathrm{d}\Omega$. We did not allow for scattering of the beam with beam or target with target particles, and we switched off gravity for the deflection test set-up. However, each beam particle can undergo multiple scatterings, with in principle any of the target particles.

As mentioned already in Sect.~\ref{subsec: Theory Deflection problem}, we consider two configurations:
\begin{itemize}
\item[(i)] no recoil is transmitted from beam to target particles. Physically this can be realised if beam and target particles are distinguishable, and target particles are formally considered to have infinite inertia. In this case the magnitude of the initial velocity is conserved for beam particles. This case is valuable due to the known exact analytical solution Eq.~\eqref{eq: Goudsmit-Saunderson} of the angular distribution function.
\item[(ii)] recoil is transmitted from beam to target particles, assuming they have equal inertial mass. This is the physically more relevant case, realised if beam and target particles are identical particle species.
\end{itemize}
Furthermore, for the test set-up, we consider the three following methods:
\begin{itemize}
\item[(a)] hSIDM scheme: The hybrid approach introduced above, for various choices of the critical angle $\theta_{\mathrm{c}}$ to divide between small and large-angle scattering regimes.
\item[(b)] anisotropic rSIDM scheme: The case for which {\it all} scatterings are sampled explicitly from the differential cross section. We often refer to this case simply as rSIDM in this section. It corresponds to the (computationally expensive) limit of hSIDM for $\theta_{\mathrm{c}}\to 0$.\footnote{This scheme should not be confused with the isotropic rSIDM scheme considered in the context of galaxy cluster mergers in Sect.~\ref{sec: Merger simulation and analysis} below, which simply refers to isotropic self-scattering.}
\item[(c)] fSIDM scheme: The case for which {\it all} scatterings are treated according to the effective drag force and momentum diffusion approach. We refer to this case as fSIDM. It corresponds to hSIDM for $\theta_{\mathrm{c}}\to \uppi/2 \ (\uppi)$ for indistinguishable (distinguishable) particles.
\end{itemize}
We furthermore consider the various examples for differential cross sections as given in Sect.~\ref{subsec: Scattering differential cross sections}.

Since the parameters of the test set-up are chosen somewhat arbitrarily, we normalise all timescales to a given reference time $t_1$, defined to be the dynamical time $t_{\rm dyn}$ on which the angular distribution is expected to evolve significantly, see Sect.~\ref{subsec: Theory Deflection problem}. Concretely, we define $t_1$ such that the opacity $\tau(t)\equiv \rho v t\sigma/m_\chi$ satisfies $\tau(t_1)=1$ for a generic cross section $\sigma/m_\chi=1\,\mathrm{cm}^2\,\mathrm{g}^{-1}$. We note that this definition is insensitive to which type of cross section $\sigma$ stands for.

As a basic check of the implementation, we also consider an artificial case where each beam particle is allowed to scatter only {\it once} during the entire simulation. In this case the angular distribution simply follows the differential cross section, see Appendix~\ref{appendix: Single scattering implementation}. After this check, we allowed each beam particle to scatter multiple times, as determined according to the chosen scheme.

\begin{figure} 
    \centering
    \includegraphics[width=\columnwidth]{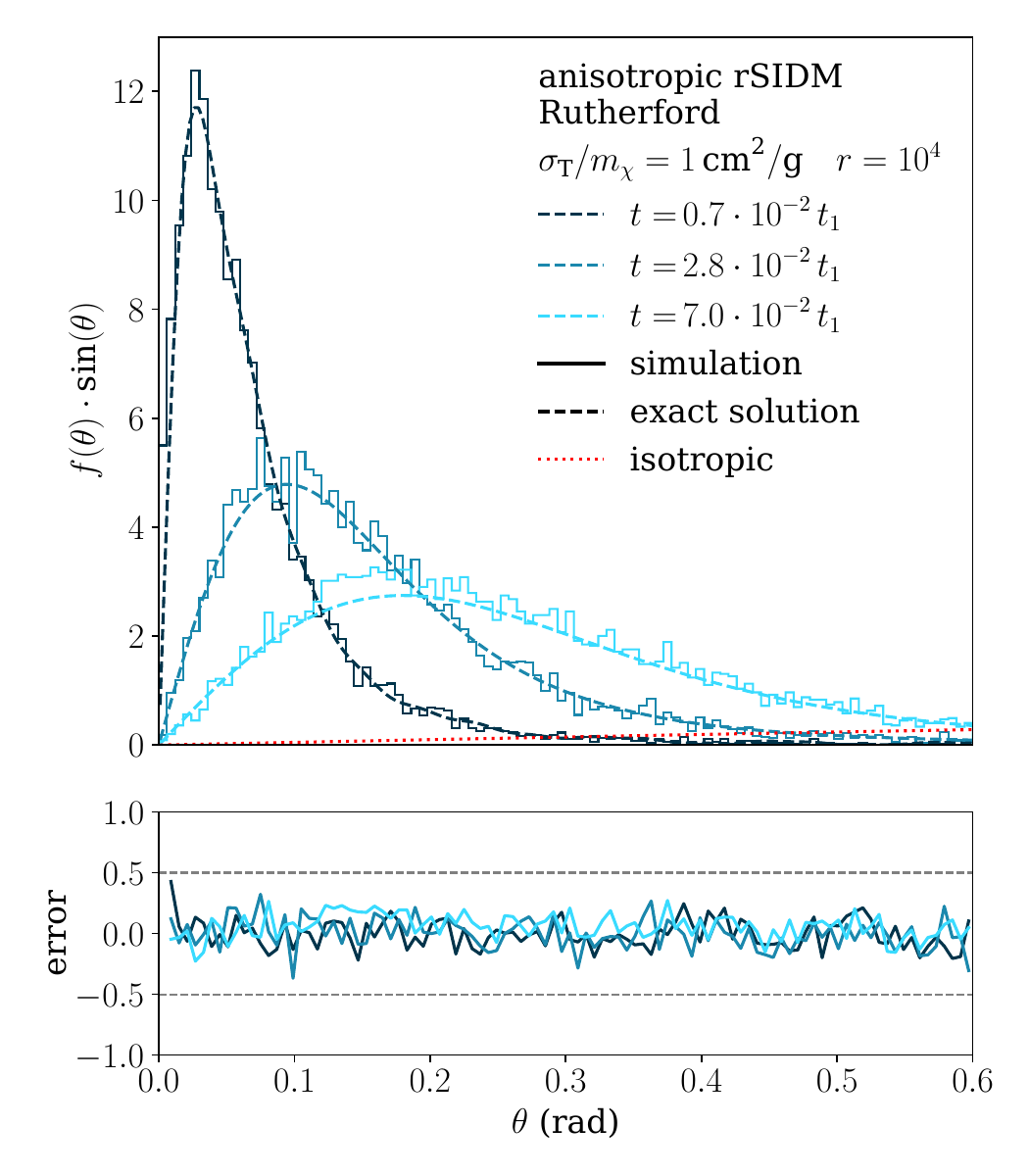}
    \caption{Validation of the implementation of angle-dependent SIDM using the explicit sampling technique to determine the deflection angles from the underlying differential cross section (anisotropic rSIDM scheme). We show the angular distribution function $f(t,\theta)\cdot\sin(\theta)$ of beam particles within the deflection test set-up for three different times $t$  (solid lines) in units of the reference time $t_1$ (see main text), and compare to the exact analytical solution for multiple scatterings in a target given in Eq.~\eqref{eq: Goudsmit-Saunderson} (dashed lines). The lower panel shows the difference, normalised to the statistical error. Here we assume the `Rutherford' cross section given in Eq.~\eqref{eq: Rutherford differential cross section} with $r=10^4$, and beam-target scattering without recoil. For illustration, also the isotropic distribution $f_\infty=1/2$ is shown (dotted) that, as we checked, is approached for $t\gg t_1$ (we note that the $x$-axis range extends only up to $\theta\leq 0.6\simeq \uppi/5$). }
    \label{fig: Goudmit Saunderson check}
\end{figure}

As a first non-trivial validity check, we consider the no-recoil case (i),\footnote{Simulating the no-recoil case requires a small modification of the SIDM implementation. The details are described in Appendix~\ref{appendix: No-recoil case implementation}.} for which the exact analytical solution is known, see Eq.~\eqref{eq: Goudsmit-Saunderson}. In Fig.~\ref{fig: Goudmit Saunderson check} we show the angular distribution function $f(t,\theta)$ of beam particles at three different times $t$ (solid lines) obtained numerically within the (anisotropic) rSIDM scheme (b), and compare them to the exact analytical solution (dashed), observing good agreement.

\begin{figure}
    \centering
    \includegraphics[width=\columnwidth]{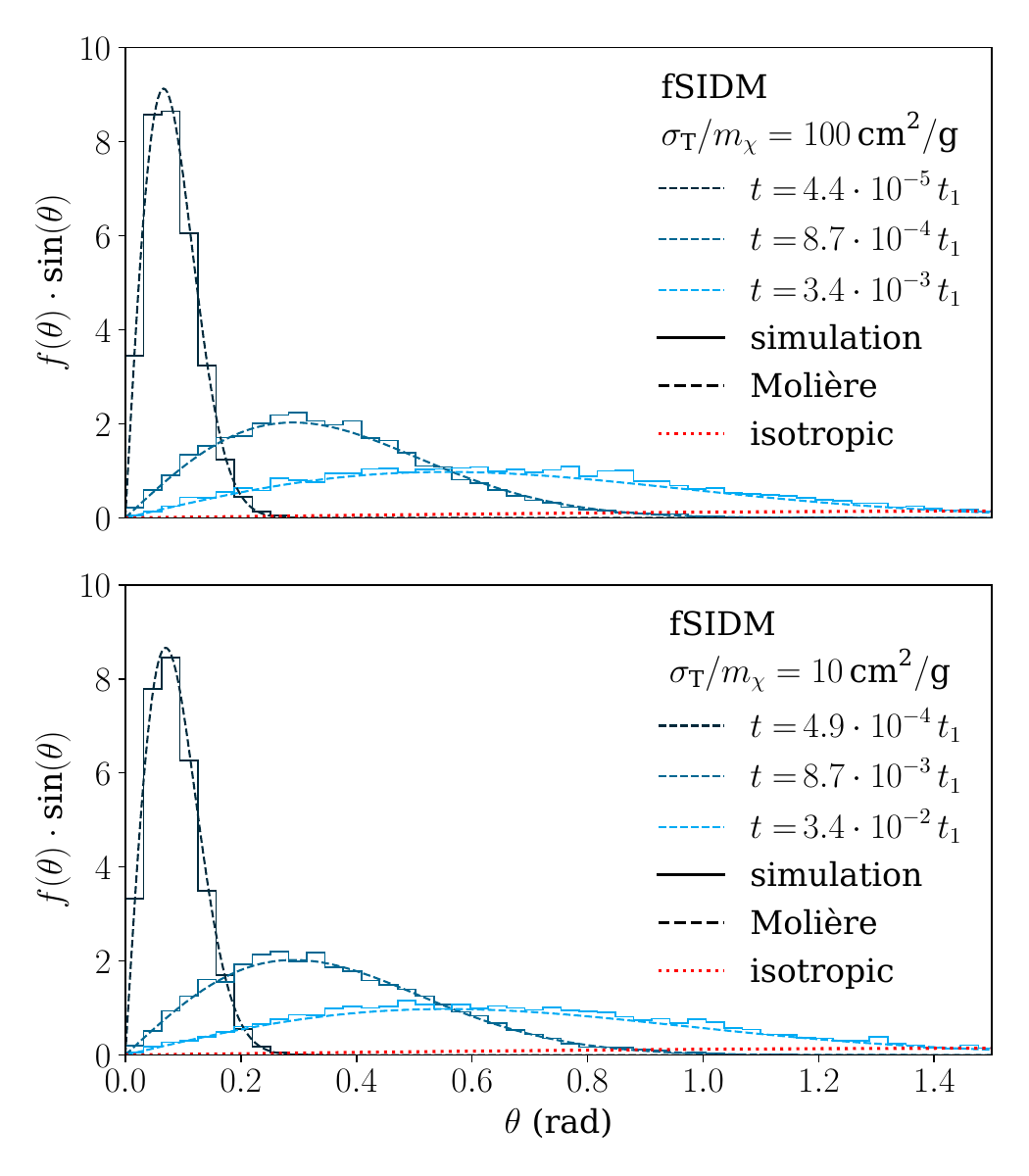}
    \caption{Validation of the implementation of angle-dependent SIDM based on the effective drag force and momentum diffusion approach (fSIDM scheme). Each panel shows the angular distribution of beam particles within the deflection test set-up at three different times $t$ (solid lines), compared to a Gaussian Moli\`ere distribution with width related to the transfer cross section (dashed lines), see  Eq.~\eqref{eq: Moliere}. Here we assume beam-target scattering without recoil. The two panels correspond to two values of the transfer cross section, differing by a factor of ten. The distribution is expected to be identical at times rescaled by a factor of one-tenth, which we observe to be the case.}
    \label{fig: Moliere vs fSIDM}
\end{figure}

In addition, we show the angular distribution obtained from the fSIDM scheme (c) within the no-recoil case (i) for three different times in Fig.~\ref{fig: Moliere vs fSIDM} (solid lines).  In this case, we compare to the Gaussian Moli\`ere distribution from Eq.~\eqref{eq: Moliere} (dashed lines). Agreement between fSIDM and Eq.~\eqref{eq: Moliere} is expected as long as the width of the Gaussian distribution is much smaller than $\uppi$, which is the case for the times $t$ shown in Fig.~\ref{fig: Moliere vs fSIDM}. As an additional check, the two panels in Fig.~\ref{fig: Moliere vs fSIDM} show results for two values of the transfer cross section that differ by a factor of ten. Due to the absence of gravitational forces in the test set-up, the solutions for these two cases are expected to match when rescaling the simulation time by a factor of one-tenth. We find that this is indeed the case, serving as an additional validation of the implementation of the time step criterion. We note that this check is similar to the one performed by~\cite{Fischer_2021a}, except that we consider the set-up without recoil (case (i)) here. This check validates that the effective treatment of small-angle scattering via a drag force and momentum diffusion leads to the expected Gaussian beam broadening. However, it does {\it not} provide a check of whether this treatment is a valid approximation for a given differential cross section and choice of critical angle. We turn to these checks next.

\subsection{Validity check of effective small-angle approach}
\label{subsec: fixed-angle scattering implementation}

The effective treatment of small-angle scattering via a drag force and momentum diffusion is valid provided the scattering is sufficiently forward-dominated, and sufficiently frequent. In this sub-section we provide a quantitative check of these intuitive validity requirements. In order to expose the validity criteria most clearly, we consider the artificial differential cross section introduced in Eq.~\eqref{eq: Delta differential cross section}, for which all individual scatterings occur with a fixed scattering angle $\theta_0$ in the centre of mass frame. Details of this fixed scattering angle case are described in Appendix~\ref{appendix: Fixed-angle deflection test}. We solve the deflection test set-up for various fixed values of $\theta_0$. Since all scatterings occur with a single angle, we expect the effective small-angle approach to work the better the smaller $\theta_0$.

In order to check this expectation, we run two simulations for each value of $\theta_0$: one using the effective small-angle approach (fSIDM, scheme (b)), and one using the explicit sampling technique (rSIDM, scheme (c)). The latter is numerically more expensive, but yields the exact angular distribution function $f(t,\theta)$ at all times $t$ (within the numerical accuracy). This allowed us to check whether the effective small-angle approach is valid by comparing the angular distribution obtained from the fSIDM simulation to the rSIDM result.

\begin{figure}
    \centering
    \includegraphics[width=\columnwidth]{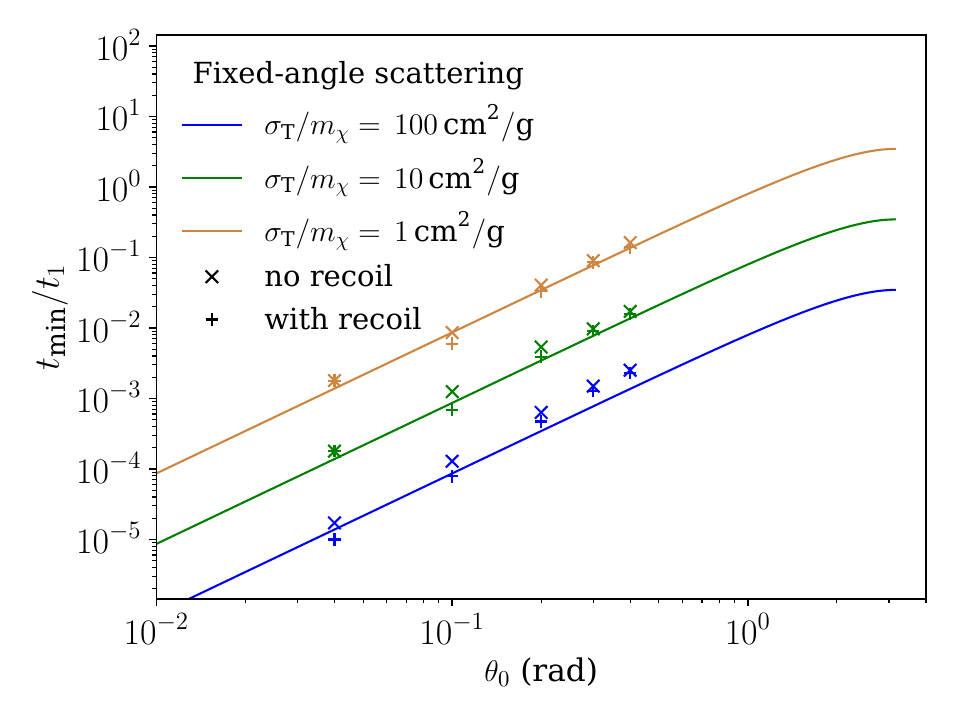}
    \caption{Validity check of the effective small-angle approach (fSIDM, scheme b) as compared to an explicit treatment of angle dependence (rSIDM, scheme c). Shown is the minimal timescale $t_{\rm min}$ after which the angular distributions obtained for the deflection test problem solved with the fSIDM and the rSIDM method agree with each other. Cross symbols show the numerically determined minimal agreement time $t_{\rm min}$, for three values of $\sigma_{\mathrm{T}}/m_\chi$ and for the set-up with and without recoil in beam-target scattering, respectively. Lines show the analytical prediction for $t_{\rm min}$ from Eq.~\eqref{eq:tmin}. Here we assume a differential cross section for which all individual scatterings occur with a fixed angle $\theta_0$ (shown on the $x$-axis). The numerically more efficient effective small-angle approach is advantageous if $t_{\rm min}$ is much smaller than the typical dynamical scale of the system, given by $t_{\rm dyn}=(1,0.1,0.01)\cdot t_1$ for $\sigma_{\mathrm{T}}/m_\chi=(1,10,100)\,\mathrm{cm}^2\,\mathrm{g}^{-1}$, respectively. This is safely the case for scatterings with $\theta\lesssim 1$. Details on the fixed-angle scattering are presented in Appendix~\ref{appendix: Fixed-angle deflection test}.}
    \label{fig: time resolution criterion delta function}
\end{figure}

We find that both angular distributions agree with each other (within numerical errors) after a finite simulation time $t_{\rm min}(\theta_0)$ which depends on the choice of $\theta_0$. By agreement we mean, that the deviation between the distribution functions is strictly smaller than a certain threshold for all angles. Numerically, we carry this out by subtracting both angular distributions and normalising it to the square root of the number of particles in each bin. In our case, we use an error threshold of $0.5$. This minimal time is shown by the cross symbols in 
Fig.~\ref{fig: time resolution criterion delta function} (relative to the common reference time $t_1$ introduced above), for three different values of the transfer cross section, and for the case (i) without recoil as well as the case (ii) with recoil, respectively.

This finding is in line with the analytical discussion in Sect.~\ref{sec:fSIDMvalidity}, based on the analysis of the exact analytical solution of the deflection problem in the no-recoil case. Indeed, our numerical results precisely follow the expected behaviour; that is, the fSIDM approach becomes valid only after some minimal time that is required to lead to sufficiently many scatterings for the effective drag force description to apply. We find that the dependence of the minimal time $t_{\rm min}$ derived analytically in Eq.~\eqref{eq:tmin} on $\theta_0$ as well as on the overall size of the cross section (shown by solid lines in Fig.~\ref{fig: time resolution criterion delta function}) agrees with our results for the numerically determined minimal time $t_{\rm min}(\theta_0)$ (cross symbols).\footnote{We note that the time $t_{\rm min}$ defined in Eq.~\eqref{eq:tmin} and the numerically determined $t_{\rm min}$ differ by a constant factor of the order of ten, that is, independent of $\theta_0$ and $\sigma_{\mathrm{T}}/m_\chi$. We attribute this to the precise definition for when the angular distributions obtained numerically within the fSIDM and rSIDM schemes agree with each other within numerical uncertainties. The solid lines in Fig.~\ref{fig: time resolution criterion delta function} are rescaled by this global common factor as compared to Eq.~\eqref{eq:tmin} in order to show that the dependence of the analytically derived and of the numerically extracted values of $t_{\rm min}$ on both $\theta_0$ and $\sigma_{\mathrm{T}}/m_\chi$ is (practically) identical.} This demonstrates that the validity criteria derived analytically in Sect.~\ref{sec:fSIDMvalidity} are applicable in practice. Moreover, our numerical results for the deflection test cases with and without recoil suggest that the validity criteria derived for the no-recoil set-up can be carried over to the case with recoil. This is an important observation, since all our later results are based on scatterings among identical particles, including recoil. 

As can be observed in Fig.~\ref{fig: time resolution criterion delta function}, the minimal time after which the fSIDM approach becomes applicable is the shorter the smaller $\theta_0$. This agrees with the intuitive notion that the fSIDM approach works the better the smaller the (typical) scattering angles are in individual collisions (which are given by $\theta_0$ for the differential cross section considered in this sub-section, and in general bounded from above by $\theta_{\mathrm{c}}$ within the hSIDM approach). As discussed in Sect.~\ref{sec:fSIDMvalidity}, the fSIDM approach is useful if $t_{\rm min}\ll t_{\rm dyn}$, where the latter is the dynamical timescale over which the angular distribution evolves significantly. For the deflection set-up, $t_{\rm dyn}=(1,0.1,0.01)\cdot t_1$ for $\sigma_{\mathrm{T}}/m_\chi=(1,10,100)\,\mathrm{cm}^2\,\mathrm{g}^{-1}$, respectively. We find
\be
  t_{\rm min}/t_{\rm dyn} \simeq 10^{-2}\cdot\left(\frac{\theta_0}{0.1}\right)^2\,,
\ee
as long as $\theta_0\lesssim 1$, independently of $\sigma_{\mathrm{T}}/m_\chi$. The dependence on $\theta_0^2$ as well as the independence of the ratio $t_{\rm min}/t_{\rm dyn}$ on the {\it overall} size of the cross section is consistent with the analytical result Eq.~\eqref{eq:tmin}, as stated above. 

Altogether, this implies scatterings by angles $\theta\lesssim 1$ can be accurately treated within the effective small-angle (fSIDM) approach. This is valid on timescales that are one to two orders of magnitude shorter than the typical dynamical scale of the system.

\subsection{Validity check of hSIDM approach}
 \label{subsec: Rutherford and Moeller scattering implementation}

Let us now return to the hSIDM approach for efficiently describing angle-dependent DM self-scattering.
We solve the deflection test problem using the hSIDM scheme (a) employing various critical angles $\theta_{\mathrm{c}}$, see Sect.~\ref{subsec: Simulation setup deflection test}. We consider first the case (i) without recoil, for which we can compare the angular distribution $f(t,\theta)$ to the analytical solution Eq.~\eqref{eq: Goudsmit-Saunderson}.
We assume a `Rutherford' cross section given in Eq.~\eqref{eq: Rutherford differential cross section}, with three values of the anisotropy parameter (related to the mediator-to-dark matter mass ratio)  $r=10^2,10^3,10^4$. 

\begin{figure*}
    \centering
    \includegraphics[width=\columnwidth]{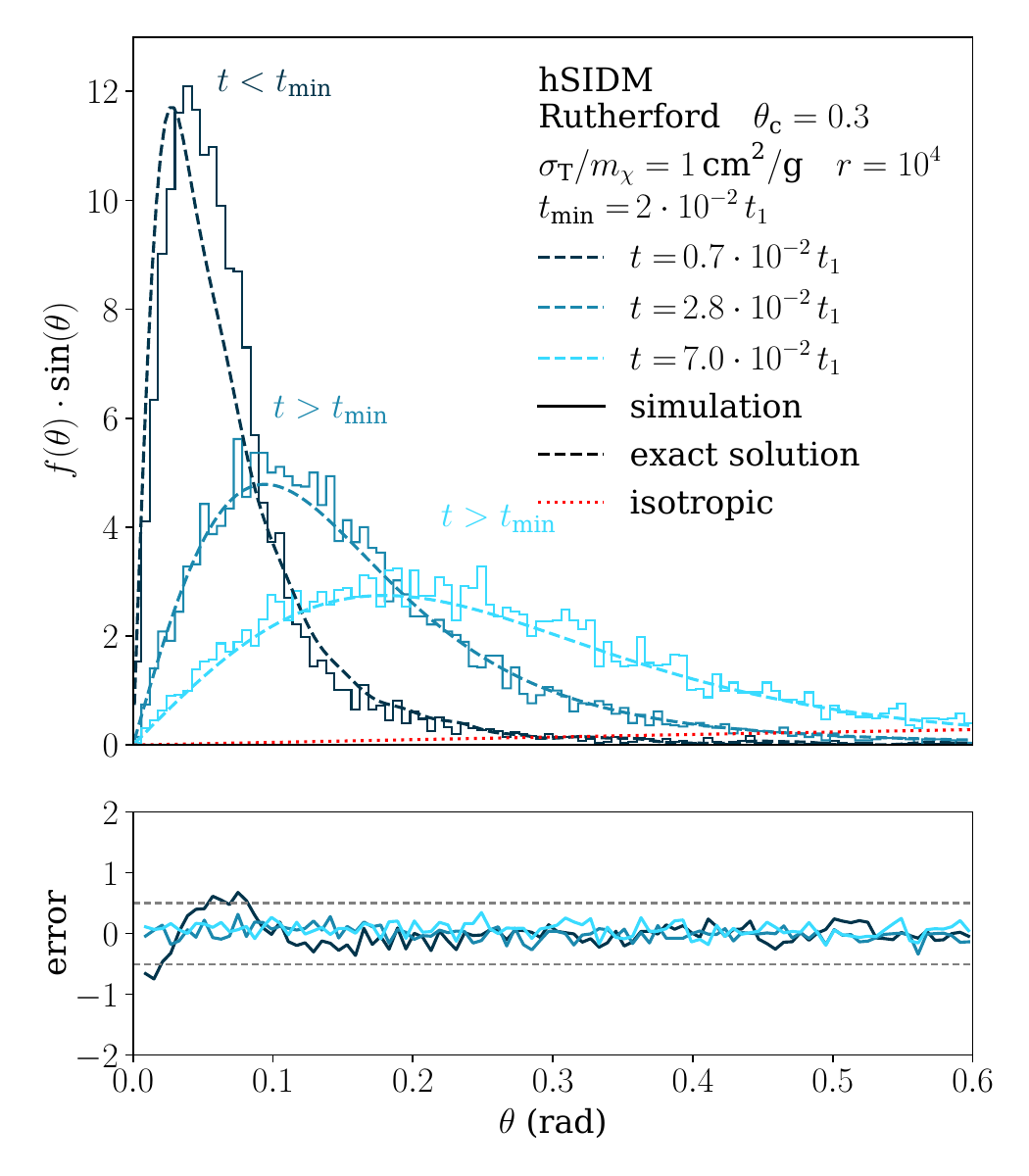}
    \includegraphics[width=\columnwidth]{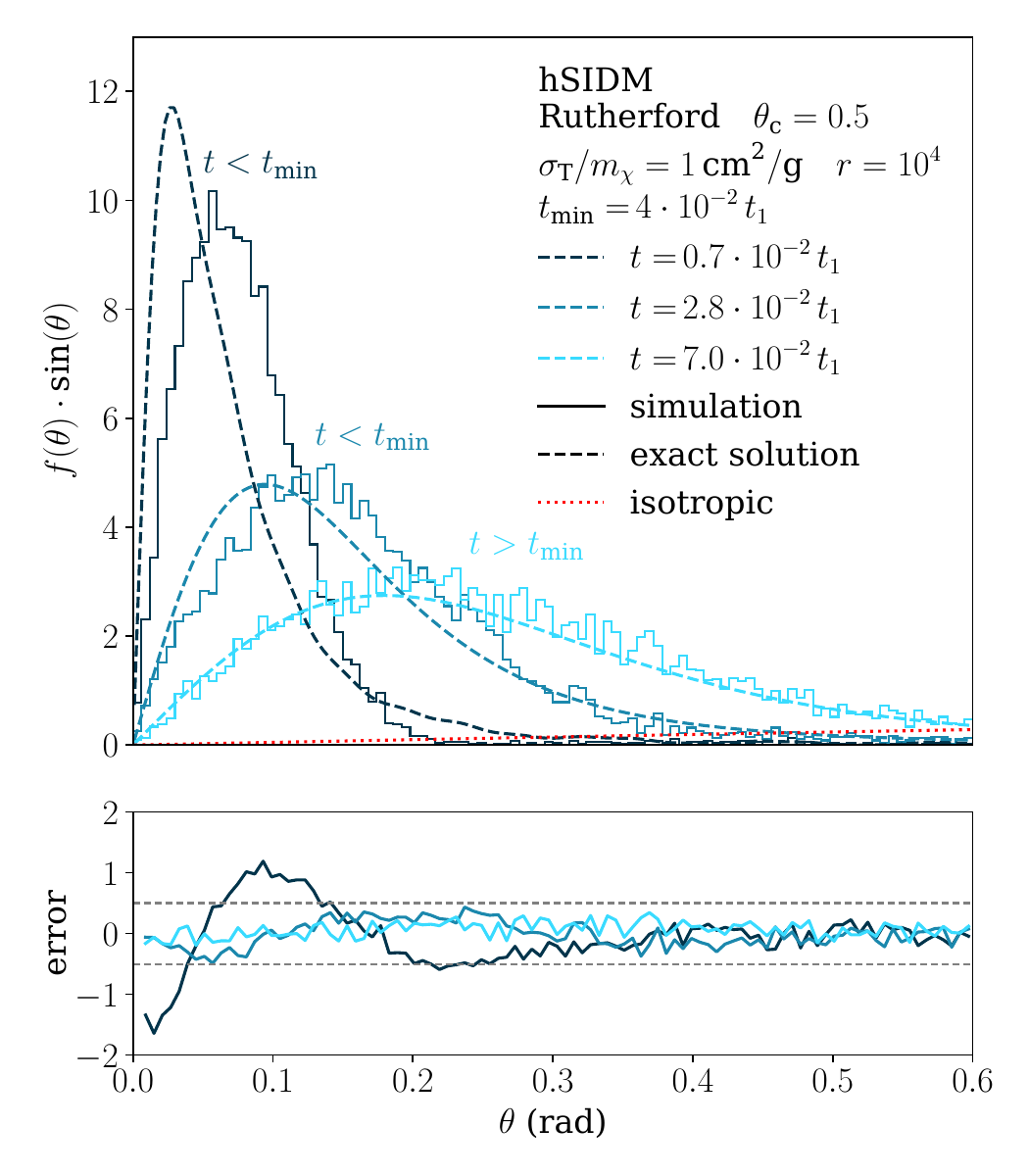}
    \caption{Angular distribution $f(t,\theta)\cdot\sin(\theta)$ of beam particles within the deflection test set-up as obtained within the hSIDM scheme at three (very early) times $t$ (solid lines) and for two values of the critical angle, $\theta_{\mathrm{c}}=0.3$ (left panel) and $\theta_{\mathrm{c}}=0.5$ (right panel), respectively. The angular distribution is expected to agree with the analytical result Eq.~\eqref{eq: Goudsmit-Saunderson} (dashed lines) after some minimal time $t>t_{\rm min}(\theta_{\mathrm{c}})$ for which scatterings by angles $\theta<\theta_{\mathrm{c}}$ are sufficiently frequent such that the effective drag force and momentum diffusion approach applies.
    The agreement is indeed reached after the expected time $t_{\rm min}(0.3)\simeq 2\cdot 10^{-2}t_1$ and $t_{\rm min}(0.5)\simeq 4\cdot 10^{-2}t_1$, respectively. We note that hSIDM is thus well applicable on the relevant dynamical timescale $t\sim t_{\rm dyn}(\equiv t_1)$. }
    \label{fig: hybrid vs rare}
\end{figure*}

The angular distribution of beam particles at three different simulation times is shown in Fig.~\ref{fig: hybrid vs rare} for $r=10^4$. As expected, the hSIDM approach yields an angular distribution that agrees with the analytical solution only after some minimal time $t>t_{\rm min}(\theta_{\mathrm{c}})$. The reason is that scatterings by small angles ($\theta<\theta_{\mathrm{c}}$) are treated with the effective drag force and momentum diffusion method within hSIDM, and this approach requires a minimal time after which scatterings are sufficiently frequent to apply, as discussed in detail above. We observe that for the smaller critical angle $\theta_{\mathrm{c}}=0.3$ (left panel in  Fig.~\ref{fig: hybrid vs rare}) the agreement is reached at earlier times as compared to $\theta_{\mathrm{c}}=0.5$ (right panel), i.e.\ $t_{\rm min}(0.3)<t_{\rm min}(0.5)$. This is also expected, since in the former case only scatterings by smaller angles are included in the effective small-angle approach, such that it becomes valid earlier on.

\begin{figure}
        \centering
        \includegraphics[width=\columnwidth]{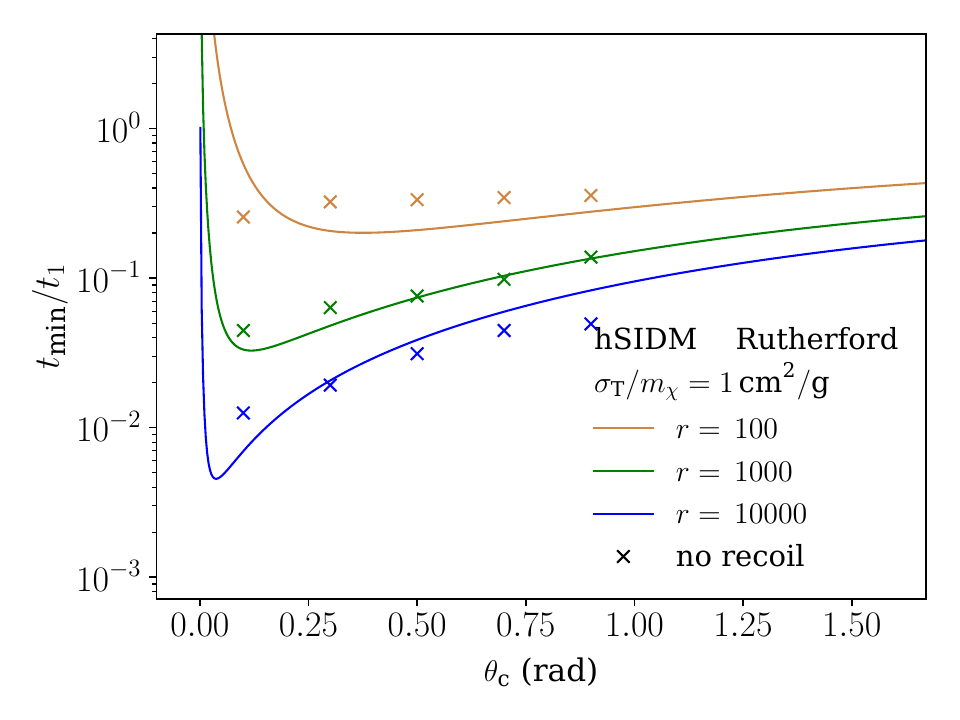}
        \includegraphics[width=\columnwidth]{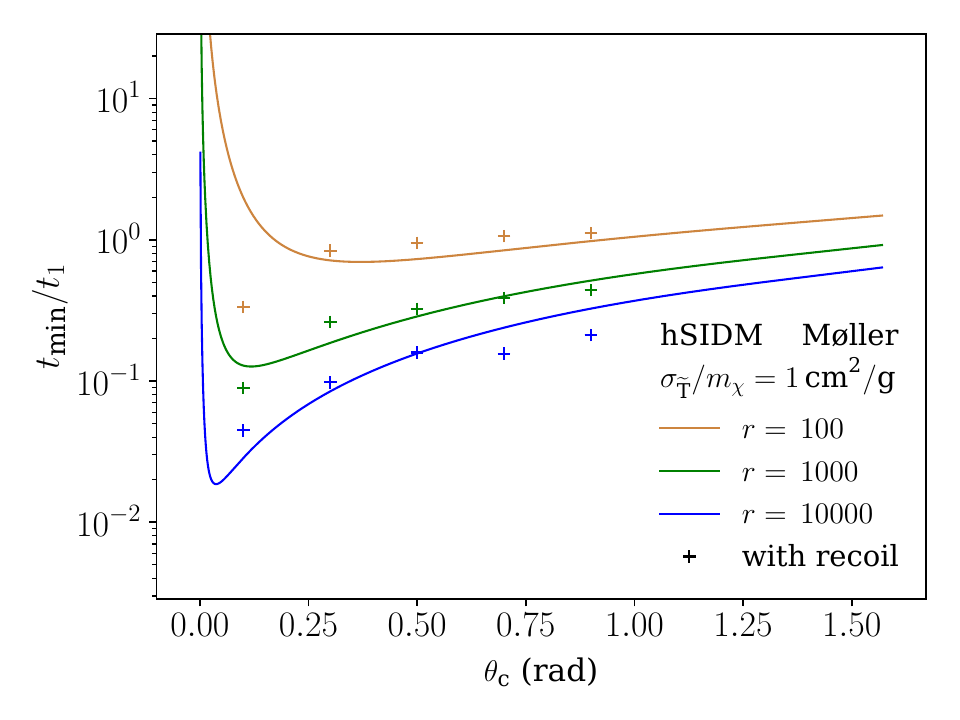}
    \caption{Minimal timescale $t_{\rm min}(\theta_{\mathrm{c}})$ after which the hSIDM approach is applicable (extracted from the deflection test set-up). The symbols are derived from a comparison of numerical solutions using the hSIDM scheme with the result obtained when using the numerically expensive explicit sampling technique for all scattering angles (anisotropic rSIDM scheme).
    We show results for $\theta_{\mathrm{c}}=0.1,0.3,0.5,0.7,0.9$, and for differential cross sections with anisotropy parameter $r=10^2,10^3,10^4$, respectively.  We furthermore consider the case without recoil and with `Rutherford' scattering cross section (upper panel) as well we the case with recoil and with `M{\o}ller' cross section (lower panel). Solid lines show the dependence of the analytical prediction Eq.~\eqref{eq:tminhSIDM} for $t_{\rm min}(\theta_{\mathrm{c}})$ on $\theta_{\mathrm{c}}$ and on $r$.}
    \label{fig: time criterion Rutherford+Moeller}
\end{figure}

We note that $t_{\rm min}(\theta_{\mathrm{c}})\ll t_{\rm dyn}(=t_1)$ is much smaller than the dynamical timescale for both values of $\theta_{\mathrm{c}}$ shown in Fig.~\ref{fig: hybrid vs rare}. This means the hSIDM approach yields a valid description on the relevant timescale of order $t_{\rm dyn}$.
In the upper panel of Fig.~\ref{fig: time criterion Rutherford+Moeller} we show $t_{\rm min}(\theta_{\mathrm{c}})$ as obtained from hSIDM simulations with various critical angles $\theta_{\mathrm{c}}=0.1,0.3,0.5,0.7,0.9$, and three values of $r$ (cross symbols). In addition, the dependence of the analytical prediction for $t_{\rm min}$ from Eq.~\eqref{eq:tminhSIDM} on $\theta_{\mathrm{c}}$ and on $r$ is shown (solid lines). We observe that the analytical result yields a useful estimate for the dependence of the minimal validity timescale on the model parameters and on the critical angle. We note that for very small values of $\theta_{\mathrm{c}}<0.1$, the analytical result for $t_{\rm min}(\theta_{\mathrm{c}})$ strongly increases. This is due to the fact that for such small values of the critical angle, only a very small range of scattering angles is treated via the drag force method, such that it formally leads to sufficiently frequent scatterings only after a rather long time. However, we notice that for $\theta_{\mathrm{c}}=0.1$ the value of $t_{\min}$ extracted from our numerical solutions is actually still rather small (left-most crosses in Fig.~\ref{fig: time criterion Rutherford+Moeller}). We attribute this to the fact that the physical impact of scatterings with $\theta>\theta_{\mathrm{c}}$ is dominant for such small values of $\theta_{\mathrm{c}}$. Since those are treated based on the explicit sampling technique within hSIDM, the potentially larger relative error of the small-angle regime is not noticeable in the complete angular distribution function for $\theta_{\mathrm{c}}=0.1$. 

Next, we considered the deflection test set-up with recoil, see case (ii) described in Sect.~\ref{subsec: Simulation setup deflection test}. We use a `M{\o}ller' cross section given in Eq.~\eqref{eq: Moeller differential cross section} (as appropriate for beam and target particles with identical mass), again with three values $r=10^2,10^3,10^4$ and for $\theta_{\mathrm{c}}=0.1,0.3,0.5,0.7,0.9$, respectively. Since no analytical result is available in this case, we instead compare the angular distribution when using the hSIDM scheme (a) to the one obtained when using the sampling technique for scatterings by all angles (anisotropic rSIDM scheme (b)), see  Sect.~\ref{subsec: Simulation setup deflection test}. Similarly as before, we find that the angular distributions $f(t,\theta)$ obtained from hSIDM and rSIDM agree after some minimal time $t_{\rm min}(\theta_{\mathrm{c}})$, shown by the plus symbols in the lower panel of Fig.~\ref{fig: time criterion Rutherford+Moeller}. The dependence of $t_{\rm min}(\theta_{\mathrm{c}})$ on $\theta_{\mathrm{c}}$ and on $r$ is similar as for the no-recoil case (i). For comparison, we also show the dependence of the analytical prediction for $t_{\rm min}(\theta_{\mathrm{c}})$  from Eq.~\eqref{eq:tminhSIDM} on $\theta_{\mathrm{c}}$ and on $r$ by the solid lines in the lower panel of Fig.~\ref{fig: time criterion Rutherford+Moeller}. We stress that this analytical result has been derived within the no-recoil case. Nevertheless, we find that it yields a useful analytical estimate of $t_{\rm min}(\theta_{\mathrm{c}})$ also in the physically relevant case with recoil, similarly as observed before for the set-up considered in Sect.~\ref{subsec: fixed-angle scattering implementation}. 

Thus, we find that the validity criteria for applying the effective small-angle approach derived in Sect.~\ref{sec:hSIDMvalidity} are applicable in practice also beyond the strict set-up for which they have been derived, and yield a useful quantitative estimate for when the hSIDM approach can be used.
Finally, we notice that $t_{\rm min}(\theta_{\mathrm{c}})$ is significantly smaller than the dynamical timescale $t_{\rm dyn}(\equiv t_1)$ as long as $\theta_{\mathrm{c}}$ is sufficiently small for both cases with (i) and without (ii) recoil. In addition, the ratio $t_{\rm min}/t_{\rm dyn}$ is the smaller the larger $r$. This implies the hSIDM approach is valid particularly for very strongly forward-dominated differential cross sections. Since the gain in numerical efficiency is also most pronounced for large $r$, this implies that hSIDM is indeed applicable in the cases for which it is expected to be most advantageous from the point of view of computational resources. We note that as an additional check, we explicitly verified that all results presented in Sect.~\ref{sec: Merger simulation and analysis} are insensitive to the choice of critical angle $\theta_{\mathrm{c}}$, being particularly relevant for moderate values of the anisotropy parameter $r\lesssim {\cal O}(10^2)$. 

For the explanations given above, we assumed that the dynamical time scale is set by the self-interactions. This is not true in general, for example, the evolution of astrophysical systems typically involves gravity setting its own time scale. There is no positive value for $\theta_\mathrm{c}$ that can always be safely used. Instead, accurately modelling the evolution of a system requires to choose the critical angle in respect to its dynamical time scale (i.e.\ $t_\mathrm{min}(\theta_\mathrm{c}) \ll t_\mathrm{dyn}$). We come back to this point in case of galaxy cluster mergers in Sect.~\ref{sec: Merger simulation and analysis}.

\subsection{Performance gain due to the hSIDM approach} \label{subsec: hSIDM implementation performance}

\begin{figure}
\centering
\text{\quad \quad \quad \quad hSIDM \quad M{\o}ller \quad $\sigma_{\mathrm{\widetilde{T}}}/m_{\chi} = 1\,\mathrm{cm}^2\,\mathrm{g}^{-1}$ \quad $r = 10^4$ }\par
        \includegraphics[width=0.5\textwidth]{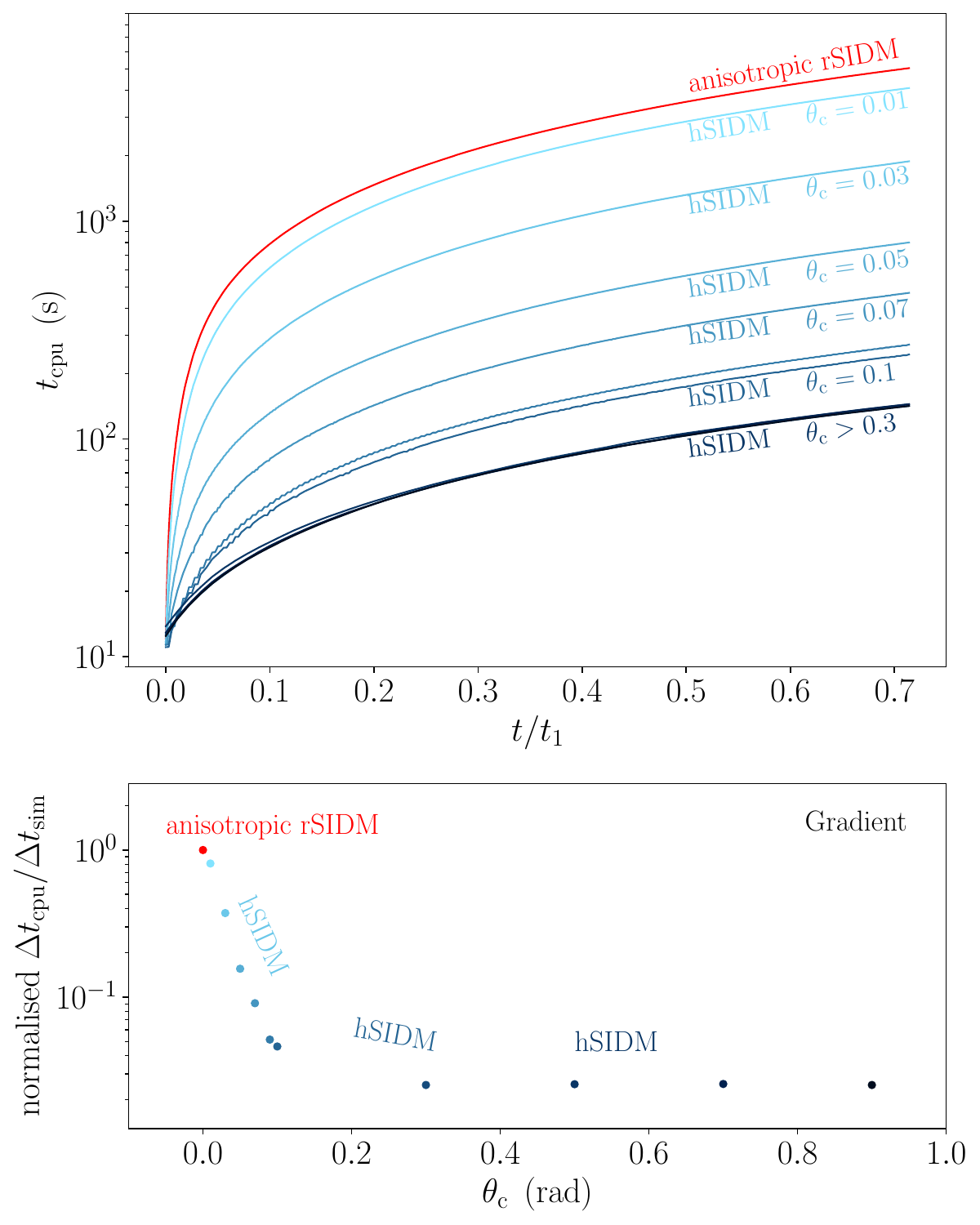}
    \caption{Performance gain of the hybrid scheme (hSIDM) developed in this work for taking the angle dependence of DM self-scattering (characteristic for e.g.\ light mediator models) into account in $N$-body simulations. We show the CPU run-time $t_{\rm cpu}$ (executed on a single CPU only) required for advancing the deflection test problem up to physical simulation time $t$ (in units of the typical reference timescale $t_1$) in the upper panel. The red line corresponds to the naive scheme for which scatterings by {\it all} angles are sampled from $\mathrm{d}\sigma/\mathrm{d}\Omega$ (anisotropic rSIDM). The CPU-time within the hSIDM approach is shown for runs with various values of the critical angle $\theta_{\mathrm{c}}$ used to separate effective small- and explicit large-angle regimes. The CPU-time is reduced by almost a factor $100$ for $\theta_{\mathrm{c}}\sim 0.3$, and already significantly smaller for $\theta_{\mathrm{c}}\sim 0.1$. The lower panel shows the amount of CPU run-time $\Delta t_{\rm cpu}$ required to simulate a given physical simulation time interval $\Delta t_{\rm sim}$, which is roughly $t$-independent. The $y$-axis is normalised to unity for the anisotropic rSIDM scheme. The CPU run-time decreases strongly for $\theta_{\mathrm{c}}\lesssim 0.1$, and saturates for $\theta_{\mathrm{c}}\sim 0.3$. For even larger critical angles, the numerical cost of the effective small-angle approach becomes relevant as compared to the explicit treatment of large-angle scatterings within hSIDM. }
    \label{fig: cpu plot}
\end{figure}

The hSIDM approach to angle-dependent SIDM allows us to simulate differential cross sections that are characteristic for realistic (e.g.\ light mediator) models in a numerically significantly more efficient way as compared to a naive approach for which scatterings by {\it all} angles are sampled from $\mathrm{d}\sigma/\mathrm{d}\Omega$. Instead, the division into small- and large-angle scattering regimes and the treatment of the former by an effective drag force and momentum diffusion approach within hSIDM speeds up the numerical performance. In this section we quantify the numerical efficiency. We consider again the deflection test set-up, 
using the variant with recoil (case (ii) in Sect.~\ref{subsec: Simulation setup deflection test}) and assuming a `M{\o}ller' cross section, see Eq.~\eqref{eq: Moeller differential cross section}, with $r=10^4$. We perform runs using the hSIDM approach, with various values for the critical angle $\theta_{\mathrm{c}}$ used to separate small and large-angle regimes. We compare to the naive approach for which all scattering angles are treated with the sampling technique (anisotropic rSIDM), which corresponds to hSIDM in the limit $\theta_{\mathrm{c}}\to 0$. 

The required CPU run-time $t_{\rm cpu}$ is shown in the upper panel of Fig.~\ref{fig: cpu plot}, versus the simulation time $t$. We observe that the CPU run-time significantly decreases when increasing $\theta_{\mathrm{c}}$. We note that the physical results for the distribution of beam particles are compatible for all shown cases of $\theta_{\mathrm{c}}$, i.e.\ hSIDM yields a valid description, but at a numerical cost that is almost a factor $100$ lower for $\theta_{\mathrm{c}}\sim 0.3$. This is comparable to (but slightly better than) the theoretically expected speed-up, see Eq.~\eqref{eq:effgain} and Fig.~\ref{fig: contribution of total and transfer cross section}.

The lower panel of Fig.~\ref{fig: cpu plot} shows the required CPU run-time $\Delta t_{\rm cpu}$ per simulation time-interval $\Delta t_{\rm sim}$, which is roughly $t$-independent. We normalise the $y$-values to unity for the anisotropic rSIDM scheme, and show the performance gain within hSIDM when increasing the value of the critical angle $\theta_{\mathrm{c}}$. The improvement is most drastic for values $\theta_{\mathrm{c}}\sim 0.1$, and saturates for $\theta_{\mathrm{c}}\gtrsim 0.3$. The saturation indicates that the numerical cost of the scatterings treated via the effective small-angle approach becomes noticeable for these values.

In summary, we verified the applicability and accuracy of the hSIDM approach. In particular, we demonstrated that it yields an improvement in computational performance by up to a factor $100$ for models of SIDM featuring a differential cross section with enhancement at small scattering angles, as is typical for light mediator models.

\section{Simulation of merging galaxy clusters}
\label{sec: Merger simulation and analysis}

\begin{table*}
    \caption{Simulated SIDM models and parameters.}
    \centering
    \begin{tabular}{|c|c|c|c|c|c|c|c|c|}
       \hline 
       & & & & & & & &\\[-1.5ex]
       numerical & $\mathrm{d}\sigma/\mathrm{d}\Omega$ & $\theta_{\mathrm{c}} $ & $r$ & $\sigma_{\mathrm{V}}/m_{\chi}$ & $\sigma_{\mathrm{\widetilde{T}}}/m_{\chi}$ & $\sigma_{\mathrm{tot}}/m_{\chi}$ & $\sigma_{\mathrm{\widetilde{T}},<}/\sigma_{\mathrm{\widetilde{T}}}$ & $\sigma_{\mathrm{tot},>}/\sigma_{\mathrm{tot}}$\\ 
       scheme  &  &  &  & ($\mathrm{cm}^2\,\mathrm{g}^{-1}$) & ($\mathrm{cm}^2\,\mathrm{g}^{-1}$) & ($\mathrm{cm}^2\,\mathrm{g}^{-1}$) & &\\[1.1ex] 
        \hline
        & & & & & & & &\\[-1.5ex]
          CDM & - & -& -& -& -& -& -& -\\
         fSIDM & forward limit & -& -& $1$ & $2/3$ & $\infty$ & -&-\\
         rSIDM & isotropic & -& -& $1$ & $1$ & $1$ & -&-\\
         hSIDM & M{\o}ller & $0.1$ & $0.1$ & $1$ & $1.0$ & $1.0$ & $2.5\times10^{-5}$ & $0.99$\\ 
         hSIDM & M{\o}ller & $0.3$ & $0.1$ & $1$ & $1.0$ & $1.0$ & $2.0\times10^{-3}$ & $0.96$\\
         hSIDM & M{\o}ller & $0.1$ & $1$ & $1$ & $0.98$ & $1.1$ & $3.8\times10^{-5}$ & $0.99$\\ 
         hSIDM & M{\o}ller & $0.3$ & $1$ & $1$ & $0.98$ & $1.1$ & $3.0\times10^{-3}$ & $0.94$\\
         hSIDM & M{\o}ller & $0.1$ & $10$ & $1$ & $0.87$ & $2.0$ & $3.5\times10^{-4}$ & $0.97$\\ 
         hSIDM & M{\o}ller & $0.3$ & $10$ & $1$ & $0.87$ & $2.0$ & $2.2\times10^{-2}$ & $0.77$\\
         hSIDM & M{\o}ller & $0.1$ & $10^2$ & $1$ & $0.77$ & $7.0$ & $8.8\times10^{-3}$ & $0.79$\\ 
         hSIDM & M{\o}ller & $0.3$ & $10^2$ & $1$ & $0.77$ & $7.0$ & $0.18$ & $0.28$\\
         hSIDM & M{\o}ller & $0.1$ & $10^3$ & $1$ & $0.73$ & $37$ & $0.11$ & $0.28$\\ 
         hSIDM & M{\o}ller & $0.3$ & $10^3$ & $1$ & $0.73$ & $37$ & $0.45$ & $3.8\times10^{-2}$\\
         hSIDM & M{\o}ller & $0.1$ & $10^4$ & $0.2$ & $0.14$ & $50$ & $0.32$ & $3.8\times10^{-2}$\\ 
         hSIDM & M{\o}ller & $0.3$ & $10^4$ & $0.2$ & $0.14$ & $50$ & $0.62$ & $4.0\times10^{-3}$\\
         hSIDM & M{\o}ller & $0.1$ & $10^4$ & $1$ & $0.71$ & $2.5\times10^{2}$ & $0.32$ & $3.8\times10^{-2}$\\  
         hSIDM & M{\o}ller & $0.3$ & $10^4$ & $1$ & $0.71$ & $2.5\times10^{2}$ & $0.62$ & $4.0\times10^{-3}$\\
         hSIDM & M{\o}ller & $0.1$ & $10^4$ & $1.4$ & $1$ & $3.5\times10^{2}$ & $0.32$ & $3.8\times10^{-2}$\\ 
         hSIDM & M{\o}ller & $0.3$ & $10^4$ & $1.4$ & $1$ & $3.5\times10^{2}$ & $0.62$ & $4.0\times10^{-3}$\\
         hSIDM & M{\o}ller & $0.1$ & $10^4$ & $3$ & $2.1$ & $7.4\times10^{2}$ & $0.32$ & $3.8\times10^{-2}$\\ 
         hSIDM & M{\o}ller & $0.3$ & $10^4$ & $3$ & $2.1$ & $7.4\times10^{2}$ & $0.62$ & $4.0\times10^{-3}$\\
         hSIDM & M{\o}ller & $0.1$ & $10^5$ & $1$ & $0.70$ & $1.8\times10^{3}$ & $0.48$ & $3.9\times10^{-3}$\\ 
         hSIDM & M{\o}ller & $0.3$ & $10^5$ & $1$ & $0.70$ & $1.8\times10^{3}$ & $0.71$ & $4.0\times10^{-4}$\\[1.1ex]
         \hline
    \end{tabular}
    \vspace*{1mm}
    \tablefoot{Summary of SIDM models and parameters used in the galaxy cluster mergers simulations. The first column gives the numerical scheme employed to model the self-interactions, followed by the second column specifying the differential cross section. For the M{\o}ller differential cross section, we use Eq.~\eqref{eq: Moeller differential cross section}. The subsequent columns give the critical angle $\theta_{\mathrm{c}}$, the anisotropy parameter $r$, the viscosity cross section, the modified transfer cross section and the total cross section. The second last column specifies the fraction of the momentum transfer that is treated with the small-angle approximation and the last column gives the fraction of scattering events that are simulated by explicitly sampling the scattering angle. See Sect.~\ref{subsec: Angle dependent scattering-cross section} for the definitions of the various angle-averaged cross sections and Sect.~\ref{subsec: Hybrid self-interaction} for discussion related to the last two columns. 
    The critical angle $\theta_{\mathrm{c}}$ used in the hSIDM scheme to separate large and small-angle regimes is a technical parameter, and we check independence of our results by comparing runs with two values. The model parameter $r$, related to the ratio of mediator and dark matter mass, determines the amount of anisotropy of the differential cross section $\mathrm{d}\sigma/\mathrm{d}\Omega$.}
    \label{tab: simulation runs}
\end{table*}

In this section, we apply the hSIDM method for efficiently treating dark matter self-interaction with angular dependence characteristic for light mediator models. Specifically, we simulate a merger of two galaxy clusters of equal mass, and compare the dynamical evolution of spatial dark matter and galaxy distributions under the influence of gravity as well as dark matter self-interactions with a differential cross section given by Eq.~\eqref{eq: Moeller differential cross section}. This corresponds to self-interaction via exchange of a light mediator in the Born approximation for indistinguishable  particles, to which we refer as `M{\o}ller'-scattering. In order to demonstrate the impact of the angular dependence, we fix the anisotropy parameter $r$ entering in Eq.~\eqref{eq: Moeller differential cross section} for all of our galaxy cluster simulations.\footnote{A complete treatment including also the dependence of the differential cross section on the relative velocity in each scattering event is left for future work.}

For the remainder of this work, we consider the following configurations:
\begin{itemize}
\item[(A)] hSIDM: Angle-dependent self-interaction with a `M{\o}ller' cross section (Eq.~\eqref{eq: Moeller differential cross section}) for various values of $r$ in the range $0.1\leq r\leq 10^5$, using the hybrid scheme characterised by a dedicated treatment of small- and large-angle scatterings separated by a critical angle $\theta_{\mathrm{c}}$. We verify that all our results are independent of $\theta_{\mathrm{c}}$, by checking agreement between runs with $\theta_{\mathrm{c}}=0.1$ and $\theta_{\mathrm{c}}=0.3$.
\item[(B)] isotropic rSIDM: Isotropic self-interaction (`billiard ball scattering'). This commonly considered case can be viewed as arising from taking the `heavy mediator' limit $r\to 0$ in Eq.~\eqref{eq: Moeller differential cross section}, for which $\sigma_{\rm tot}/\sigma_{\mathrm{\widetilde{T}}}\to 1$. We often simply refer to this case as rSIDM in this section.\footnote{This scheme should not be confused with the anisotropic rSIDM scheme considered in the context of the deflection test in Sect.~\ref{sec: Deflection test} above, which refers to hSIDM in the limit $\theta_{\mathrm{c}}\to 0$, i.e.\ the computationally expensive explicit sampling technique for scatterings of {\it all} angles from a given differential cross section.}
\item[(C)] fSIDM: Purely forward-dominated scattering \citep[see e.g.][]{Fischer_2021a}. Here {\it all} dark matter self-interactions are modelled by the effective drag force and momentum diffusion technique. It can be considered as the formal `massless mediator' limit obtained for $r\to \infty$ in Eq.~\eqref{eq: Moeller differential cross section} (while keeping the modified transfer cross section $\sigma_{\mathrm{\widetilde{T}}}$ fixed), for which $\sigma_{\rm tot}/\sigma_{\mathrm{\widetilde{T}}}\to \infty$.
\item[(D)] CDM: For comparison, we also consider the case of collisionless dark matter.
\end{itemize}
The parameters for each run considered in this work are summarised in Tab.~\ref{tab: simulation runs}.

Next, we describe the set-up of the simulations, explain their analysis and demonstrate that the results are insensitive to the choice of the critical angle (Sect.~\ref{subsec: Simulation Setup}). It follows a general description of the merger evolution (Sect.~\ref{subsec: Peak Positions}) as well as a discussion how well models of different angular dependencies can be mapped onto each other in the case of merging galaxy clusters (Sect.~\ref{subsec: Matching of angular dependencies}). A detailed investigation and discussion of the DM-galaxy offsets follow in Sect.~\ref{sec:offsets}.

\subsection{Merger simulation set-up, analysis  and validation}\label{subsec: Simulation Setup}

\begin{figure}
        \centering
\text{hSIDM \quad M{\o}ller \quad $r = 10^4$ \quad $\sigma_{\mathrm{V}}/m_{\chi}=1\,\mathrm{cm}^2\,\mathrm{g}^{-1}$}\par
        \includegraphics[width=\columnwidth]{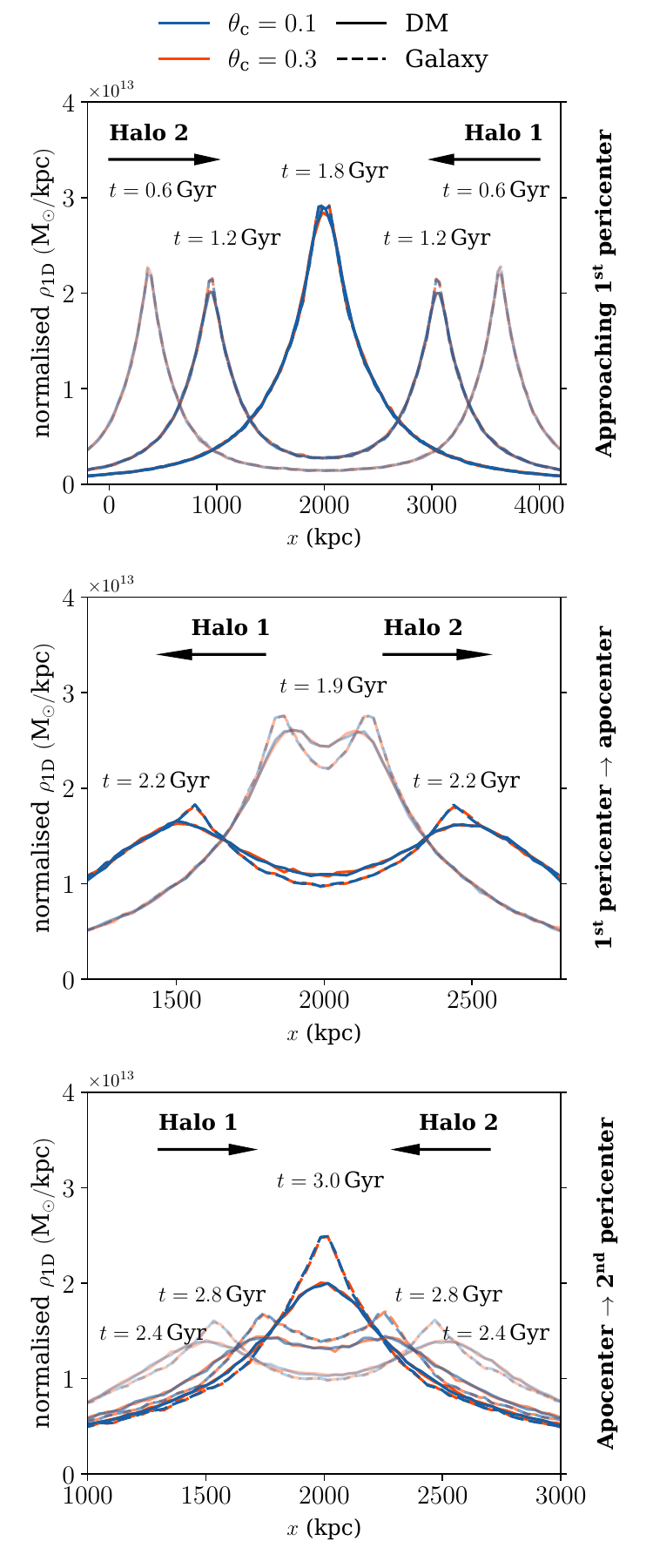}
    \caption{One-dimensional density profile along the merger axis for DM (solid) and galaxies (dashed) at various times, as indicated. The top panel shows the evolution before the first pericentre passage, and the middle panel shows it shortly afterwards. In the bottom panel, the evolution following the first apocentre passage is displayed. The hSIDM scheme allows us to simulate a cluster merger for the strongly anisotropic M{\o}ller cross section with $r=10^4$. We validate this approach by noting that the results are (practically) independent of the choice of the critical angle, for $\theta_{\mathrm{c}} \in \{0.1,\, 0.3\}$ (blue and orange curves, respectively). The full time evolution movie is available online.}
    \label{fig: Evolution of normalized 1D density profile of the merging haloes}
\end{figure}

Following \cite{Fischer_2021b}, we consider a merger of two galaxy clusters of equal mass, moving towards each other with an initial relative velocity of $1\,000\, \mathrm{km\,s}^{-1}$ along the merger axis (head-on collision), separated by $4\,000\,\mathrm{kpc}$. The initial dark matter content of each cluster is modelled by a radially symmetric Navarro-Frenk-White \citep[NFW, ][]{Navarro_1996a} profile, with a concentration parameter $c = 5.4$ and a scale radius $r_s = 3.9 \times 10^2 \,\mathrm{kpc}$,  corresponding to a virial mass of $M_{\mathrm{vir}} = 1\times10^{15}\,\mathrm{M}_{\sun}$. We represent each halo using $10.1\times10^{6}$ DM particles with a particle mass of $m_{\mathrm{DM}} = 2\times10^{8}\,\mathrm{M}_{\sun}$. The halo is sampled up to a radius of $7.8\,\mathrm{Mpc}$. We furthermore set the gravitational softening length to $\epsilon = 1.2\,\mathrm{kpc}$ and use $N_{\mathrm{ngb}}=48$ neighbouring particles to determine the kernel overlap entering the implementation of dark matter self-interactions, see Sect.~\ref{sec: implementation}. 

Apart from the DM particles, we include a population of collisionless particles as tracers representing the galaxy distribution, with each halo hosting  $10.1\times10^{6}$  particles as well, but with a particle mass of $m_{\mathrm{Gal}} = 4\times10^{6}\,\mathrm{M}_{\sun}$. We use an equivalent number of particles for galaxies as for DM to sample their distributions with approximately equal resolution. These particles, which are more abundant than galaxies in clusters, do not represent individual galaxies but rather represent a smoothed-out galaxy distribution. It is noteworthy that while `galaxy' particles in our simulation are approximated to be collisionless, real galaxies may not behave entirely like that~\citep[see][]{Kummer_2018}. We also include an additional particle with mass  $m_{\mathrm{BCG}} = 7\times10^{10}\,\mathrm{M}_{\sun}$ positioned at the centre of each halo to simulate the BCG. It is acknowledged that this representation entails an idealised treatment of BCGs, neglecting their spatial extension. The BCG mass is selected to be much lower than typical BCG masses to reduce numerical artefacts. This is in line with previous studies of merging galaxy clusters~\citep[e.g.][]{Kim_2017b, Fischer_2023, Sabarish_2024}; see also~\cite{Valdarnini_2024} for an alternative approach. 

A characteristic signature of dark matter self-interactions are offsets between the dark matter and galaxy populations of the individual galaxy clusters, that are dynamically generated during the merger process.
To quantify such offsets, it is necessary to identify a suitable `peak' location of the dark matter and
galaxy distributions, respectively, for which various methodologies have been explored \citep[e.g.][]{Power_2003, Kim_2017b, Robertson_2017a}. Here we adopt the peak identification algorithm proposed by \citet{Fischer_2021b}, which tracks for each particle whether it belonged originally to the first or the second halo,\footnote{We note that for M{\o}ller scattering the scattering angle $\theta$ larger than $\uppi/2$ are exchanged to $\uppi - \theta$ to ensure consistent association of the particles in the two halos. See also Sect.~2.1.2 by \cite{Fischer_2021b}.} and utilises a peak search approach based on the gravitational potential. To compute the peak position for a component, for example, the galaxies of the first halo (association based on the ICs), the gravitational potential sourced by these particles only is computed and the peak position is set by the potential minimum.
For the head-on collision set-up studied in this work, the peaks are always located along the merger axis. 
We denote the cluster moving initially in the negative (positive) direction of the merger axis as `halo 1' (`halo 2'). The peak positions along the merger axis are denoted by $d_{\rm DM}^{{\rm halo}\,n}$ and
$d_{\rm Galaxies}^{{\rm halo}\,n}$ for the DM and galaxy populations and the two halos ($n=1,2$), respectively.
In addition, we consider the location $d_{\rm BCG}^{{\rm halo}\,n}$ of the BCG for each cluster.
We define the offsets as\footnote{This definition matches the one used by~\citep{Fischer_2021b}, \cite{Fischer_2023} and \cite{Sabarish_2024} except that the offset has a different sign. We note that the definition by \cite{Fischer_2021a} differs due to the use of absolute distances between the peaks of the same component.}
\begin{equation}
    d_\mathrm{DM} - d_i \equiv \frac{-(d^\mathrm{halo\,1}_\mathrm{DM}- d^\mathrm{halo\,1}_i) + d^\mathrm{halo\,2}_\mathrm{DM} - d^\mathrm{halo\,2}_i }{2} \,,
\end{equation}
for $i=$Galaxies or $i=$BCG, respectively. The sign convention ensures that the offsets for both halos have the same sign. We quote the mean value over the two halos unless stated otherwise.

The hSIDM set-up involves the critical $\theta_{\mathrm{c}}$, that determines the separation between the regimes treated using the dedicated small-angle and large-angle approaches, respectively. As stated above, we checked that {\it all} our results are independent of the choice of $\theta_{\mathrm{c}}$ within numerical uncertainties. While we provide more details for specific quantities of interest below, we give a first example in Fig.~\ref{fig: Evolution of normalized 1D density profile of the merging haloes}, showing the one-dimensional density profile within a small region\footnote{In practice we use a cuboid centred around the x-axis (merger axis) with depth of $100\,\mathrm{kpc}$ and width of $30\,\mathrm{kpc}$.} around the merger axis, for various times, and for the DM density (solid lines) as well as the galaxy density (dashed lines). Here we used a strongly anisotropic M{\o}ller cross section with $r=10^4$. The results obtained from the hSIDM runs for $\theta_{\mathrm{c}}=0.1$ and $\theta_{\mathrm{c}}=0.3$ are shown by different colours, and are hardly distinguishable from each other in Fig.~\ref{fig: Evolution of normalized 1D density profile of the merging haloes}. This already indicates that peak positions and the resulting offsets between DM and galaxy distributions are independent of $\theta_{\mathrm{c}}$, as we also verify later on. 

We also stress that the hSIDM scheme allow us to simulate the strongly forward-dominated model for self-interactions described by $r=10^4$. It would be numerically prohibitively expensive when using an algorithm that does not employ an effective treatment of small-angle scatterings (i.e.\ for $\theta_{\mathrm{c}}=0$).

\subsection{Peak positions} \label{subsec: Peak Positions}

\begin{figure*}
    \centering
    \includegraphics[width=\textwidth]{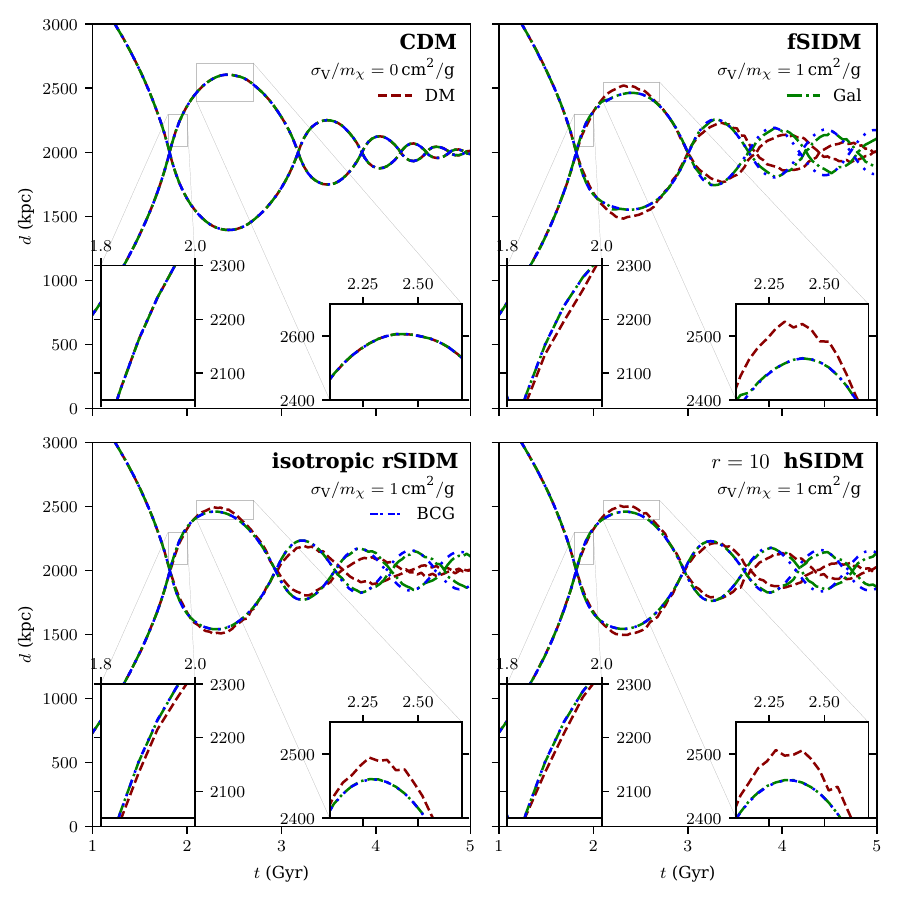}
    \caption{Time evolution of the DM (red dashed) and galaxy (green dot-dashed) peak positions, as well as the BCG position (blue short dot-dashed), for both merging galaxy clusters. The upper left panel corresponds to CDM, while the other panels show three different models for SIDM: purely forward-dominated (fSIDM, upper right panel), isotropic (rSIDM, lower left panel), and a light mediator model described by a M{\o}ller cross section with $r=10$ (simulated using the hSIDM approach, lower right panel). The absolute self-interaction strength is matched in all cases by requiring a viscosity cross section $\sigma_{\mathrm{V}}/m_{\chi}= 1\,\mathrm{cm}^2\,\mathrm{g}^{-1}$. In each panel we also show the peak positions shortly after the first pericentre passage (left inset) and those around apocentre (right inset)
    }
    \label{fig: DM, Gal, BCG absolute posisition using existing approaches}
\end{figure*}  

In Fig.~\ref{fig: DM, Gal, BCG absolute posisition using existing approaches} we present the time evolution of the peak positions of DM and galaxy distributions, as well as the BCG position, along the merger axis. The two sets of lines within each panel correspond to the first and second halo, respectively. For comparison, the upper left panel shows the purely collisionless DM case, that is, without self-interactions. The peaks of DM and galaxies as well as the BCG position are then indistinguishable from each other for all times, as expected. Since we consider an equal-mass merger that occurs head-on, the peaks oscillate along the merger axis (around the centre of mass ) in a symmetrical way, with an amplitude that is damped due to dynamical friction. The other three panels in Fig.~\ref{fig: DM, Gal, BCG absolute posisition using existing approaches} show the evolution when assuming various models for DM self-interactions, specifically purely forward-dominated scattering (fSIDM), isotropic scattering (isotropic rSIDM), as well as an anisotropic M{\o}ller cross section with $r=10$ (hSIDM), respectively. We fixed the absolute self-interaction strength in all these cases such that they correspond to the same viscosity cross section, specifically $\sigma_{\mathrm{V}}/m_\chi=1\,{\rm cm}^2\,{\rm g}^{-1}$. While we discuss the rationale for this choice in more detail below (see Sect.~\ref{subsec: Matching of angular dependencies}), we already notice that it leads to an evolution of peak positions that is broadly similar in all cases. Nevertheless, significant differences between the models remain, see Sect.~\ref{subsec: Anisotropy of scattering}.

Before going in details, we briefly review some general features for convenience. 
An initial observation is that, due to the presence of self-interactions, the DM component merges faster, i.e.\ coalesces on a shorter timescale compared to the galaxy component. Consistent with previous findings by for example~\citet{Kim_2017b} and~\citet{Fischer_2021a}, we observe that while the DM component experiences a rapid merging process, the galactic components undergo relatively stable and prolonged oscillations. These can be attributed to lower central DM densities and thus reduced dynamical friction as compared to the collisionless DM case. Famously, self-interactions also give rise to a spatial separation between DM and galaxy/BCG positions at earlier times, for example shortly after the first pericentre passage at $t\simeq 1.8$\,Gyr or around the first apocentre at $t\simeq 2.3$\,Gyr. The insets in Fig.~\ref{fig: DM, Gal, BCG absolute posisition using existing approaches} show the evolution for these times in more detail. Shortly after the first pericentre, namely for $t\lesssim 2$\,Gyr, the DM peaks are behind those for galaxies (left inset in each panel), providing a classic signature originating from the deceleration of the DM halos due to self-interaction. Subsequently, the galaxies are decelerated in their outward motion due to the gravitational pull of the DM. When approaching the first apocentre at $t\simeq 2.3$\,Gyr (right inset), the galaxies therefore turn around at a somewhat smaller distance from the centre of mass  than the DM peaks. Thus, during this stage, the offset is inverted, and the DM peaks are more outwards directed. A detailed comparison of the right insets for the three models of SIDM also indicate quantitative differences in the relative offsets between DM and galaxies/BCG, see Sect.~\ref{subsec: Anisotropy of scattering} for details.

\subsection{Matching of models with different angular dependence} \label{subsec: Matching of angular dependencies}

\begin{figure*}
\centering
\text{hSIDM \quad M{\o}ller}\par
        \includegraphics[width=\textwidth]{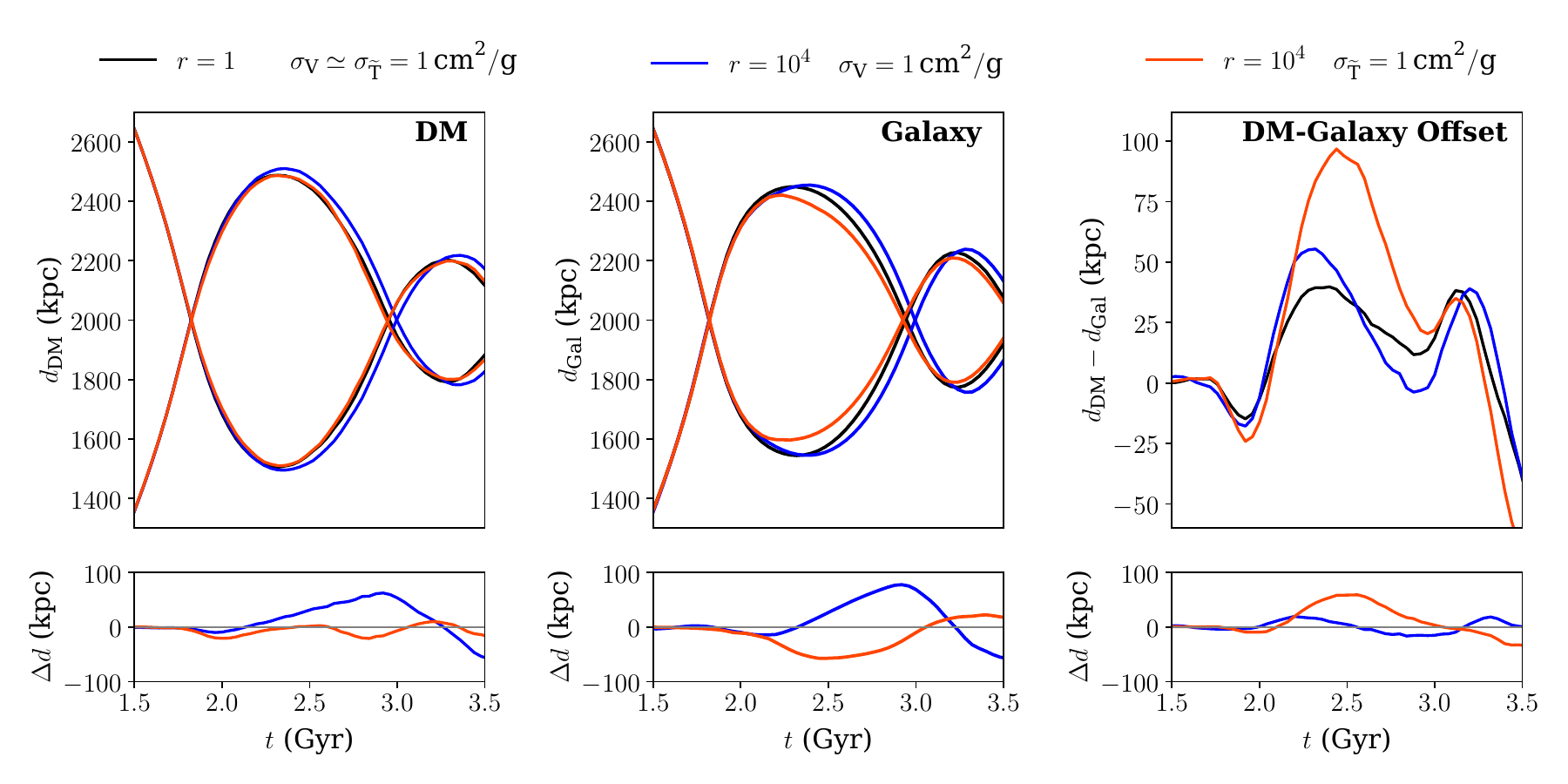}
    \caption{Comparison of the evolution of DM peak positions (left column), galaxy density peak positions (middle column), and the DM-galaxy offset (right column) for three different models considered in Sect.~\ref{subsec: Matching of angular dependencies}. The result for the almost isotropic model (1) is shown in black, the result for the anisotropic model (2) with matched transfer cross section is shown in orange, and the result for the anisotropic model (3) with matched viscosity cross section is shown in blue.
    In the lower panels the difference (2)-(1) (orange) and (3)-(1) (blue) is shown. While the evolution of the DM peak position can be approximately matched among (1) and (2), i.e.\ for constant transfer cross section, this is not the case for the evolution of galaxy positions and the DM-galaxy offset. In particular, the offset is larger for anisotropic models, by a factor of $\sim 4$ ($\sim 2$) when comparing to an isotropic model with the same transfer (viscosity) cross section. Thus, a simulation including the angle dependence is required in general to achieve reliable results.}
    \label{fig: Matching cross section of hSIDM}
\end{figure*}
The hSIDM approach allows us to simulate SIDM models with strongly anisotropic differential cross section.
Using these results, we are able to scrutinise the question whether certain characteristic features of SIDM can be related among models with distinct {\it differential} cross sections by considering a suitable {\it angle-averaged} cross sections. Specifically, in this context, we are interested in the question whether the time evolution of DM and galaxy peak positions during the merger process, as well as their offset, can be `mapped' across different models. If such a mapping exists, it would imply that models with different differential cross section but identical angle-averaged cross section yield indistinguishable results for one particular quantity (or ideally even several) of interest.

To investigate this question we consider two well-known candidates to define such a mapping, being the {\it modified transfer cross section} $\sigma_{\mathrm{\widetilde{T}}}$ as well as the {\it viscosity cross section} $\sigma_{\mathrm{V}}$, see Sect.~\ref{subsec: Angle dependent scattering-cross section}.
We note that it is clear that the total cross section $\sigma_{\rm tot}$ is {\it not} a good candidate for identifying a viable `mapping', since it  strongly increases for forward-dominated scattering, while the physical effect of self-interactions does not increase correspondingly due to the tiny momentum transfer incurred by each individual scattering event.
The viscosity cross section has for example been proposed to provide a `mapping' in the context of gravothermal collapse dynamics of isolated halos in~\citet{Yang_2022D}, see also \cite{Sabarish_2024}. In the context of galaxy cluster mergers,  \citet{Fischer_2021a} showed that when matching the modified transfer cross section, offsets for purely forward-dominated scattering are typically larger than for isotropic scattering, indicating that $\sigma_{\mathrm{\widetilde{T}}}$ may not provide a viable mapping for the DM-galaxy or DM-BCG offset in general. Here we come back to this question, and provide an in-depth investigation.

To be concrete, we considered three specific models, and we simulated all of them with the hSIDM scheme:
\begin{itemize}
\item[(1)] Differential M{\o}ller cross section Eq.~\eqref{eq: Moeller differential cross section} with $r=1$, being {\it almost isotropic}, with $\sigma_{\mathrm{\widetilde{T}}}/m_\chi\simeq\sigma_{\mathrm{V}}/m_\chi=1\,{\rm cm}^2\,{\rm g}^{-1}$.
\item[(2)] Differential M{\o}ller cross section Eq.~\eqref{eq: Moeller differential cross section} with $r=10^4$, being {\it strongly anisotropic}, with modified transfer cross section being matched to (1), i.e.\ $\sigma_{\mathrm{\widetilde{T}}}/m_\chi=1\,{\rm cm}^2\,{\rm g}^{-1}$. This implies $\sigma_{\mathrm{V}}/m_\chi\simeq 1.4\,{\rm cm}^2\,{\rm g}^{-1}$.
\item[(3)] As (2), but with viscosity cross section being matched to (1), i.e.\ $\sigma_{\mathrm{V}}/m_\chi=1\,{\rm cm}^2\,{\rm g}^{-1}$. This implies $\sigma_{ \mathrm{\widetilde T}}/m_\chi\simeq 0.7\,{\rm cm}^2\,{\rm g}^{-1}$.
\end{itemize}

Let us now turn to the DM and galaxy peak positions, as well as their offset. The time evolution is shown in the three columns of Fig.~\ref{fig: Matching cross section of hSIDM}, respectively, with each panel including lines for model (1) in black, model (2) in orange, and model (3) in blue. In addition, the lower panels show the difference (2)-(1) (orange) and (3)-(1) (blue). From the left panel in Fig.~\ref{fig: Matching cross section of hSIDM}, we see that the DM peak position for model (1) and (2) is similar at all times. This indicates that matching via the transfer cross section is a reasonable approximation for the evolution of the DM peaks. On the other hand, for the galaxy peak position (middle panel in Fig.~\ref{fig: Matching cross section of hSIDM}) neither model (2) nor (3) are close to (1) at all times. However, during the most relevant stage from the first pericentre ($t\simeq 1.8$\,Gyr) until around the first apocentre ($t\simeq 2.3$\,Gyr), the galaxy peak position of model (3) is closer to model (1). This suggests that galaxy positions can be better matched using the viscosity cross section, but only until around the first apocentre. Afterwards, no clear matching seems possible.

Finally, we turn to the offset between the DM and galaxy peak positions, shown in the right panel of Fig.~\ref{fig: Matching cross section of hSIDM}. We observe that the offsets for both the anisotropic models (2) and (3) are significantly larger than for the isotropic model (1). In particular, this is the case shortly after the first pericentre passage at $t\simeq 2$\,Gyr (when the DM peaks of the two clusters are closer to each other than the galaxy peaks, corresponding to negative offset) as well as around the first apocentre ($t\simeq 2.3$\,Gyr, positive offset). The maximal absolute offsets in the two stages differ by a factor of around two between model (3) and (1), and even by a factor of around four for model (2) and (1). The finding of {\it larger} absolute values for the offsets in anisotropic models versus the isotropic case is in line with~\citet{Fischer_2021a}. Thus we find that neither the transfer nor the viscosity cross section yield a useful matching between models when regarding the DM-galaxy offset during the most relevant stages of the merger evolution. 

In conclusion, while an approximate mapping of DM peak positions with the transfer cross section is reasonable, the galaxy positions as well as the DM-galaxy offset cannot be mapped with this prescription, with offsets differing by a factor of four between isotropic and strongly anisotropic cases. When comparing models with different angular dependence but identical viscosity cross section, we observe that the galaxy peak position can be roughly mapped until about the first apocentre. However, in that case the DM peak positions differ for the isotropic and anisotropic models, such that the DM-galaxy offsets differ by a factor of around two. Altogether, this implies that the complex merger dynamics requires to include the actual angle dependence for a given model of self-interactions in order to achieve reliable predictions.  

\section{Model dependence of dark matter-galaxy offsets}
\label{sec:offsets}

In this section we discuss implications of the results obtained from the simulation of  galaxy cluster mergers in the previous section. In particular, a main quantity of interest in this context is the offset between the DM halo and either the galaxy population or the brightest galaxy of the cluster (i.e.\ the BCG). Here we discuss our findings on how these offsets depend on the underlying model describing DM self-interactions. Specifically, our focus is on the characteristic angle dependence of the differential cross section as predicted in realistic (e.g.\ light mediator) models, that we take into account in the simulations based on the hSIDM scheme developed in this work.

\subsection{Dependence on viscosity cross section for fixed anisotropy} \label{subsec: Strength of scattering}

\begin{figure*}
        \centering
\text{hSIDM \quad M{\o}ller \quad $r = 10^4$ }\par
        \includegraphics[width=0.49\textwidth]{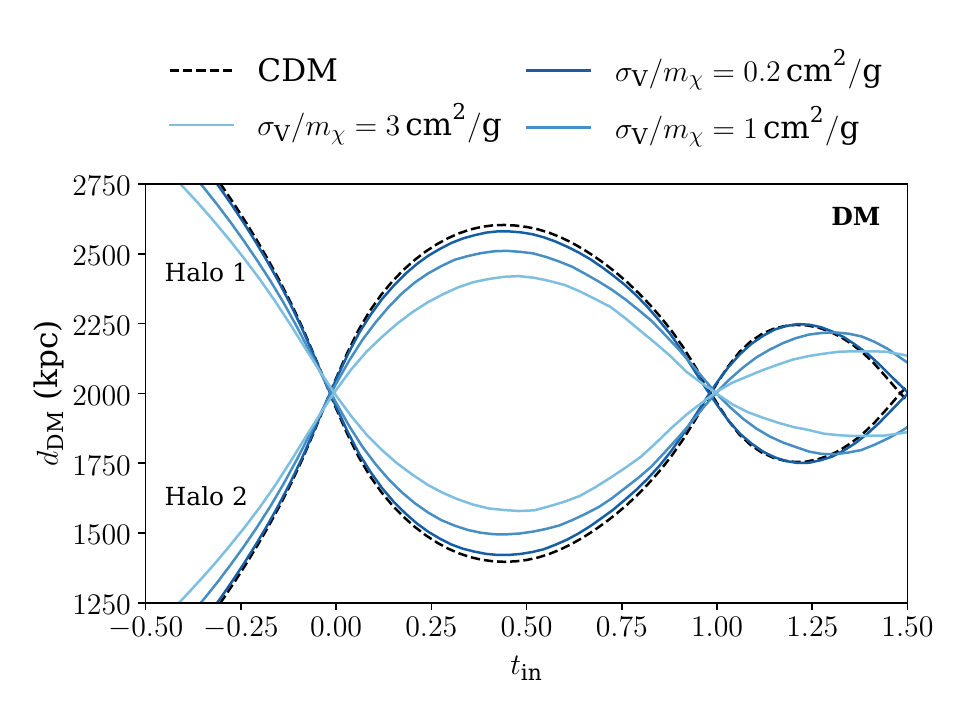}
        \includegraphics[width=0.49\textwidth]{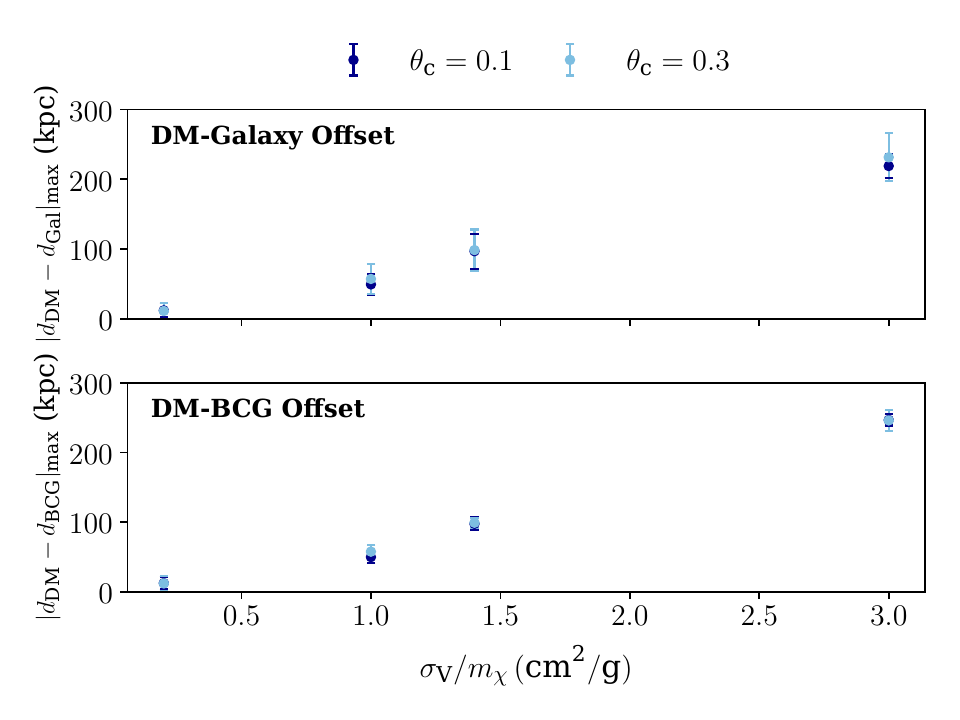}
    \caption{Dependence of the DM peak positions (left) as well as maximal DM-galaxy (upper right) and DM-BCG (lower right) offsets on the viscosity cross section $\sigma_{\mathrm{V}}/m_\chi$ while keeping the amount of anisotropy of the differential cross section $\mathrm{d}\sigma/\mathrm{d}\Omega$ fixed (`M{\o}ller' cross section Eq.~\eqref{eq: Moeller differential cross section} with $r=10^4$). The time-axis on the left is rescaled such as to align the first and second pericentre passage for all models. The DM peaks remain the closer to each other the larger  $\sigma_{\mathrm{V}}/m_\chi$. The offsets increase (roughly linearly) with $\sigma_{\mathrm{V}}/m_\chi$ within the considered range (right column). We note that our results are independent of the choice of critical angle $\theta_{\mathrm{c}}$ entering the hSIDM scheme, with offsets for $\theta_{\mathrm{c}}=0.1$ and $0.3$ being compatible with each other. The error bars are computed following~\citet{Fischer_2021b}.}
    \label{fig: sigma_dependence hSIDM}
\end{figure*}

As a first step, we verify the expectation that the offsets increase when increasing the overall DM self-interaction rate. To demonstrate this dependence, we simulate galaxy cluster mergers for models with various values of the viscosity cross section in the range $0.2\,{\rm cm}^2\,{\rm g}^{-1}\leq \sigma_{\mathrm{V}}/m_\chi\leq 3\,{\rm cm}^2\,{\rm g}^{-1}$, but keeping the angle dependence of the cross section fixed (we choose a `M{\o}ller' cross section Eq.~\eqref{eq: Moeller differential cross section} with $r=10^4$). The time evolution of the DM peak positions for the two halos is shown in Fig.~\ref{fig: sigma_dependence hSIDM} (left panel). Compared to collisionless DM (CDM, black dashed lines), the peaks remain the more close to each other the larger $\sigma_{\mathrm{V}}$, as expected. We note that, following~\citet{Fischer_2021b}, we use a linearly re-scaled time variable, $t_{\rm in}$, here, defined such that the first (second) pericentre occurs at $t_{\rm in}=0\ (1)$, respectively.
The maximal absolute DM-galaxy and DM-BCG offsets between the first and second pericentre are shown in the right panels of Fig.~\ref{fig: sigma_dependence hSIDM}, versus $\sigma_{\mathrm{V}}/m_\chi$. We observe that the offsets increase roughly linearly within the range of shown viscosity cross sections. To illustrate the magnitude, we extract an approximate estimate for both DM-galaxy and DM-BCG offsets given by
\be
{\rm maximal\ offset} \approx 80\,{\rm kpc}\cdot\frac{\sigma_{\mathrm{V}}/m_\chi}{1\,{\rm cm}^2\,{\rm g}^{-1}}\qquad ({\rm for}\ r=10^4)\,,
\ee
for the equal-mass merger considered in this work.

\subsection{Dependence on anisotropy for fixed viscosity cross section} \label{subsec: Anisotropy of scattering}

\begin{figure*} 
\centering
        \includegraphics[width=0.49\textwidth]{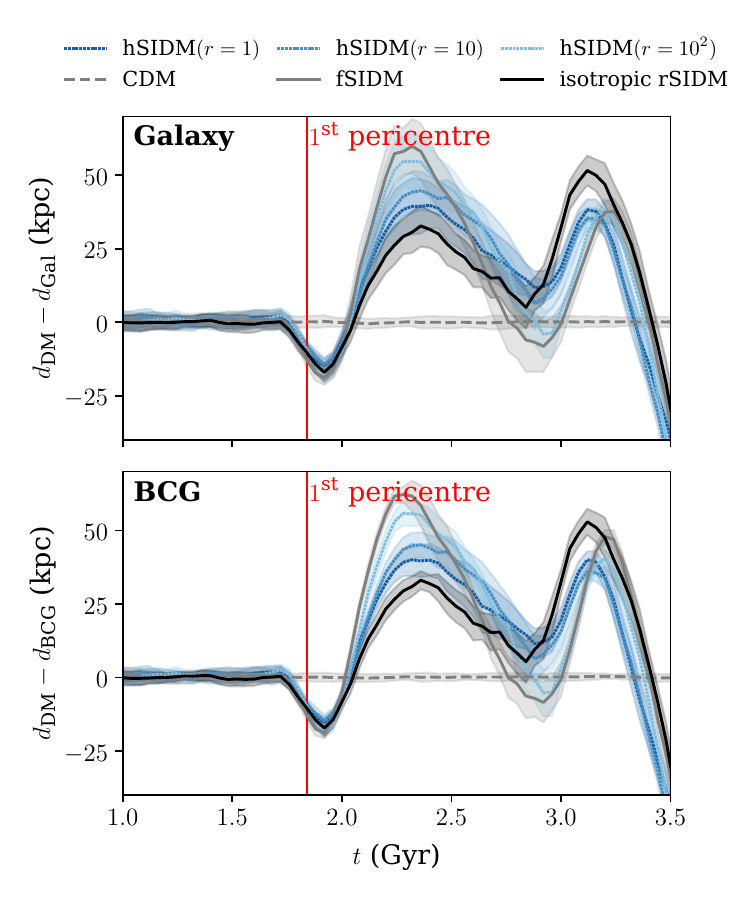}
        \includegraphics[width=0.49\textwidth]{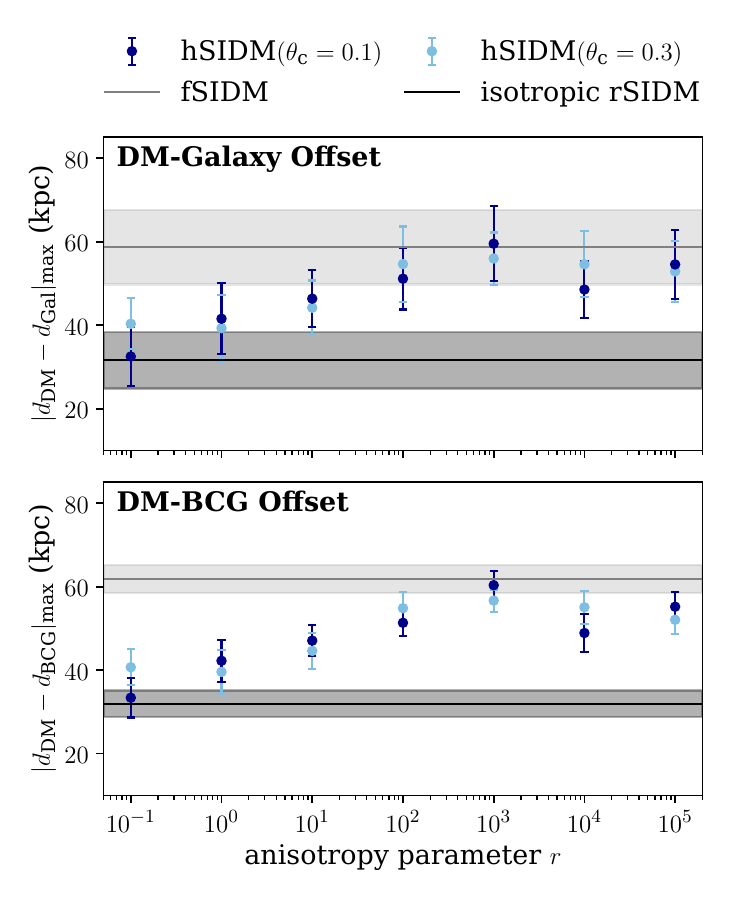}
    \caption{Dependence of the DM-galaxy (upper row) and DM-BCG (lower row) offsets on the amount of anisotropy of the differential cross section $\mathrm{d}\sigma/\mathrm{d}\Omega$ while keeping the viscosity cross section $\sigma_{\mathrm{V}}/m_\chi=1\,{\rm cm}^2\,{\rm g}^{-1}$ fixed. We show results obtained within the hSIDM scheme, assuming the `M{\o}ller' cross section Eq.~\eqref{eq: Moeller differential cross section} with various values of the anisotropy parameter $r$ (see Eq.~\eqref{eq:rdef}) related to the ratio of mediator and DM particle masses. For comparison, we also show the cases of purely forward-dominated scattering (fSIDM) and isotropic scattering (isotropic rSIDM), respectively. The left column shows the time evolution, the right column the maximal absolute offsets within $t \in [ 1.5\,\mathrm{Gyr}, 3\,\mathrm{Gyr} ]$. The maximal offsets for models with realistic angle dependence approach those for purely forward-dominated scattering for $r\gtrsim 10^2$ (corresponding to $m_\phi/m_\chi\lesssim 0.1v/c$, where $v$ is the typical velocity scale), and those for isotropic scattering for $r\lesssim 1$ ($m_\phi/m_\chi\gtrsim v/c$). The error bars and error bands are computed following~\citet{Fischer_2021b}.}
    \label{fig: r-dependence hSIDM}
\end{figure*}

The differential cross section of DM self-scattering within light mediator models is the more forward-dominated the lighter the mediator mass $m_\phi$, while it approaches the isotropic limit for a sufficiently heavy mediator. The level of anisotropy is captured by the parameter $r$ defined in Eq.~\eqref{eq:rdef}, with $r\lesssim {O}(1)$ ($r\gg 1$) corresponding to the almost isotropic (strongly forward-dominated) limits, respectively. To assess the dependence of the DM-galaxy offset on the anisotropy, we simulate galaxy cluster mergers as described above using the hSIDM scheme with an angle dependence described by the `M{\o}ller' cross section Eq.~\eqref{eq: Moeller differential cross section}, and for $r \in \{0.1,\, 1,\, 10,\, 10^2,\, 10^3,\, 10^4,\, 10^5 \}$. For comparison, we also consider the purely forward-dominated case (fSIDM) as well as isotropic scattering (isotropic rSIDM). We fix the overall size of the differential cross section such that the angle-averaged viscosity cross section is identical in all cases, with $\sigma_{\mathrm{V}}/m_\chi=1\,{\rm cm}^2\,{\rm g}^{-1}$.

The left panels of Fig.~\ref{fig: r-dependence hSIDM} show the time evolution of the DM-galaxy (top) and DM-BCG (bottom) offset, respectively. As expected from Sect.~\ref{sec: Merger simulation and analysis}, the offsets are negative shortly after the first pericentre passage, since the DM halos were slowed down when passing through each other, while the offsets are positive around the first apocentre, since the galaxy populations are subsequently slowed down by the gravitational force exerted by the DM mass distribution. The overall size of the offsets tend to be the larger the more anisotropic the scattering is, as can be seen by comparing the lines showing the evolution for $r=1,10,10^2$, respectively. This is in line with the findings of~\cite{Fischer_2021a}, where offsets for fSIDM were compared to those for isotropic rSIDM.\footnote{We note, however, that the modified transfer cross section was kept fixed in that case, leading somewhat larger differences; compare to Sect.~\ref{subsec: Matching of angular dependencies} and Fig.~\ref{fig: Matching cross section of hSIDM}.} Indeed, we find that our results for $r=1,10,10^2$ yield offsets that typically lie between those for fSIDM and isotropic rSIDM within the numerical accuracy. To highlight this point, we show the maximal absolute offset within the time interval $t \in [ 1.5\,\mathrm{Gyr}, 3\,\mathrm{Gyr} ]$ in the right panels of Fig.~\ref{fig: r-dependence hSIDM}, versus the model parameter $r$. These results allow us to infer the relevance of model-dependence, being one of the main results to be discussed next.

\subsection{Discussion} \label{subsec: Discussion}

Models of dark matter self-interactions typically feature angle-dependent scattering. Here we use our results to assess in how far the limits of purely forward-dominated (fSIDM) or isotropic (isotropic rSIDM) scattering yield an accurate approximation to a true angle dependence described by the `M{\o}ller' cross section Eq.~\eqref{eq: Moeller differential cross section}. Its anisotropy is determined by the parameter $r$  (see Eq.~\eqref{eq:rdef}), with fSIDM (isotropic rSIDM) being approached in the limiting cases $r\to\infty$ ($r\to 0$), respectively.
Here, we discuss for which finite values of $r$ these limits are approached in practice. In particular, we focus on the maximal DM-galaxy and DM-BCG offsets shown in the right column of Fig.~\ref{fig: r-dependence hSIDM}.
We notice several features:
\begin{itemize}
\item The maximal DM-galaxy and DM-BCG offsets that we obtain by simulating self-interactions described by the `M{\o}ller' cross section Eq.~\eqref{eq: Moeller differential cross section} with the hSIDM scheme converge towards those obtained for purely forward-dominated scattering (fSIDM, shown by the light grey horizontal band in Fig.~\ref{fig: r-dependence hSIDM}) for $r\gtrsim 10^2$. In terms of the model parameters, this corresponds to a mass ratio of mediator and dark matter mass \be
  m_\phi/m_\chi\lesssim 0.1v/c\,,
\ee
where $v$ stands for the typical velocity scale of the considered system.
Equivalently, the condition for approaching the fSIDM limit can be expressed as $\sigma_{\rm tot}/\sigma_{\mathrm{\widetilde{T}}}\gtrsim 10$. We note that this ratio is formally assumed to be infinite for fSIDM. The inequality from above can thus be viewed as a criterion for when this limiting case is applicable in practice for realistic models of self-interacting dark matter, and constitutes one of the main results of this work.
\item On the other hand, for $r\lesssim 1$ the maximal offsets are compatible with those obtained for isotropic scattering (isotropic rSIDM, shown by the dark grey horizontal band in Fig.~\ref{fig: r-dependence hSIDM}). This corresponds to 
\be
 m_\phi/m_\chi\gtrsim v/c\,,
\ee
or equivalently $\sigma_{\rm tot}/\sigma_{\mathrm{\widetilde{T}}}\lesssim 1.1$. We note that the ratio is unity for the limit of exactly isotropic scattering. Thus, our result shows by `how much' a model may quantitatively deviate from isotropy, while still leading to offsets that are compatible with those obtained when assuming isotropic scattering. 
\item Consequently, for a mediator mass in the range $0.1v/c\lesssim m_\phi/m_\chi\lesssim v/c$, the precise angular dependence of DM self-scattering needs to be taken into account to obtain accurate predictions for DM-galaxy and DM-BCG offsets. This corresponds to $1.1\lesssim \sigma_{\rm tot}/\sigma_{\mathrm{\widetilde{T}}}\lesssim 10$.
\item Finally, coming back to the discussion of validity checks from Sect.~\ref{subsec: Rutherford and Moeller scattering implementation} and Sect.~\ref{subsec: Simulation Setup}, we note that (like all other considered quantities) the offsets shown in Fig.~\ref{fig: r-dependence hSIDM} are independent of the critical angle $\theta_{\mathrm{c}}$ entering the hSIDM scheme (within numerical errors), as demonstrated by showing results for $\theta_{\mathrm{c}}=0.1$ and $0.3$, respectively.
\end{itemize}
Let us stress that the angle dependence enters the merger dynamics in general in a complex way, affecting galaxy and DM distributions differently, see Sect.~\ref{subsec: Matching of angular dependencies}. Nevertheless, when considering only the DM-galaxy offset, one may ask how large the error on the inferred self-interaction cross section would be when assuming an incorrect model for the angle dependence. Given that the offsets differ by about a factor of two between the limits of small and large $r$, see Fig.~\ref{fig: r-dependence hSIDM}, and that their dependence on the viscosity cross section is approximately linear, see Sect.~\ref{subsec: Strength of scattering}, we estimate that the deviation in $\sigma_{\mathrm{V}}/m_\chi$ would amount to about a factor of two as well. For the modified transfer cross section, this number would go up to a factor of approximately three (see Sect.~\ref{subsec: Matching of angular dependencies}). However, we stress that these estimates should be taken as indicative only. \cite{Fischer_2021b} found that the difference in the size of offsets between isotropic and forward-dominated scattering is larger in unequal-mass mergers compared to equal-mass mergers. Accordingly, this would imply a larger estimate for the error than the factor of two for the viscosity cross section in the case of merging galaxy clusters with a significantly unequal mass ratio. Modelling an observed galaxy cluster merger with detailed simulations of various models for dark matter self-interactions and astrophysical settings could reveal more information than when just considering (maximal) offsets based on a peak identification method that is not applicable to observations. 

We note that our study considers head-on equal mass mergers only and does not model the intra-cluster medium. It would be interesting to extend this analysis to unequal-mass merges and allow for a non-central collision. This as well as modelling the intra-cluster medium and including the velocity-dependence of the cross section is left to future work. Nevertheless, our results can give a valuable hint for whether a given model for DM self-interactions can be approximately described by the purely forward-dominated fSIDM or the isotropic rSIDM limits, respectively.

\section{Conclusions}
\label{sec: Conclusions}

In this work, we have proposed, validated, and employed a numerical scheme to 
efficiently perform $N$-body simulations of SIDM while taking into account the angle dependence predicted in realistic particle physics models.
Most previous works considered either the case of relatively rare scatterings by large angles (`rSIDM', typically assuming an isotropic distribution) or very frequent and purely forward-dominated scatterings (`fSIDM'). The hybrid scheme developed in this work, dubbed `hSIDM', combines the methods underlying both of these approaches, yielding an efficient algorithm for incorporating DM self-interactions described by an (in principle) arbitrary differential cross section $\mathrm{d}\sigma/\mathrm{d}\Omega$.

As a first step, we extensively validated the approach based on an analytically solvable test set-up. In this context, we also derived a quantitative validity criterion for the effective treatment of small-angle scattering by a drag force combined with a diffusive transverse motion. The insights were used to show under which conditions the hSIDM scheme is expected to improve over a naive sampling of scattering angles in each collision of $N$-body particles, and we demonstrated a speed-up by several orders of magnitude in computation time. The theoretically expected speed-up was estimated in terms of a ratio of angle-weighted cross sections in Eq.~\eqref{eq:effgain}. We checked the independence of our results of the auxiliary `critical angle' entering the hSIDM scheme within a validity regime derived from the condition given in Eq.~\eqref{eq:hSIDMvalidity}.

Next, we applied hSIDM to simulate equal-mass galaxy cluster mergers including DM self-interactions with the characteristic angle dependence predicted by light mediator models. Concretely, we assumed a dependence of the scattering probability on the deflection angle given by Eq.~\eqref{eq: Moeller differential cross section}. In this parameterisation, the model parameter $r$ (related to the ratio of mediator and DM mass; see Eq.~\eqref{eq:rdef}) controls the amount of anisotropy, with $r\gg 1$ corresponding to strongly forward-dominated scattering, and $r\ll 1$ corresponds to the isotropic limit. We first investigated whether simulations with a different $r$ but an identical transfer or viscosity cross section can be mapped onto each other, and we found that this is in general {\it not} the case, especially when jointly considering the time evolution of the DM distribution, the galaxy population, the position of the BCG, and their relative offsets. Nevertheless, the dependence of the offset on $r$ is somewhat weaker when fixing the viscosity cross section as compared to the case when keeping the (modified) transfer cross section fixed. 

Finally, we investigated the dependence of the DM-galaxy and DM-BCG offsets on the properties of the model for DM self-interactions. We find that the maximal offset approaches the (larger) value predicted by purely forward-dominated scattering for $r\gtrsim 10^2$. This corresponds to a mediator to DM mass ratio of $m_\phi/m_\chi\lesssim 0.1v/c$, where $v$ is the typical velocity scale. Equivalently, this implies $\sigma_{\rm tot}/\sigma_{\mathrm{\widetilde{T}}}>10$ for the ratio of the total and modified momentum transfer cross sections. Conversely, the maximal offset converges to the (smaller) value obtained for isotropic self-scattering when $r\lesssim 1$ (i.e.\ $m_\phi/m_\chi\gtrsim v/c$ or $\sigma_{\rm tot}/\sigma_{\mathrm{\widetilde{T}}}\lesssim 1.1$). Our results thus yield a quantitative criterion for when the fSIDM and rSIDM limits are applicable for extracting offsets. Beyond that, we find that the angle dependence of DM self-scattering influences the time evolution of DM and galaxy populations in a distinct way, suggesting that it should be taken into account when modelling observed galaxy cluster mergers with simulations of self-interacting DM.

Our work can be extended by considering broader classes of galaxy cluster merger set-ups (unequal mass, non-central) with a more realistic modelling (e.g.\ including the intracluster medium) and by also taking the interplay of the angle- and velocity-dependence of the differential cross section into account. Furthermore, the hSIDM approach could be applied to other situations, such as the evolution of satellite populations, for a precision study of the impact of self-interactions with pronounced angle dependence on isolated halos or for studying the approach of gravothermal collapse.

\begin{acknowledgements}
    The authors thank Tobias Binder, Felix Kahlhoefer, Sowmia Balan, Marc Wiertel and all participants of the Darkium SIDM Journal Club for discussion. CA thanks Christoph Selbmann for helpful discussion.
    This work is funded by the \emph{Deutsche Forschungsgemeinschaft (DFG, German Research Foundation)} under Germany’s Excellence Strategy -- EXC-2094 `Origins' -- 390783311.
    Software: NumPy \citep{NumPy}, Matplotlib \citep{Matplotlib}.
\end{acknowledgements}


\bibliographystyle{aa}
\bibliography{library}

\begin{appendix}

\section{Implementation of anisotropic large-angle scattering for hSIDM}
\label{appx: Implementing large-angle anisotropic scattering in rSIDM scheme}

The contribution of large-angle scattering within the hSIDM scheme is implemented by explicitly sampling deflection angles from the differential cross section, as discussed in Sect.~\ref{subsec: Implementing hSIDM}, following the method used in~\citet{Robertson_2017b}. However, there is a small difference. Within the hSIDM scheme, this requires the cumulative probability function for scattering between an angle $\theta_{\mathrm{c}}$ and $\theta$ given in Eq.~\eqref{eq: cumulative distribution function hSIDM}.
We find that it can be computed analytically for the models considered in this work.
For the `M{\o}ller' scattering cross section Eq.~\eqref{eq: Moeller differential cross section} the cumulative distribution function reads
\begin{equation}
\begin{split}
    P(\theta) &=  \frac{ \sigma_0}{4 \sigma_{\mathrm{tot},>}(\theta_{\mathrm{c}})} \frac{a(\theta)}{b(\theta)}\,, \ {\rm with}\\
    a(\theta)&=  4 r (2+r) c_2(\theta_{\mathrm{c}}) \cos (\theta) \\
    & \;\;\;\;\;\;\; -c_3(\theta) \left(8 r (2+r) \cos (\theta_{\mathrm{c}})+c_2(\theta_{\mathrm{c}}) \log \left(c_1(\theta)\right)\right)\,,\\
    b(\theta)&= r (2+r) c_3(\theta)c_3(\theta_{\mathrm{c}})\,,\\
    c_1(\theta)&= \frac{(2+r-r \cos (\theta)) (2+r+r \cos (\theta_{\mathrm{c}}))}{(2+r+r
   \cos (\theta)) (2+r-r \cos (\theta_{\mathrm{c}}))}\,,\\
   c_2(\theta_{\mathrm{c}})&= -8-r (8+r)+r^2 \cos (2\theta_{\mathrm{c}})\,,\\
   c_3(\theta)&= -(2+r)^2+r^2 \cos ^2(\theta)\,,
\end{split}
\end{equation}

\begin{equation}
     \sigma_{\mathrm{tot},>}(\theta_{\mathrm{c}}) = \sigma_0  \left(\frac{4 \cos (\theta_{\mathrm{c}})}{(2+r)^2-r^2 \cos ^2(\theta_{\mathrm{c}})}+\frac{\log \left(-1+\frac{2
   (2+r)}{2+r+r \cos (\theta_{\mathrm{c}})}\right)}{r (2+r)}\right).
\end{equation}
For the `Rutherford' scattering cross section Eq.~\eqref{eq: Rutherford differential cross section}, the cumulative distribution function is given by
\begin{equation}
\begin{split}
    P(\theta) &=  \frac{2 \sigma_0}{\sigma_{\mathrm{tot},>}(\theta_{\mathrm{c}})} \frac{\left(\cos(\theta_{\mathrm{c}})- \cos(\theta)\right)}{\left(2+r-r\cos(\theta_{\mathrm{c}})\right) \left(2+r-r \cos(\theta)\right)}\,,\\
\end{split}
\end{equation}
with
\begin{equation}
     \sigma_{\mathrm{tot},>}(\theta_{\mathrm{c}}) = \sigma_0 \frac{ (1+\cos (\theta_{\mathrm{c}}))}{(1+r) (2+r-r \cos (\theta_{\mathrm{c}}))}\,.
\end{equation}

\section{Implementation of frequent small-angle scattering for hSIDM}
\label{appx: Implementation of frequent small-angle scattering for hSIDM}

The contribution from small-angle scattering within the hSIDM scheme is implemented using an effective description based on a drag force as well as transverse momentum diffusion, following~\citet{Fischer_2021a}. The drag force is determined by the (modified) transfer cross section. For the hSIDM scheme, we use the contribution to this quantity from small-angle scatterings only, see Sect.~\ref{subsec: Implementing hSIDM}.

For the `M{\o}ller' scattering cross section Eq.~\eqref{eq: Moeller differential cross section}  the modified transfer cross section for scattering by angles $\theta\leq\theta_{\mathrm{c}}$ or $\theta\geq \uppi-\theta_{\mathrm{c}}$ reads
\begin{equation}
\begin{split}
    \sigma_{\mathrm{\widetilde T},<}(\theta_{\mathrm{c}}) &= 2 \!\!\!\!\!\int\displaylimits_{\theta \le \theta_{\mathrm{c}} \lor \, \theta \ge \uppi -\theta_{\mathrm{c}}} \!\!\!\!\! \mathrm{d}\Omega \, (1-\left|\cos(\theta)\right|) \, \frac{\mathrm{d}\sigma}{\mathrm{d}\Omega}  \\
    &= 2 \sigma_0 \frac{a(\theta_{\mathrm{c}})}{b(\theta_{\mathrm{c}})}\,,
\end{split}
\end{equation}
where the numerator is given by
\begin{equation}
    \begin{split}
      a(\theta) = &24\log\big(c_1(\theta)\big) \\
    & - 2r\bigg[-4\log\big(c_2(\theta)\big)+r \Big[2(2+r)\cos(\theta)\\
   &-(7+r)\log\big(2+r-r\cos(\theta)\big)-(11+2r)\log\big(c_3(\theta)\big)\\
  &+\cos^2(\theta)\Big(-4+3\log\big(c_4(\theta)\big) \\
  & +r\Big(-2 +\log\big(c_5(\theta)\big)\Big) -6\log(2)\Big) +3(6+r)\log(2) \Big]\bigg],
    \end{split}
\end{equation}
and the denominator reads
\begin{equation}
    b(\theta) =r^2 (2+r) (2+r-r\cos(\theta)) (2+r+r\cos(\theta)).
\end{equation}
Here we defined several terms:
\begin{equation}
\begin{split}
    c_1(\theta) &= \frac{\left(2+r-r \cos (\theta)\right)\left(2+r+r \cos (\theta)\right)}{4(1+r)}\,,\\
    c_2(\theta) &= \frac{\left(2+r-r \cos (\theta)\right)^4\left(2+r+r \cos (\theta)\right)^5}{512(1+r)^5}\,,\\
    c_3(\theta) &= \frac{\left(2+r+r \cos (\theta)\right)}{(1+r)}\,,\\
    c_4(\theta) &= \frac{\left(2+r-r \cos (\theta)\right)\left(2+r+r \cos (\theta)\right)}{(1+r)}\,,\\
    c_5(\theta) &= \frac{\left(2+r-r \cos (\theta)\right)\left(2+r+r \cos (\theta)\right)^2}{8(1+r)^2}\,.
\end{split}
\end{equation}

The `Rutherford' scattering cross section Eq.~\eqref{eq: Rutherford differential cross section} is used for cases with distinguishable particles, for which the transfer cross section is relevant. In addition, only scatterings by $\theta\leq\theta_{\mathrm{c}}$ are included in the small-angle regime. We thus consider
\begin{equation}
\begin{split}
    \sigma_{\mathrm{T},<}(\theta_{\mathrm{c}}) &=  \int\displaylimits_{\theta \le \theta_{\mathrm{c}} } \mathrm{d}\Omega \, (1-\cos(\theta)) \frac{\mathrm{d}\sigma}{\mathrm{d}\Omega}  \\
    &= \sigma_0 \frac{c_1(\theta_{\mathrm{c}})+2 \log \left(c_2(\theta_{\mathrm{c}})\right)-2}{r^2}\,,
\end{split}
\end{equation}
where
\begin{equation}
\begin{split}
    c_1(\theta_{\mathrm{c}}) &= \frac{4}{2+r-r \cos (\theta_{\mathrm{c}})}\,,\\
    c_2(\theta_{\mathrm{c}}) &= \frac{(2+r-r \cos (\theta_{\mathrm{c}}))}{2}\,.
\end{split}
\end{equation}

\section{Vector calculation for anisotropic scattering}
\label{appendix: Vector calculation for anisotropic scattering}

For large-angle anisotropic scattering, the deflection angle is defined in the centre of mass frame of the two scattering particles. In this section, we describe an algorithm how to obtain the velocity vectors after the scattering. We consider the most general case where both particles can have different masses (relevant for distinguishable particles), denoted by $m_1$ and $m_2$, respectively. 

\begin{figure*}
    \centering
    \includegraphics[width=0.7\textwidth]{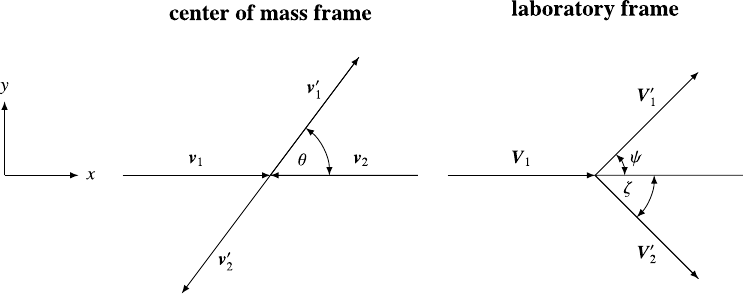}
    \caption{Illustration of a scattering event in the centre of mass frame (left) and in a frame where the second particle is initially at rest (right).}
    \label{fig: COM illu}
\end{figure*}

Given two arbitrary initial velocity vectors of particles that  scatter off each other $\vec{u}_{1},\vec{u}_{2} \in \mathbb{R}^3$, we first transform to the rest frame of the second particle using (see illustration in Fig.~\ref{fig: COM illu})
\begin{equation}
\begin{split}
    \vec{V}_1 &\equiv \vec{u}_1 -\vec{u}_2\,,\\
    \vec{V}_2 &\equiv \vec{u}_2 -\vec{u}_2 =  0\,.
\end{split}
\end{equation}
Next, we consider the centre of mass frame, for which
\begin{equation}
\begin{split}
    &\vec{v}_1 = \vec{u}_1 - \vec{v}_{\mathrm{com}} =  \frac{m_2}{m_1+m_2} ( \vec{u}_1 -\vec{u}_2) \,,\\ 
    &\vec{v}_2 = \vec{u}_2 - \vec{v}_{\mathrm{com}} =  - \frac{m_1}{m_1+m_2}  ( \vec{u}_1 -\vec{u}_2)\,, \\ 
\end{split}
\end{equation}
where
\begin{equation}
\begin{split}
    &\vec{v}_{\mathrm{com}} \equiv \frac{m_1 \vec{u}_1 + m_2 \vec{u}_2}{m_1+m_2}  \,.
\end{split}
\end{equation}
To simplify the calculation, we consider in addition a three-dimensional rotation, defined by the orthogonal matrix
\begin{equation}
    R =
    \begin{pmatrix}
\cos(\alpha) & -\sin(\alpha)\cos(\beta) & \sin(\alpha)\sin(\beta)\\
\sin(\alpha) & \cos(\alpha)\cos(\beta) & -\cos(\alpha)\sin(\beta) \\
0 & \sin(\beta) & \cos(\beta)
\end{pmatrix}\,,
\end{equation}
with
\begin{equation}
\begin{split}
    \beta&= \arctan\left(-\frac{\mathrm{z}}{\mathrm{y}} \right)\,,\\
    \alpha&= \arctan\left(-\frac{\cos(\beta)\mathrm{y}-\sin(\beta)z}{\mathrm{x}} \right)\,.
\end{split}
\end{equation}
In this rotated frame, the initial velocities in the centre of mass frame are aligned along the $x$-axis. The velocity vectors before (un-primed) and after (primed) scattering read as
\begin{equation}
\begin{split}
    \vec{v}_1 &= v_1 \hat{e}_{\mathrm{x}} \\ 
    \vec{v}_2 &= -\frac{m_1}{m_2} v_1 \hat{e}_{\mathrm{x}} \\ 
    \vec{v}_1' &= v_1 \left(\cos(\theta)\hat{e}_{\mathrm{x}}+\sin(\theta)\hat{e}_{\mathrm{y}}\right) \\
    \vec{v}_2' &= \frac{m_1}{m_2} v_1  \left(-\cos(\theta)\hat{e}_{\mathrm{x}}-\sin(\theta)\hat{e}_{\mathrm{y}}\right) \\
    \vec{V}_1 &= \left(1+\frac{m_1}{m_2}\right)v_1 \hat{e}_{\mathrm{x}} \\
    \vec{V}_1' &= v_1  \left(\left(\cos(\theta)+\frac{m_1}{m_2}\right)\hat{e}_{\mathrm{x}}+\sin(\theta)\hat{e}_{\mathrm{y}}\right) \\
    \vec{V}_2' &= \frac{m_1}{m_2} v_1  \left((1-\cos(\theta))\hat{e}_{\mathrm{x}}-\sin(\theta)\hat{e}_{\mathrm{y}}\right), \\
\end{split}
\end{equation}
where $\theta$ is the scattering angle in the centre of mass frame, which in practice is chosen according to the cumulative distribution function in Appendix~\ref{appx: Implementing large-angle anisotropic scattering in rSIDM scheme}.
We note that the angles $\psi$ and $\zeta$ indicated in Fig.~\ref{fig: COM illu} are related to $\theta$ by
\begin{equation}
    \tan(\psi) = \frac{\sin(\theta)}{\cos(\theta)+\frac{m_1}{m_2}}\,,
\end{equation}
and $\zeta=(\uppi-\theta)/2$.
For the choice from above the velocity vectors lie in the $xy$-plane after scattering. To capture the most general case, we rotate them around the $x$-axis by an arbitrary rotation angle $\varphi \in [0,2\uppi]$ using
\begin{equation}
    R_{\mathrm{x}} =
    \begin{pmatrix}
1 & 0 & 0\\
0 & \cos(\varphi) & -\sin(\varphi) \\
0 & \sin(\varphi) & \cos(\varphi)
\end{pmatrix}.
\end{equation}
Now we rotate the vectors back to the initial frame with the inverse matrix $R^{-1}= R^{\mathrm{T}}$ and finally add back the initial velocity $\vec{u}_2$ of the second particle, giving
\begin{equation}
\begin{split}
    \vec{u}_1' &:= \vec{V}_1' + \vec{u}_2\,,\\
    \vec{u}_2' &:= \vec{V}_2' + \vec{u}_2\,.
\end{split}
\end{equation}

\section{Single scattering test}
\label{appendix: Single scattering implementation}

\begin{figure}
    \centering
    \includegraphics[width=\columnwidth]{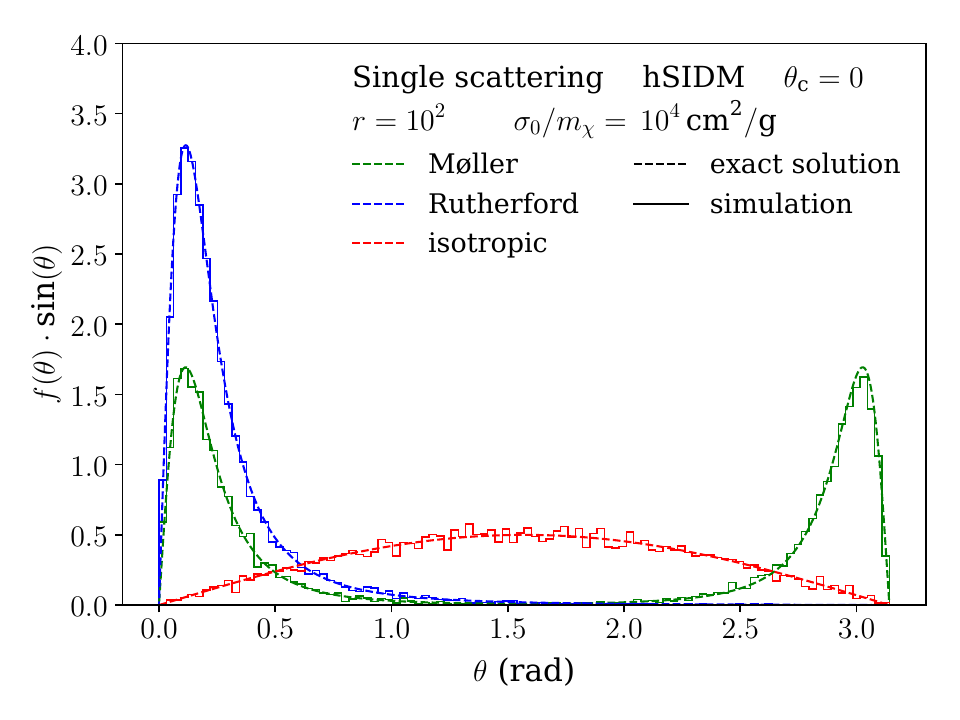}
    \caption{Single scattering test. Within the deflection set-up (see Sect.~\ref{sec: Deflection test}), we restrict the beam particles to undergo just one single scattering throughout the entire simulation, and let it run until all beam particles have scattered once. The solid line corresponds to the hSIDM implementation (which is essentially the anisotropic rSIDM case since we set $\theta_{\mathrm{c}} =0$ here) while the dashed line corresponds to the theoretical prediction. The colours represent the scattering model we use: M{\o}ller (green), Rutherford (blue) and isotropic (red).}
    \label{fig: single scattering test}
\end{figure}

As an initial sanity check of the implementation of anisotropic (large-angle) scattering, we consider the case in which each beam particle within the deflection set-up discussed in Sect.~\ref{sec: Deflection test} is allowed to scatter only exactly {\it once} during the entire duration of the simulation. In this artificial case the angular distribution $f(t,\theta)$ has to approach the underlying differential cross section $\mathrm{d}\sigma/\mathrm{d}\Omega$ for $t\to\infty$. We performed this test for various choices of the differential cross section, and setting $\theta_{\mathrm{c}}=0$ to capture the entire range of scattering angles for this test purpose. The result is shown in Fig.~\ref{fig: single scattering test}. We find good agreement between the simulation and the expected shape in all cases.

\section{No-recoil case implementation}
\label{appendix: No-recoil case implementation}

For some of the validation tests, and in particular to be able to compare to the exact analytical Goudsmit--Saunderson solution Eq.~\eqref{eq: Goudsmit-Saunderson} of the deflection tests problem, we also implement scattering of distinguishable types of particles. Specifically, we consider the case in which `beam' particles have some (finite) mass $m_\chi$, while `target' particles have a much larger mass $M\to\infty$.\footnote{We note that gravity is not included in this test set-up.} In this limit, the recoil transmitted to the target vanishes, i.e.\ target particles remain at rest even after scattering.

Here we describe how we adapt the implementation of both the small- and large-angle scattering regimes to this situation.
For large-angle scattering, we follow the steps described in Appendices~\ref{appx: Implementing large-angle anisotropic scattering in rSIDM scheme} and~\ref{appendix: Vector calculation for anisotropic scattering}, setting $m_1\equiv m_\chi$ and taking the limit $m_2\equiv M\to\infty$. We note that the centre of mass frame and the frame where the heavy test particle is at rest formally coincide in this limit, and that the magnitude of the velocity of the beam particle is unchanged before and after the scattering, such that energy is conserved.

For small-angle scattering, we infer the drag force by considering the deceleration rate of an individual physical DM particle travelling with velocity $v_0$ through a background density $\rho_j$, and undergoing small-angle scatterings. This is the same ansatz as by \cite{Kahlhoefer_2014} in their Appendix~A for the case of equal masses. 

In the no recoil case the modified drag force\footnote{We note that the prefactor differs from Eq.~\eqref{eq:drag} by a factor of four. This arises from the combination of a factor of two due to no-recoil kinematics and another factor 2 due to the replacement $\sigma_{\mathrm{\widetilde{T}}} \to \sigma_{\mathrm{T}}$ involving a different normalisation factor.} is given by
\begin{equation}
    F_{\mathrm{drag}}= |\Delta \vec{v}_{ij}|^2 \frac{\sigma_{\mathrm{T}}}{m_{\chi}} m_i m_j \Lambda_{ij} .
\end{equation}
Furthermore, in re-adding the energy, one should use
\begin{equation}
    \frac{2\Delta E}{m} = |\Delta \vec{v}_{\mathrm{drag}}| \left(2 |\Delta \vec{v}_{ij}| - |\Delta \vec{v}_{\mathrm{drag}}|\right).
\end{equation}
With this, we can modify the fSIDM routine and transform it from recoil to no-recoil case.

\section{Fixed-angle deflection test}
\label{appendix: Fixed-angle deflection test}

\begin{figure*}
    \centering
    \includegraphics[width=\columnwidth]{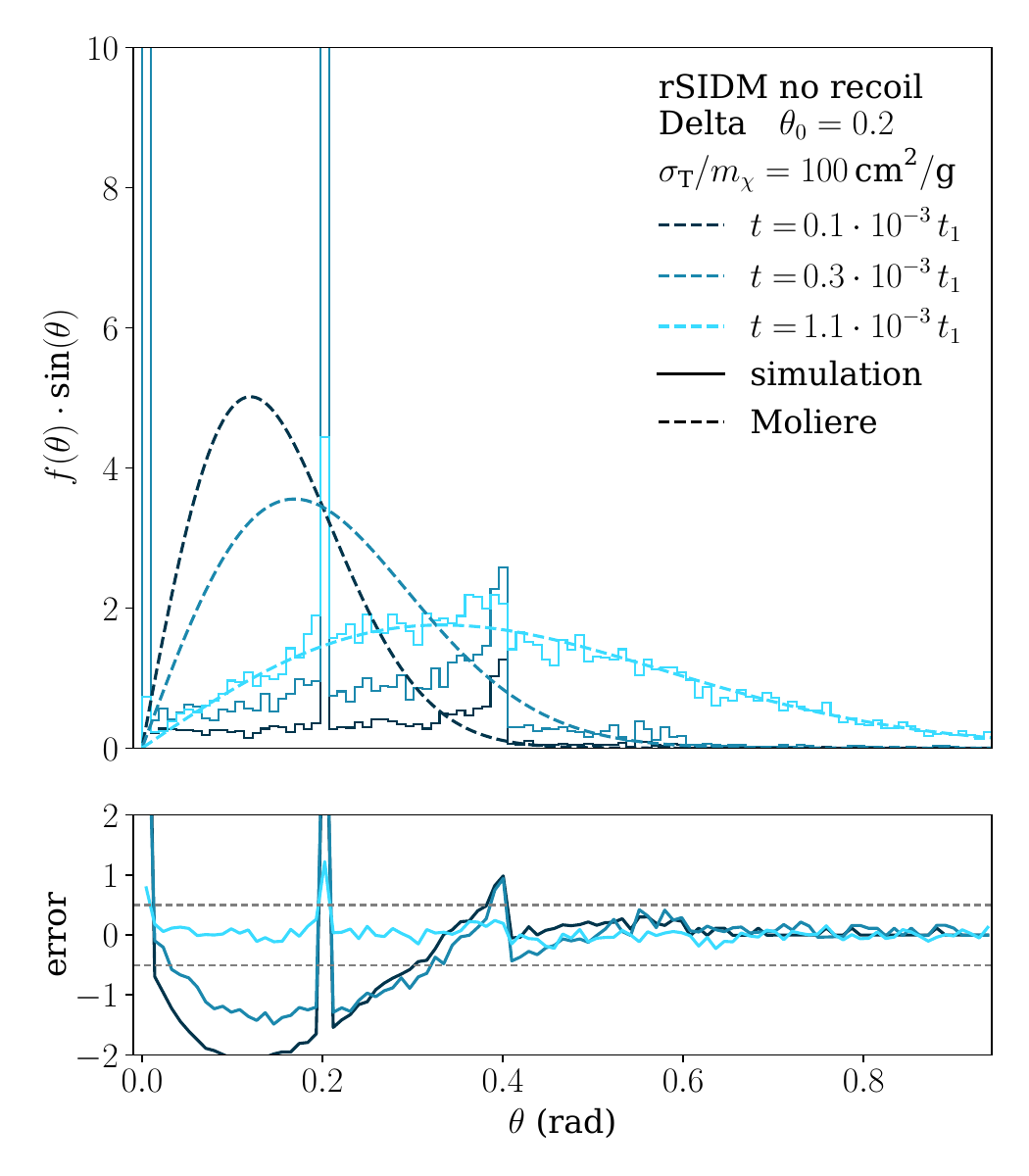}
     \includegraphics[width=\columnwidth]{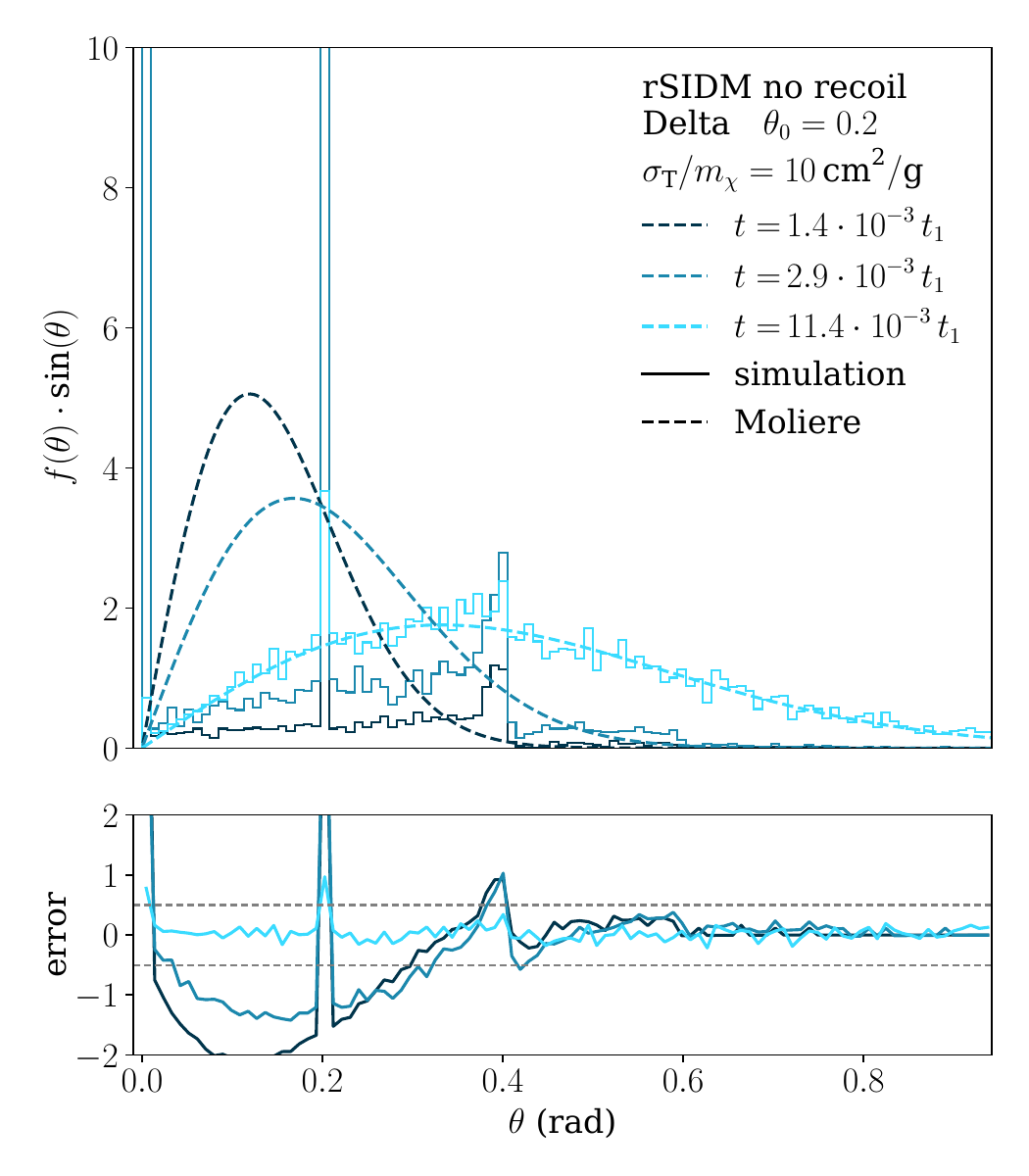}
    \caption{Numerical solutions of the deflection test set-up when assuming a differential cross section Eq.~\eqref{eq: Delta differential cross section} that is peaked at a single value of the deflection angle, given by $\theta_0=0.2$ here. Both panels show the angular distribution function $f(t,\theta)\cdot\sin(\theta)$ at various (early) times $t$, for $\sigma_{\mathrm{T}}/m_\chi=100 (10)\,{\rm cm}^2\,{\rm g}^{-1}$ on the left (right) side. In both cases, the simulation result agrees with the Gaussian Moli\`ere approximation Eq.~\eqref{eq: Moliere} only after some minimal time $t>t_{\rm min}(\theta_0)$. This minimal time is shown in Fig.~\ref{fig: time resolution criterion delta function} in the main text, and compared to the theoretical expectation, finding good agreement. }
    \label{fig: one angle scattering test}
\end{figure*}
In Sect.~\ref{subsec: fixed-angle scattering implementation} we discussed a validation test of the hSIDM approach, using the differential cross section Eq.~\eqref{eq: Delta differential cross section} that is peaked at a single value of the deflection angle $\theta=\theta_0$. Thus, all individual scatterings occur with this fixed angle. We used this approach to check the validity criterion  for the effective treatment of small-angle scattering by a drag force and momentum diffusion derived in Sect.~\ref{sec:fSIDMvalidity}. In the context of the deflection test set-up from Sect.~\ref{subsec: fixed-angle scattering implementation}, this corresponds to the minimal time $t_{\rm min}(\theta_0)$ (see Eq.~\eqref{eq:tmin}) after which the distribution function $f(t,\theta)$ agrees with the Gaussian Moli\`ere approximation Eq.~\eqref{eq: Moliere}. While $t_{\rm min}(\theta_0)$ is shown in Fig.~\ref{fig: time resolution criterion delta function} in the main text, we additionally provide more detailed simulations results of the deflection test problem when assuming a differential cross section given by Eq.~\eqref{eq: Delta differential cross section}. Specifically, we show the resulting angular distribution functions  $f(t,\theta)$ at various times $t$ in Fig.~\ref{fig: one angle scattering test}, and compare them to the Moli\`ere approximation Eq.~\eqref{eq: Moliere}. We see that both agree for late enough times, as stated in the main text.

\section{Multiple scattering and time step}\label{appendix: Multiple scattering and time step}

The sampling of the scattering angle in the rSIDM scheme is based on the assumption that the physical particles represented by the numerical particles do not scatter more than once per pairwise interaction per time step. This assumption is implicitly made when sampling the scattering angle for the two numerical particles from the differential cross section instead of a distribution that takes multiple scatterings into account such as Eq.~\eqref{eq: Goudsmit-Saunderson} for the no-recoil case. In the following, we test which role this assumption plays in the case of our fixed-angle deflection test, see Sect.~\ref{subsec: fixed-angle scattering implementation}.

When considering the pairwise interaction of numerical particles, the size of the time step affects how many of the represented physical particles would scatter. It also alters how many physical particles scatter multiple times relative to only once or not at all. We note that this refers to (1) scatterings of physical particles with in a single time step and (2) scatterings of physical particles belonging to the same pair of numerical particles.

To estimate the probabilities for no and one scatter, we assume that the scattering probability is constant over time: 
\begin{equation}
    \frac{\mathrm{d}P}{\mathrm{d}t} = \mathrm{const} > 0.
\end{equation}
The probability not to scatter if we have only one time step over an interval $\Delta t$ is 
\begin{equation}
    P_{\mathrm{ns}} = 1 - \frac{\mathrm{d}P}{\mathrm{d}t} \Delta t.
\end{equation}
Given that we have $n$ equally spaced time steps of size $\Delta t / n$, it becomes (because of Bernoulli's inequality)
\begin{equation}
    P_{\mathrm{nsm}} = \left( 1 - \frac{\mathrm{d}P}{\mathrm{d}t} \frac{\Delta t}{n} \right)^n > P_{\mathrm{ns}}.
\end{equation}
Here $P_{\mathrm{nsm}} > P_{\mathrm{ns}} $ for $n > 1$. Hence, smaller time steps slightly increase the number of unscattered particles. The probability to scatter once can be expressed as 
\begin{equation}
    P_{\mathrm{os}} = \frac{\mathrm{d}P}{\mathrm{d}t} \Delta t.
\end{equation}
Given that we have $n$ equally spaced time steps, it becomes
\begin{equation} \label{eq:prop_onescattermulti}
    P_{\mathrm{osm}} = n \frac{\mathrm{d}P}{\mathrm{d}t} \frac{\Delta t}{n}  \left( 1 - \frac{\mathrm{d}P}{\mathrm{d}t} \frac{\Delta t}{n} \right)^{n-1}<  P_{\mathrm{os}}.
\end{equation}
Here $P_{\mathrm{osm}} < P_{\mathrm{os}} $ for $n > 1$. Hence, smaller time steps slightly decrease the number of particles that have scattered only once. Furthermore, there is a recent study by \citet{Fischer_2024b} that gives the limit for $n \to \infty$.
We note that these equations are not straightforwardly applicable to the simulations, for example, a numerical particle can interact with multiple neighbours per time step and $N_\mathrm{ngb}$ is not taken into account in the equations. However, the equations provide a qualitative understanding and thus provide insight into the limitations of the simulations.

\begin{figure}
    \centering
    \includegraphics[width=\columnwidth]{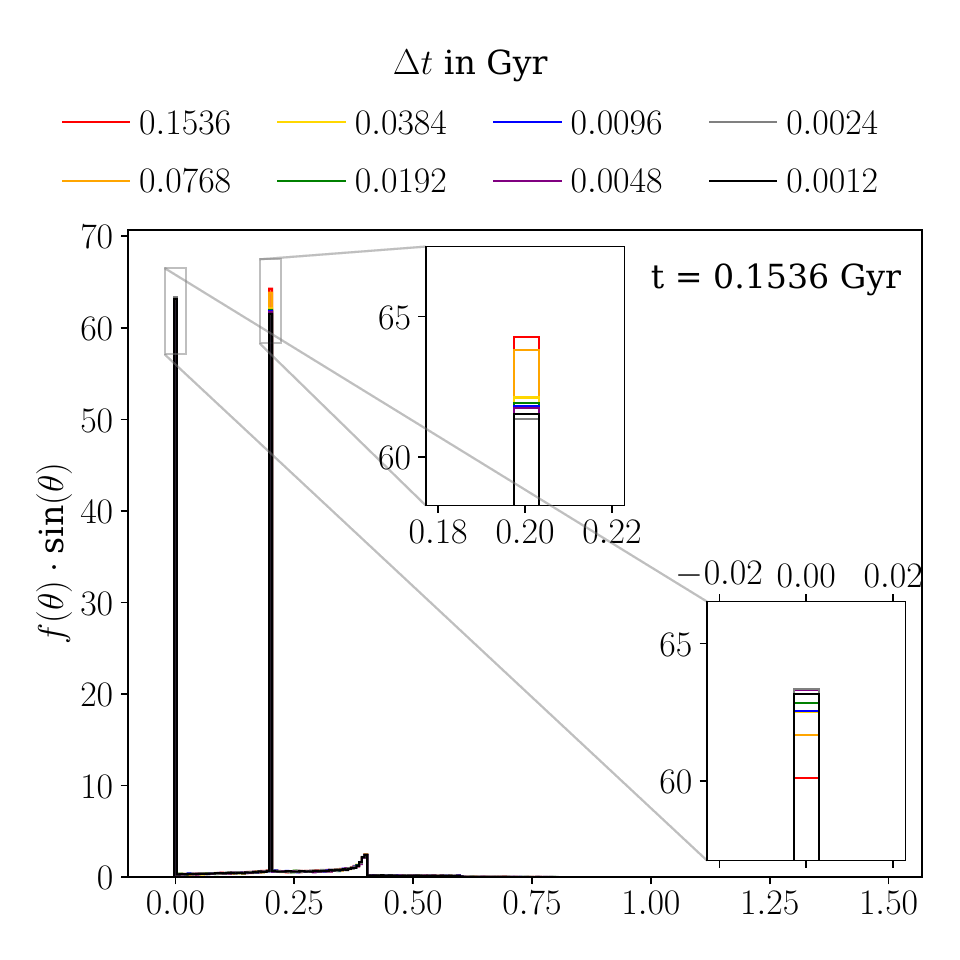}
    \includegraphics[width=\columnwidth]{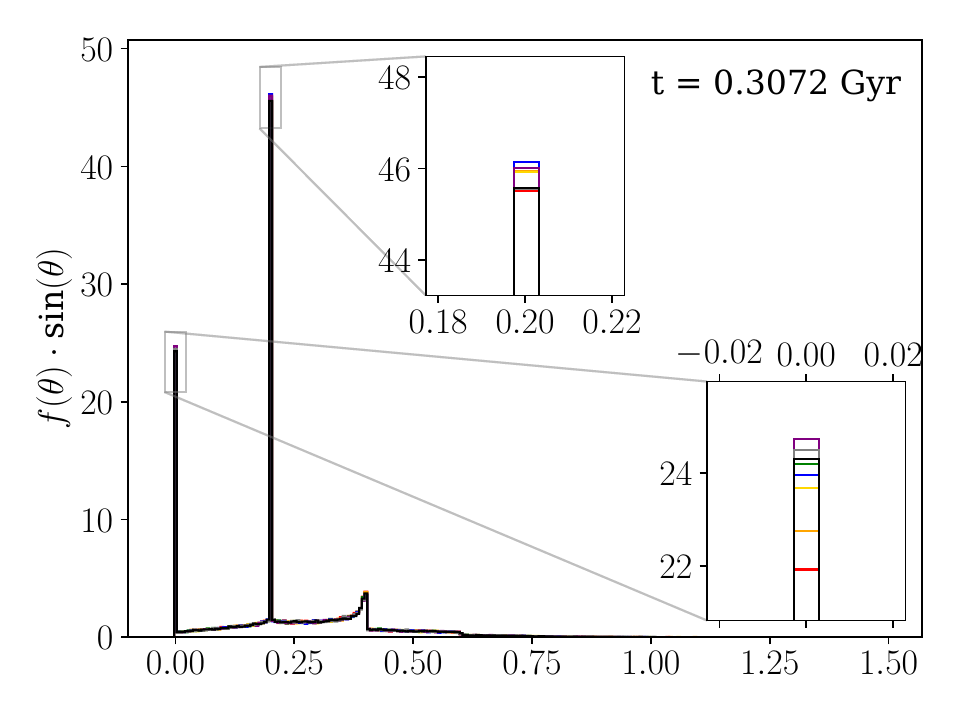}
    \includegraphics[width=\columnwidth]{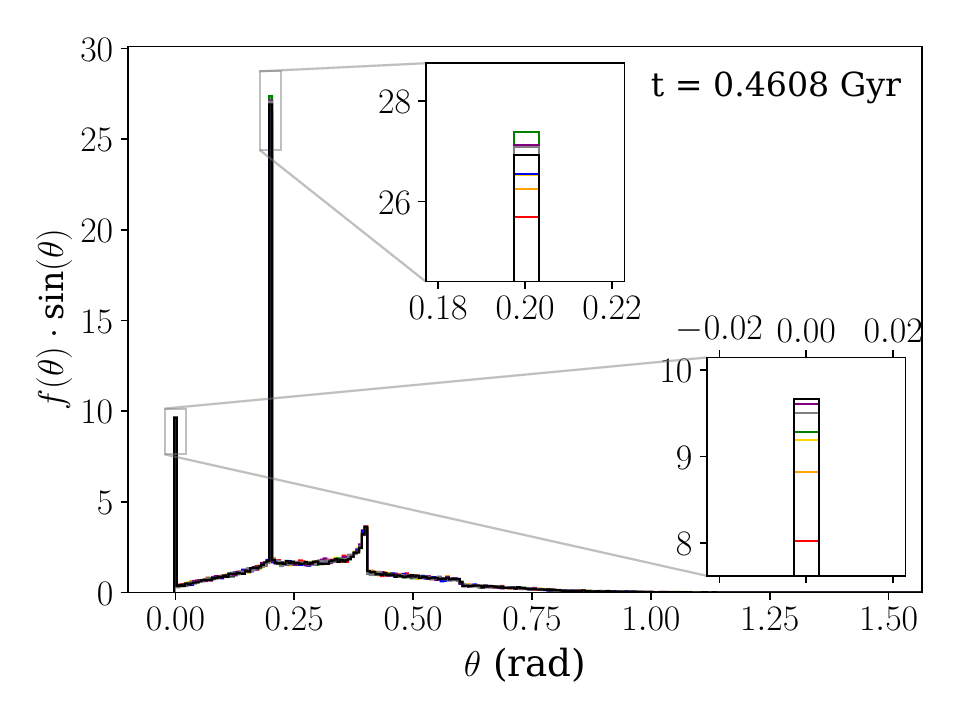}
    \caption{ Fixed-angle deflection test for several time resolutions. The distribution of the deflection angles for runs using a different number of time steps to reach the solution is shown. The three panels give the results for different points in time.}
    \label{fig: Multiple scattering}
\end{figure}

To illustrate these limitations, we show in Fig.~\ref{fig: Multiple scattering} the fixed-angle deflection test simulated with a varying number of time steps for a given time $t$. More precisely, we simulate this deflection test with a fixed angle of $\theta_0 = 0.2$ for a reference time step of $\Delta t = 0.0012$ Gyr and other equally spaced time steps of $\Delta t \in \{ 0.1536, 0.0768, 0.0384, 0.0192, 0.0096, 0.0048, 0.0024\}$ Gyr. We use the same simulation set-up described in Sect.~\ref{subsec: Simulation setup deflection test}, a transfer cross section of $\sigma_{\mathrm{T}}/m_\chi=100 \,{\rm cm}^2\,{\rm g}^{-1}$ and $N_{\mathrm{ngb}}=16$ neighbouring particles. The angle distribution is the (already normalised) mean value of $8$ simulations employing different seeds for the pseudo-random number generator.

The three panels of Fig.~\ref{fig: Multiple scattering} show the distribution of the deflection angle at different times.
The first peak in the plot (from left to right) corresponds to the unscattered particles, the second peak to the particles that have scattered once, and the third peak is mainly particles that have scattered twice.
The height of the first peak decreases over the course of the simulation, while the height of the second peak first increase and later decreases.
It is clearly visible that the height of the peaks depends on the time resolution, that is, how many time steps we use to arrive at the solution. In line with the previous estimates, we find that the number of particles that have not scattered yet increases with increasing time resolution. In contrast, for the particles that have scattered once the picture is more complicated, in an initial phase a better time resolution leads to a smaller peak while at a later stage, it leads to a higher peak. At that late time also the peak height is decreasing with time. We note that these findings are in line with Eq.~\eqref{eq:prop_onescattermulti}.

\begin{figure}
    \centering
    \includegraphics[width=1.1\columnwidth]{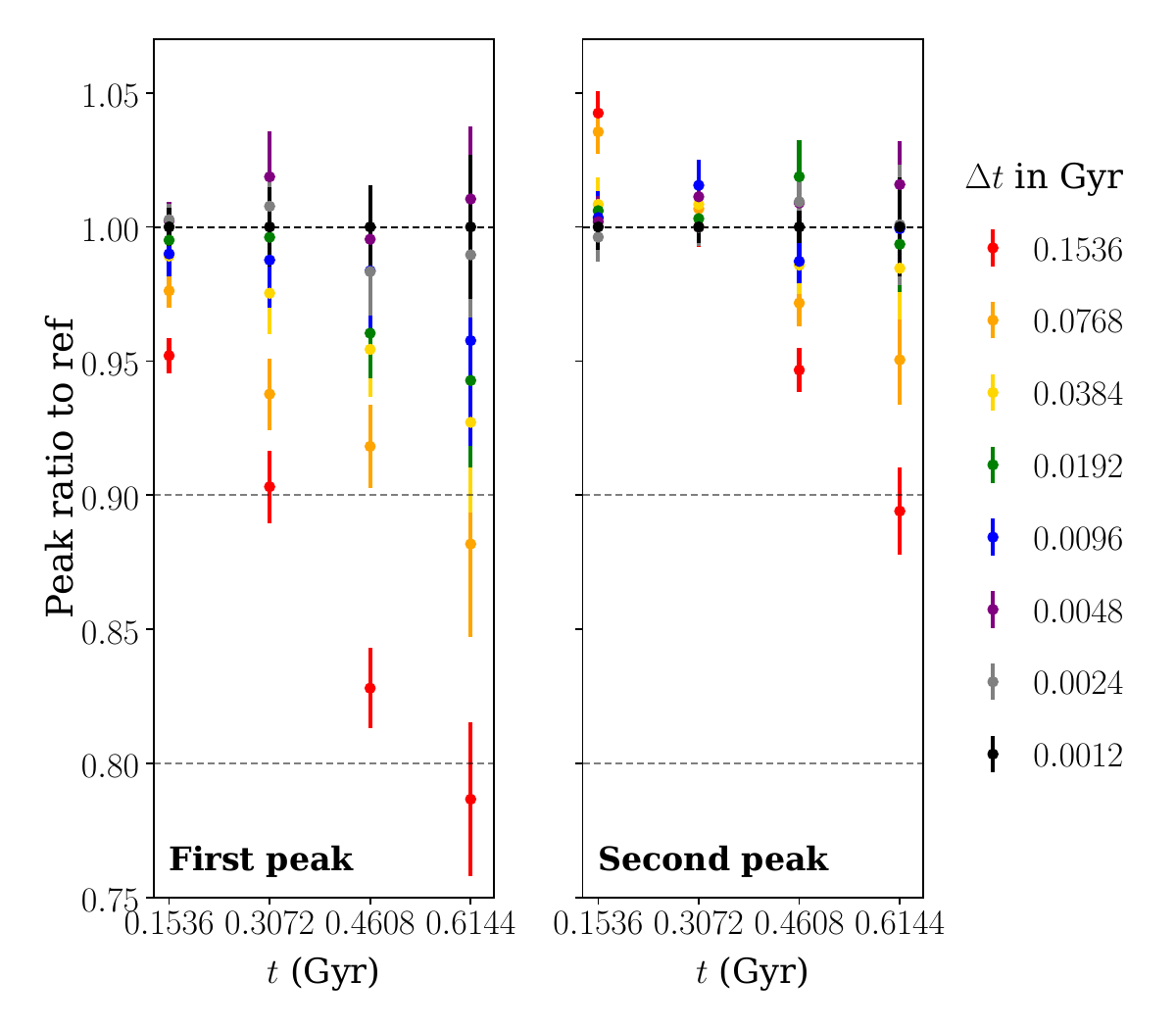}
    \caption{Fixed-angle deflection test with a varying number of time steps to the solution. The ratio of the number of particles in the first peak (i.e.\ particles that have not yet scattered) for the various time resolutions compared to the reference simulation is displayed as a function of time in the left panel. The results for the second peak (i.e.\ particles that have scattered once) are shown in the right panel.
    }
    \label{fig: Multiple scattering_2}
\end{figure}

For each of our simulation times (set by multiples of the largest time step), $t \in \{ 0.1536, 0.3072, 0.4608, 0.6144\}$ Gyr we can compute the ratio of peak heights between the different time resolutions and the reference ($\Delta t = 0.0012$ Gyr) for both peaks to better see the multiple scattering effect. We show the result for the first (second) peak in the left (right) panel of Fig.~\ref{fig: Multiple scattering_2} as a function of time. Again, we see from the first peak that the number of time steps we use clearly influences the number of unscattered particles. For both peaks, we find that the relative difference is increasing with time at the later times we show here.

We can conclude that one needs time steps not only small enough to ensure $P_\mathrm{os} < 1$ but $P_\mathrm{os} \ll 1$ to obtain accurate results, because numerically we can only describe the single scatterings but not the multiple ones of physical particles.
How important this is for a simulation, depends on how crucial resolving the precise angular dependence for the evolution of the SIDM distribution is. While we have chosen a sensitive test, this problem might be less relevant for typical astrophysical simulations. In consequence, the optimal choice of the time step to achieve a good trade-off between accuracy and performance might be problem dependent. Thus the value for $\kappa$ in the time step criterion (Eq.~\eqref{eq: time steps in rSIDM}), which effectively limits the interaction probability, could be chosen depending on the problem at hand.

\end{appendix}

\end{document}